\documentclass{theoretics}

\usepackage{dsfont}
\usepackage{verbatim}
\usepackage[capitalise,noabbrev,nameinlink]{cleveref}

 \crefname{appsec}{Appendix}{Appendices}
\usepackage{tikz}

 \crefname{lemma}{Lemma}{Lemmas}
 \crefname{definition}{Definition}{Definitions}
 \crefname{fact}{Fact}{Facts}
\crefname{claim}{Claim}{Claims}
 \crefname{proposition}{Proposition}{Propositions}

\DeclareMathOperator{\odd}{odd}

\newcommand{\one}{\mathds{1}}

\newcommand{\norm}[1]{\left\lVert #1 \right\rVert}

\newcommand{\grad}{\nabla}

\newcommand{\poly}{\mathrm{poly}}
\newcommand{\dist}{\mathrm{dist}}
\newcommand{\sgn}{\mathrm{sgn}}
\newcommand{\fpras}{\mathsf{FPRAS}}

\newcommand{\fptas}{\mathsf{FPTAS}}

\newcommand{\eps}{\varepsilon}
\newcommand{\ccomp}{\mathsf{c}}

\newcommand{\N}{\mathbb{N}}

\newcommand{\R}{\mathbb{R}}
\newcommand{\C}{\mathbb{C}}

\newcommand{\D}{\mathbb{D}}
\newcommand{\HP}{\mathbb{H}}

\newcommand{\boldlam}{\boldsymbol{\lambda}}
\newcommand{\boldone}{\boldsymbol{1}}
\newcommand{\weight}{w}
\newcommand{\VSpairs}{\mathcal{P}}
\newcommand{\Pinning}{\mathcal{T}}

\newcommand{\interior}[1]{#1^{\mathrm{o}}}
\newcommand{\closure}[1]{\overline{#1}}
\newcommand{\boundary}[1]{\partial #1}

\newcommand{\eigmax}{\mathsf{EigMax}}

\newcommand{\image}{\mathrm{image}}

\newcommand{\Arg}{\mathrm{Arg}}

\newcommand{\pp}{b}
\newcommand{\spa}{j}
\newcommand{\spb}{k}

\renewcommand{\epsilon}{\varepsilon}

\title{Spectral Independence via Stability and Applications to Holant-Type Problems}

 \ThCSshorttitle{Spectral Independence via Stability and Applications to Holant-Type Problems}

 \ThCSshortnames{Z.~Chen, K.~Liu, E.~Vigoda}

\ThCSauthor[1]{Zongchen Chen}{chenzongchen@gatech.edu}[]
\ThCSauthor[2]{Kuikui Liu}{liukui@mit.edu}[]
\ThCSauthor[3]{Eric Vigoda}{vigoda@ucsb.edu}[]

\ThCSaffil[1]{Georgia Institute of Technology, Atlanta, GA, USA} 
\ThCSaffil[2]{Massachusetts Institute of Technology, Cambridge, MA, USA}
\ThCSaffil[3]{University of California, Santa Barbara, Santa Barbara, CA, USA}

\ThCSthanks{An extended abstract of this work appeared at the 62nd IEEE Annual Symposium on Foundations of Computer Science (FOCS 2021)~\cite{ChenLV21}. Kuikui Liu is supported by NSF grant CCF-1907845 and ONR-YIP grant N00014-17-1-2429. Eric Vigoda is supported by NSF grant CCF-2147094. }

\ThCSyear{2024}
\ThCSarticlenum{16}
\ThCSreceived{Sep 18, 2022}
\ThCSrevised{Feb 4, 2024}
\ThCSaccepted{Mar 15, 2024}
\ThCSpublished{Jul 12, 2024}
\ThCSdoicreatedtrue

\ThCSkeywords{Glauber dynamics, spectral independence, stability of polynomials, partition function, Holant problem}

\begin{document}

\maketitle

\begin{abstract}
This paper formalizes connections between stability of polynomials and convergence rates of Markov Chain Monte Carlo (MCMC) algorithms. We prove that if a (multivariate) partition function is nonzero in a region around a real point $\lambda$ then spectral independence holds at $\lambda$. As a consequence, for Holant-type problems (e.g., spin systems) on bounded-degree graphs, we obtain optimal $O(n\log n)$ mixing time bounds for the single-site update Markov chain known as the Glauber dynamics. Our result significantly improves the running time guarantees obtained via the polynomial interpolation method of Barvinok (2017), refined by Patel and Regts (2017).

There are a variety of applications of our results. In this paper, we focus on Holant-type (i.e., edge-coloring) problems, including weighted edge covers and weighted even subgraphs. For the weighted edge cover problem (and several natural generalizations) we obtain an $O(n\log{n})$ sampling algorithm on bounded-degree graphs. The even subgraphs problem corresponds to the high-temperature expansion of the ferromagnetic Ising model. We obtain an $O(n\log{n})$ sampling algorithm for the ferromagnetic Ising model with a nonzero external field on bounded-degree graphs, which improves upon the classical result of Jerrum and Sinclair (1993) for this class of graphs. We obtain further applications to antiferromagnetic two-spin models on line graphs, weighted graph homomorphisms, tensor networks, and more.


\end{abstract}

\section{Introduction}

A fundamental problem in a variety of settings, such as the study of spin systems in statistical physics or Bayesian inference in undirected graphical models, is the counting problem of estimating a partition function or 
the sampling problem of generating a random sample from the associated Gibbs distribution.
The classical tool for approximate counting/sampling problems is the Markov Chain Monte Carlo (MCMC) method.
There are now several alternative algorithmic approaches, most notably the 
correlation decay approach presented by Weitz~\cite{Wei06} and the polynomial interpolation method 
presented by Barvinok~\cite{Bar17book} and refined by Patel and Regts~\cite{PR17}.

Anari, Liu, and Oveis Gharan~\cite{ALO20} presented a powerful new tool for analyzing MCMC methods known as spectral independence.
Spectral independence yields optimal mixing time bounds for the Glauber dynamics (which updates a randomly chosen vertex in each step)~\cite{CLV21},
and more generally yields optimal mixing time bounds for any block dynamics and the Swendsen-Wang dynamics~\cite{BCCPSV21}.

Spectral independence serves as an important technical bridge: for many results established with different approaches, their proofs also 
imply spectral independence and as a consequence we obtain much stronger running time guarantees.
In particular, in~\cite{CLV20} it was shown that potential function techniques used for analyzing correlation decay algorithms also
imply spectral independence; see also \cite{CGSV21,FGYZ21}. More recent works~\cite{BCCPSV21,Liu21} prove that path coupling arguments for a broad class of chains,
including any local chain, yields spectral independence.  

In this paper we prove, in an almost black-box fashion,
that methods for establishing large zero-free regions needed for the polynomial interpolation method also yield spectral independence. 
The polynomial interpolation method is a mathematically elegant approach which works in the following manner. 
To approximate a partition function at a positive real value $\lambda$, one needs to prove there is a zero-free region around $\lambda$ in the complex plane
which means that the partition function has no roots in an open connected region (in the complex plane) containing $\lambda$. 
If such a zero-free region also contains an ``easy" point (usually 0) at which the partition function and its derivatives can be efficiently evaluated, 
then this implies that one can approximate the Taylor series of a simple transformation of the partition function using 
a logarithmic number of terms, which yields a polynomial-time algorithm to approximate the partition function at $\lambda$.

We prove that a zero-free region implies spectral independence. 
This immediately yields several new rapid mixing results for MCMC methods. 
We also obtain significantly improved running times in many instances. 
For a spin system on a graph with $n$ vertices and constant
maximum degree~$\Delta$, the polynomial interpolation method~\cite{PR17} yields a running time of $O(n^C)$ where the constant
$C$ depends on $\Delta$ and parameters of the model.  In contrast, spectral independence implies an optimal mixing time bound of  $O(n\log{n})$ 
for the Glauber dynamics~\cite{CLV21} (and more generally optimal mixing for the block dynamics~\cite{BCCPSV21}). 

The study of zero-free regions has a long and rich history in the analysis of phase transitions in statistical physics models. 
The well-known work of Lee and Yang~\cite{LY} utilizes zeros of the partition function to study phase transitions for the ferromagnetic Ising model; see \cref{rem:LY} for a discussion of the Lee-Yang Theorem.

Before stating our results we formally define the Glauber dynamics, also known as the Gibbs sampler.  
Let $V$ be a finite set and let $Q$ be a finite label/spin set. 
Consider a distribution $\mu$ over $Q^V$ containing all labelings of elements in $V$. 
Note that this includes spin systems and Holant problems as special cases: 
for spin systems $\mu$ is the distribution over all spin assignments of vertices of the underlying graph, and for Holant problems $\mu$ is the distribution over all subsets of edges ($\{0,1\}$-labelings of edges).
A transition of the Glauber dynamics chooses a random element $x\in V$, and then resamples the label of $x$ from the marginal distribution at $x$ conditional on the configuration outside $x$. 
The mixing time is the number of steps, from the worst initial state, which is guaranteed to be close (in total variation distance) to the desired distribution $\mu$.

\subsection{Applications}

We state here three sample applications of our techniques;
further applications are stated later in the paper.

For a graph $G=(V,E)$, we say a vertex $v$ is covered by a subset $S\subseteq E$ of edges if $v$ is incident to at least one edge in $S$. 
The subset $S\subseteq E$ is called an \emph{edge cover} if all vertices are covered by $S$. 
Note there is always a trivial edge cover by setting $S=E$. 
An $\fpras$ (fully polynomial randomized approximation scheme) was presented for counting the number of edge covers for 3-regular graphs~\cite{BR09}. 
In~\cite{LLL} an $\fptas$ (deterministic analog of an $\fpras$) for counting edge covers was presented for all graphs using the correlation decay approach, and the running time was $O(m^{1+\log_2{6}}n^2)$ where $m$ is the number of edges and $n$ is the number of vertices. 
An $\fpras$ for all graphs using MCMC was presented in~\cite{HLZ16}.

The correlation decay algorithm of~\cite{LLL} was extended to
weighted (partial) edge covers (with worse running time guarantees) in~\cite{LLZ14}. 
In the weighted version, each edge has a weight $\lambda>0$ and each vertex receives a penalty $\rho\in [0,1]$ for being uncovered.
Every subset $S \subseteq E$ is associated with the weight $w(S) = \rho^{|\mathrm{unc}(S)|} \lambda^{|S|}$, where $\mathrm{unc}(S)$ denotes the set of vertices that are not covered by $S$. 
The Gibbs distribution over all subsets of edges is given by $\mu(S) \propto w(S)$.
Note, the case $\lambda = 1$ and $\rho = 0$ corresponds to uniformly random exact edge covers. 

Finally, an $\fptas$ using the polynomial interpolation algorithm was presented for graphs with constant maximum degree~\cite{GLLZ21}, see also \cite{BCR20}. 
Using the zero-free results in~\cite{GLLZ21} with our new technical contributions we immediately obtain an $\fpras$ using a simple MCMC algorithm and with significantly faster running time guarantees. 

\begin{theorem}[Weighted Edge Covers]
\label{thm:edge_covers}
Let $\Delta \ge 3$ be an integer and let $\lambda > 0$, $\rho \in [0,1]$ be reals. 
Then for any $n$-vertex graph $G=(V,E)$ of maximum degree $\Delta$, 
the Glauber dynamics for sampling random weighted edge covers of $G$ with parameters $\lambda,\rho$ mixes in $C n \log n$ steps where $C = C(\Delta,\lambda,\rho)$ is a constant independent of $n$. 
\end{theorem}

One of the seminal results in the field of approximate counting is the work of Jerrum and Sinclair~\cite{JSising} presenting an $\fpras$
for the partition function of the ferromagnetic Ising model on any graph.  
The Ising model on a graph $G=(V,E)$ is described by two parameters, the edge activity $\beta_{\mathrm{Ising}} >0$ and the vertex activity $\lambda_{\mathrm{Ising}} >0$. The Gibbs distribution of the Ising model is over all $\{+,-\}$ spin assignments to vertices. Every configuration $\sigma: V \to \{+,-\}$ has density $\mu_{\mathrm{Ising}}(\sigma) \propto \beta_{\mathrm{Ising}}^{m(\sigma)} \lambda_{\mathrm{Ising}}^{|\sigma^{-1}(+)|}$ where $m(\sigma)$ denotes the number of monochromatic edges in $\sigma$ and $\sigma^{-1}(+)$ is the set of vertices assigned spin $+$. The model is ferromagnetic when $\beta_{\mathrm{Ising}}>1$, in which case neighboring vertices are more likely to have the same spin.

The central task of the Jerrum-Sinclair algorithm is sampling from the Gibbs distribution for the high-temperature expansion of the Ising model which is defined on all subsets of edges weighted to prefer subgraphs with more even degree vertices.
For a graph $G=(V,E)$, an edge weight $\lambda > 0$, and a vertex penalty $\rho \in [0,1]$, the Gibbs distribution $\mu$ for weighted (partial) even subgraphs is defined on all subsets of edges; a subset $S\subseteq E$ has weight $w(S) = \rho^{|\odd(S)|} \lambda^{|S|}$ where $\odd(S)$ is the set of odd-degree vertices in the subgraph $(V,S)$, and $\mu(S) \propto w(S)$. 
The weighted even subgraphs model where $\lambda \in (0,1)$ is closely related to the ferromagnetic Ising model by $\beta_{\mathrm{Ising}} = \frac{1+\lambda}{1-\lambda}$ and $\lambda_{\mathrm{Ising}} = \frac{1+\rho}{1-\rho}$. 
Note that if $\rho = 0$ then $\mu$ is the distribution over all weighted exact even subgraphs, corresponding to the ferromagnetic Ising model without external fields (i.e., $\lambda_{\mathrm{Ising}} = 1$). 
Using the Jerrum-Sinclair algorithm, Randall and Wilson \cite{RW99} first gave an approximate sampler for the ferromagnetic Ising model via an equivalence with the random cluster model \cite{FK72} and using self-reducibility.
Grimmett and Janson discovered a direct coupling between weighted even subgraphs and the random cluster model \cite[Theorem 3.5]{GJ09}, which together with the coupling between random cluster and Ising models \cite{ES88} yields a simpler and more efficient sampler for Ising configurations. 
In a recent work \cite{FGW23}, a grand coupling among the ferromagnetic Ising model, random cluster model, and weighted even subgraphs is presented when external fields exist.

In~\cite{JSising}, an MCMC algorithm is presented to sample weighted even subgraphs of an arbitrary (unbounded-degree) graph in time $O(m^3\poly(1/\rho))$ where $m$ is the number of edges. 
In another direction, \cite{LSS19} presents an $\fptas$ for approximating the partition function of the ferromagnetic Ising model with nonzero fields on bounded-degree graphs, using Barvinok's polynomial interpolation method and the Lee-Yang theory.
As is common for this type of approach, the running time of \cite{LSS19} is $n^C$ for a constant $C$ depending on the maximum degree of the graph and the parameters of the Ising model.

Here we use our results relating zero-free regions and spectral independence to obtain a faster MCMC algorithm 
for bounded-degree graphs when $\rho>0$.

\begin{theorem}[Weighted Even Subgraphs]
\label{thm:even_subgraphs}
Let $\Delta \ge 3$ be an integer and let $\lambda > 0$, $\rho \in (0,1]$ be reals. 
Then for any $n$-vertex graph $G=(V,E)$ of maximum degree $\Delta$, 
the Glauber dynamics for sampling random weighted even subgraphs of $G$ with parameters $\lambda,\rho$ mixes in $C n \log n$ steps where $C = C(\Delta,\lambda,\rho)$ is a constant independent of $n$. 

In particular, this gives an approximate sampling algorithm with running time $O(n\log n)$ for the ferromagnetic Ising model with edge activity $\beta_{\mathrm{Ising}} = \frac{1+\lambda}{1-\lambda}$ and vertex activity $\lambda_{\mathrm{Ising}} = \frac{1+\rho}{1-\rho}$. 
\end{theorem}

\begin{remark}
In \cite{JSising}, the MCMC method can actually be used to obtain a sampler for $\rho = 0$ corresponding to weighted exact even subgraphs.
This is achieved by taking $\rho = 1/n$ and using rejection sampling. 
Notice that the running time of \cite{JSising} is polynomial in $1/\rho$, and therefore this gives a $\poly(n)$ time algorithm for sampling weighted exact even subgraphs and hence for the ferromagnetic Ising model without fields. 
Unfortunately, \cref{thm:even_subgraphs} cannot be used to obtain a sampler for $\rho = 0$, since our bound on the mixing time of the Glauber dynamics (the constant $C$ from \cref{thm:even_subgraphs}) depends exponentially on $1/\rho$.
\end{remark}

Finally, we simultaneously generalize \cite{JSmatchings, DHJM21, BCR20} to all antiferromagnetic two-spin edge models, i.e., antiferromagnetic two-spin models on the class of line graphs. Again, in the bounded-degree regime we obtain optimal mixing times. Before we state the result, let us define the model more precisely. For a graph $G=(V,E)$ and fixed parameters $\beta \ge 0$, $\gamma > 0$, $\lambda > 0$, the Gibbs distribution of the corresponding two-spin edge model on $G$ is given by
\begin{align}\label{eq:twospinmodel}
    \mu(\sigma) \propto \beta^{m_{1}(\sigma)}\gamma^{m_{0}(\sigma)}\lambda^{|\sigma^{-1}(1)|}, \quad \forall \sigma \in \{0,1\}^{E}
\end{align}
where $m_{i}(\sigma)$ denotes the number of pairs of edges $e,f$ sharing a single endpoint such that $\sigma(e) = \sigma(f) = i$, for each $i=0,1$. We say the system is antiferromagnetic if $\beta\gamma < 1$ and ferromagnetic if $\beta\gamma > 1$ (note that $\beta\gamma = 1$ corresponds to a trivial product measure). The case $\beta = 0$ and $\gamma = 1$ recovers the monomer-dimer model for matchings weighted by $\lambda$, and the case $\beta = \gamma$ recovers the Ising model on the line graph of $G$.

\begin{theorem}[Antiferromagnetic Two-Spin Edge Models]
\label{thm:anti_2-spin_edge}
Let $\Delta \ge 3$ be an integer and let $\beta \ge 0$, $\gamma > 0$, $\lambda > 0$ be reals such that $\beta\gamma < 1$. 
Then for any $n$-vertex graph $G=(V,E)$ of maximum degree $\Delta$, the Glauber dynamics for sampling from the antiferromagnetic two-spin edge model on $G$ with parameters $\beta,\gamma,\lambda$ mixes in $C n\log n$ steps where $C = C(\Delta,\beta,\gamma,\lambda)$ is a constant independent of $n$.
\end{theorem}

We present further applications of our methods in \cref{sec:homtensornetwork,sec:fourier}.

\subsection{Spectral Independence via Stability of the Partition Function}

We need a few preliminary definitions before formally stating our technical results. 
Our results hold for an arbitrary distribution on a discrete product space;
this general setup contains spin systems as a special case.
Let $V$ be a finite set and we refer to the elements in $V$ as vertices. 
For an integer $q\ge 2$, the set of spins is $Q = \{0\} \cup Q_1$ where $Q_1 = \{1,\dots,q-1\}$.
The state space is $\Omega = Q^V$, the collection of all spin assignments of vertices.
Finally, let $\weight: \Omega \to \R_{\ge 0}$ be a nonnegative weight function that is not always zero; i.e., $\weight(\sigma) > 0$ for at least one $\sigma \in \Omega$.

Let $\boldlam: V \times Q_1 \to \C$ be a vector of (complex) external fields; each $\lambda_{v,\spb}$ represents the weight of vertex $v$ receiving spin $\spb$. 
Without loss of generality, we assume spin~$0$ does not have associated external fields (this can be obtained by normalizing).
Given $\weight$, the \emph{partition function} is a multivariate polynomial of $\boldlam$ defined as:
\begin{equation}\label{eq:partition_func}
Z_\weight(\boldlam)
= \sum_{\sigma \in \Omega} \weight(\sigma) \boldlam^\sigma,
\quad\text{where}~~
\boldlam^\sigma = \prod_{v \in V:\, \sigma_v \neq 0} \lambda_{v, \sigma_v}.
\end{equation}
If $\boldlam$ is real and positive (i.e., every $\lambda_{v,\spb} \in \R_+$), then the \emph{Gibbs distribution} $\mu = \mu_{\weight,\boldlam}$ is given by:
\begin{equation}\label{eq:gibbs}
\mu(\sigma) = \frac{\weight(\sigma) \boldlam^\sigma}{Z_\weight(\boldlam)}, 
\quad \forall \sigma \in \Omega.
\end{equation}
Note that $Z_\weight(\boldlam)>0$ since $\weight$ is not identically zero. 

To establish spectral independence we need to consider the model with an arbitrary ``pinning'' which is a fixed configuration on 
an arbitrary subset of vertices. 
We formally define pinnings and the associated notions in \cref{subsec:pinning}, and introduce the relevant notation here.

A configuration $\sigma \in \Omega$ is said to be valid or feasible if $\weight(\sigma) > 0$.
For $\Lambda\subseteq V$, let $\Omega_\Lambda$ denote the set of pinnings on $\Lambda$; this is the set of configurations on
$\Lambda$ which have a valid extension to the remaining vertices $V\setminus\Lambda$. 
For $\Lambda\subseteq V$ and $\tau\in\Omega_\Lambda$, let $V^\tau = V \setminus \Lambda$ denote the set of unpinned vertices, 
let $Z_\weight^\tau(\boldlam)$ be the multivariate conditional partition function under the pinning $\tau$, 
and let $\mu^\tau$ be the corresponding conditional distribution.

We can now define the notion of spectral independence.
Let $\Pinning = \bigcup_{\Lambda \subseteq V} \Omega_\Lambda$ be the collection of all pinnings. 
For $\tau \in \Pinning$ let $\VSpairs^\tau = \{(v,\spb) \in V \times Q: v \in V^\tau, \spb \in \Omega_v^\tau\}$ 
be the collection of feasible vertex-spin pairs under $\tau$, where $\Omega_v^\tau$ represents the set of feasible spins at $v$ conditioned on $\tau$. 
The following definition is taken from \cite{CGSV21}; see also \cite{ALO20,FGYZ21}.

\begin{definition}[Influence Matrix]
\label{def:inf-matrix}
Let $\tau \in \Pinning$ be an arbitrary pinning. 
For every $(u,\spa),(v,\spb) \in \VSpairs^\tau$, the (pairwise) influence of $(u,\spa)$ on $(v,\spb)$ under the pinning $\tau$ is given by $\Psi_\mu^\tau(u,\spa; v,\spb) = 0$ for $u = v$ and
\[
\Psi_\mu^\tau(u,\spa; v,\spb) = \mu(\sigma_v = \spb \mid \sigma_u = \spa, \sigma_\Lambda = \tau) - \mu(\sigma_v = \spb \mid \sigma_\Lambda = \tau) \quad \text{for $u \neq v$}.
\] 
The (pairwise) influence matrix $\Psi_\mu^\tau$ is a $|\VSpairs^\tau| \times |\VSpairs^\tau|$ matrix with entries given above. 
\end{definition}

All eigenvalues of the influence matrix $\Psi_\mu^\tau$ are real since $\Psi_\mu^\tau$ becomes symmetric after left-multiplication by a suitable diagonal matrix; see \cite{ALO20,CGSV21,FGYZ21}. 
For a square matrix $M$ with real eigenvalues, let $\eigmax(M)$ denote the maximum eigenvalue of $M$. 

\begin{definition}[Spectral Independence]
\label{def:SI}
We say $\mu$ is spectrally independent with constant $\eta$ if for every pinning $\tau \in \Pinning$ one has
\[
\eigmax(\Psi_\mu^\tau) \le \eta. 
\]
\end{definition}

For a non-empty region $\Gamma$ of the complex plane, we say that a multivariate polynomial $P(z_1,\dots,z_n)$ is \emph{$\Gamma$-stable} if $P(z_1,\dots,z_n) \neq 0$ whenever $z_j \in \Gamma$ for all $j$, 
see \cref{def:stable}.
We present a sequence of results connecting spectral independence of the distribution with stability of the partition function.
Our first result holds when the zero-free region of the partition function is sufficiently ``large'', e.g., containing the whole positive real axis. 
Below for $\Gamma \subseteq \C$ let $\closure{\Gamma}$ denote the closure of $\Gamma$ and let $\boundary{\Gamma}$ be the boundary of $\Gamma$; for $\lambda \in \C$ let $\dist(\lambda, \partial \Gamma) = \inf_{z \in \partial \Gamma} |z - \lambda|$; see \cref{subsec:plane_complex}.

\begin{theorem}
\label{thm:main_R+}
Let $\Gamma \subseteq \C$ be a non-empty open connected region such that $\Gamma$ is unbounded and $0 \in \closure{\Gamma}$. 
If the multivariate partition function $Z_\weight$ is $\Gamma$-stable, then for any $\lambda \in \R_+ \cap \Gamma$ the Gibbs distribution $\mu = \mu_{\weight,\lambda}$ with the uniform external field $\lambda$ is spectrally independent with constant
\[
\eta = \frac{8}{\delta}
\] 
where $\delta = \frac{1}{\lambda} \dist(\lambda, \partial \Gamma)$.

In particular, the statement is true when $\Gamma$ is a non-empty open connected region containing the positive real axis; i.e., $\R_+ \subseteq \Gamma$. 
\end{theorem}

We can also obtain bounds when $\Gamma$ is bounded or $0 \in \closure{\Gamma}$, under the assumption that $\Gamma$ contains a part of the positive real axis. In this case we also need to further assume that all conditional partition functions under pinnings are stable. 
Moreover, our bound on spectral independence depends on the marginal bound of the distribution $\mu$, which is defined as
\[
\pp = \min_{\substack{ \tau \in \Pinning\\(v,\spb) \in \VSpairs^\tau}} \mu^\tau(\sigma_v = \spb).
\]
Note that $\pp>0$ since $\VSpairs^\tau$ contains only feasible vertex-spin pairs. 

\begin{theorem}
\label{thm:main_lambda_c}
Let $\lambda^* \in \R_+$ and let $\Gamma \subseteq \C$ be a non-empty open connected region such that $(0,\lambda^*) \subseteq \Gamma$ (respectively, $(\lambda^*,\infty) \subseteq \Gamma$). 
If for every pinning $\tau \in \Pinning$ the multivariate conditional partition function $Z_\weight^\tau$ is $\Gamma$-stable, 
then for any $\lambda \in (0,\lambda^*)$ (respectively, $\lambda \in (\lambda^*, \infty)$) the Gibbs distribution $\mu = \mu_{\weight,\lambda}$ with the uniform external field $\lambda$ is spectrally independent with constant
\[
\eta = \frac{8}{\delta}  \min\left\{ \frac{1-\pp}{\pp},\, \frac{\lambda}{\pp(\lambda^*-\lambda)} + 1 \right\} \\
\]
\[
\left( 
\text{respectively,~~} 
\eta = \frac{8}{\delta}  \min\left\{ \frac{1-\pp}{\pp},\, \frac{\lambda^*}{\pp(\lambda-\lambda^*)} + 1 \right\}
\right)
\]
where $\pp$ is the marginal bound for $\mu$ and $\delta = \frac{1}{\lambda} \dist(\lambda, \partial \Gamma)$. 
\end{theorem}

\begin{remark}\label{rmk:robust}
The first term $(1-\pp)/\pp$ is better when $\lambda$ is close to $\lambda^*$, while the second term is better when $\lambda$ is close to $0$ (respectively, $\infty$), because usually $\pp/\lambda$ is bounded from below when $\lambda \to 0$ (respectively, $\pp \lambda$ is bounded from below when $\lambda \to \infty$).  
\end{remark}
\begin{remark}\label{rem:LY}
We point out here that \cref{thm:main_lambda_c} does not apply to the ferromagnetic Ising model. 
The celebrated Lee-Yang theorem states that the partition function for the ferromagnetic Ising model is $\D(0,1)$-stable and $\closure{\D}(0,1)^\ccomp$-stable where $\D(0,1)$ denotes the open unit ball centered at $0$ on the complex plane and $\closure{\D}(0,1)^\ccomp$ denotes the exterior of $\D(0,1)$. 
However, when a pinning is applied, particularly when some vertices are pinned to $+$ and some are $-$, we do not have either $\D(0,1)$-stability or $\closure{\D}(0,1)^\ccomp$-stability for the conditional partition function. 
To see this, notice that such a pinning can result in inconsistent external fields; some fields are $<1$ (hence in $\D(0,1)$) while others are $>1$ (hence in $\closure{\D}(0,1)^\ccomp$), and the Lee-Yang theorem does not apply.

Meanwhile, one should not expect spectral independence to hold for the ferromagnetic Ising model at all temperatures and for all external fields, since the Glauber dynamics is slow mixing when the parameters lie in the tree non-uniqueness region (see, for instance, \cite{GM07}).
\end{remark}

If limited information about the zero-free region is given, then spectral independence still holds with a worse bound.

\begin{theorem}
\label{thm:main_arb}
Let $\Gamma \subseteq \C$ be a non-empty open connected region. 
If for every pinning $\tau \in \Pinning$ the multivariate conditional partition function $Z_\weight^\tau$ is $\Gamma$-stable, then for any $\lambda \in \R_+ \cap \Gamma$ the Gibbs distribution $\mu = \mu_{\weight,\lambda}$ with the uniform external field $\lambda$ is spectrally independent with constant
\[
\eta = \frac{2}{\pp\delta^2}
\] 
where $\pp$ is the marginal bound for $\mu$ and $\delta = \frac{1}{\lambda} \dist(\lambda, \partial \Gamma)$.
\end{theorem}

It is unclear if the marginal bound $b$ is needed or not.
Our results \cref{thm:main_R+,thm:main_lambda_c,thm:main_arb} also hold for non-uniform external fields, i.e., each pair $(v,\spb)$ is assigned a distinct field $\lambda_{v,\spb}$, 
and the zero-free regions are allowed to be distinct for different pairs. 
See \cref{thm:multi-spin} for a formal statement which implies \cref{thm:main_R+,thm:main_lambda_c,thm:main_arb} as special cases.
Our proof of \cref{thm:multi-spin}
uses tools from Complex Analysis, such as the Riemann Mapping Theorem (see \cref{thm:Riemann_mapping}). 
We remark that for the simple but slightly worse bound \cref{thm:main_arb} the proof does not rely on the Riemann Mapping Theorem.

\subsection{Relations with Previous Works}

Our work builds upon the recent work of Alimohammadi, Anari, Shiragur, and Vuong~\cite{AASV21}. 
Theorem 16 of \cite{AASV21} established spectral independence for any distribution over $\{0,1\}^V$ assuming that the generating polynomial is sector-stable (that is, $\Gamma$-stable where $\Gamma = \{z\in\C: |\Arg(z)| \le \theta \}$ is a sector for some $\theta \in (0, \pi/2)$). 
Our results \cref{thm:main_R+,thm:main_lambda_c,thm:main_arb} strengthen theirs in the sense that we do not have any restriction on the zero-free region $\Gamma$, and the results hold for any open connected region. 
This allows us to apply our results in a much broader setting. 
See \cref{sec:zerofree} for more details. 

To establish zero-free regions for our main applications, we utilize the approach in \cite{GLLZ21}, which reduces the problem via Asano-Ruelle contractions \cite{Asa70, Rue71} to showing a sufficiently large zero-free region for a collection of bounded-degree univariate polynomials, one for each vertex of the input graph. These univariate polynomials are referred to as the local polynomials, since they only depend on the configuration restricted to edges incident to the given vertex. We note a very similar idea was also used in \cite{Wag09, BCR20} to establish zero-free regions, although their methods do not go through Asano-Ruelle contractions. See \cref{sec:binarysymmetric} for more details.

It was also shown in a sequence of papers \cite{Bar16, BS16, BS17, Bar17perm, Bar17boolean, Reg18} that one can establish large zero-free regions via an inductive approach based on conditioning the distribution. This method of establishing zero-free regions also works nicely for us, as spectral independence requires a bound on the pairwise influences for all conditional distributions. We show that one can deduce rapid mixing of the Glauber dynamics in a nearly black-box fashion from these zero-free methods for several problems in \cref{sec:homtensornetwork,sec:fourier}.

Algorithmically, our results have several advantages over prior works utilizing zero-free regions. In particular, the polynomial interpolation method pioneered by Barvinok \cite{Bar17book} typically only yields quasi-polynomial time algorithms in general, and polynomial time algorithms with exponent depending on the maximum degree for problems arising from graphs \cite{PR17}. In contrast, we obtain fast algorithms for sampling and counting. Another feature of our approach is that we only need the zero-free region to be sufficiently large. This is in contrast to the polynomial interpolation technique, which needs the zero-free region to also contain a point at which the partition function is easily computable.
On the other hand, our approach is
fundamentally restricted to nonnegative real parameters whereas Barvinok's approach can be extended to complex parameters, see, for instance~\cite{Bar17book}.

\paragraph{Outline of Paper}
In \cref{sec:zerofree} we prove our general technical results \cref{thm:main_R+,thm:main_lambda_c,thm:main_arb} connecting zero-free regions with spectral independence.
We prove \cref{thm:edge_covers,thm:even_subgraphs,thm:anti_2-spin_edge} in \cref{sec:binarysymmetric} regarding binary symmetric Holant problems.
In \cref{sec:homtensornetwork} we prove results for weighted graph homomorphisms and tensor networks. 
Finally, in \cref{sec:fourier} we study arbitrary measures on the discrete cube as studied in the analysis of Boolean functions.

\section{Preliminaries}\label{sec:prelim}

\subsection{Pinnings}
\label{subsec:pinning}
Let $q\ge 2$ be an integer and write $q_1 = q-1$.
Let $V$ be a finite set of vertices and let $Q = \{0\} \cup Q_1$ be the set of spins where $Q_1 = \{1,\dots,q_1\}$.
Every spin assignment $\sigma: V \to Q$ is called a \emph{configuration}. 
The state space $\Omega = Q^V$ is the collection of all configurations and let $w: \Omega \to \R_{\ge 0}$ be a nonnegative weight function that is not identically zero.
A configuration $\sigma \in \Omega$ is said to be \emph{valid} or \emph{feasible} if $\weight(\sigma) > 0$. 
For $\Lambda \subseteq V$, define set of \emph{pinnings} on $\Lambda$ by  
\[
\Omega_\Lambda = \left\{\tau \in Q^\Lambda: \text{$\exists$ valid $\sigma \in \Omega$ s.t. $\sigma_\Lambda = \tau$} \right\}.
\]
Note $\Omega_V$ is the set of all valid configurations. 
Let $\Pinning = \bigcup_{\Lambda \subseteq V} \Omega_\Lambda$ be the collection of all pinnings. 

For a pinning $\tau \in \Pinning$, let $V^\tau$ denote the set of unpinned vertices; so if $\tau \in \Omega_\Lambda$ then $V^\tau = V \setminus \Lambda$. 
For $v \in V^\tau$, let $\Omega_v^\tau$ be the set of valid spins at $v$ under $\tau$:
\[
\Omega_v^\tau = \left\{\spb \in Q: \text{$\exists$ valid $\sigma \in \Omega$ s.t. $\sigma_\Lambda = \tau$ and $\sigma_v = \spb$} \right\}. 
\]
Define the collection of feasible vertex-spin pairs under $\tau$ by
$ \VSpairs^\tau = \{(v,\spb) \in V \times Q: v \in V^\tau, \spb \in \Omega_v^\tau \} $,
and the collection of pairs with nonzero spins by
$ \VSpairs_1^\tau = \{(v,\spb) \in \VSpairs^\tau: \spb \neq 0 \} $.
We write $\Omega_v$, $\VSpairs$, and $\VSpairs_1$ when no pinning is applied. 

Let $\boldlam: \VSpairs_1 \to \C$ be a vector of complex external fields. 
Given a pinning $\tau \in \Pinning$ on $\Lambda \subseteq V$ with $U = V^\tau = V \setminus \Lambda$, the \emph{conditional partition function} $Z_\weight^\tau$ under $\tau$ is a multivariate polynomial of $\boldlam$ defined as
\[
Z_\weight^\tau(\boldlam) = \sum_{\sigma \in \Omega: \, \sigma_\Lambda = \tau} \weight(\sigma) \boldlam^{\sigma_U}, 
\quad\text{where}~~
\boldlam^{\sigma_U} = \prod_{v \in U: \, \sigma_v \neq 0} \lambda_{v,\sigma_v}.
\]
When there is no pinning, this matches \cref{eq:partition_func} from the introduction. 
Observe that $Z_\weight^\tau$ depends only on the variables $\{ \lambda_{v,\spb}: (v,\spb) \in \VSpairs_1^\tau \}$, 
and that $Z_\weight^\tau$ is not identically zero since $\tau$ is a pinning. 
If $\boldlam$ is real and positive, then $Z_\weight^\tau(\boldlam) > 0$ and we obtain the \emph{conditional Gibbs distribution}:
\[
\mu^\tau(\sigma)
= \mu(\sigma \mid \sigma_\Lambda = \tau) 
= \frac{\weight(\sigma) \boldlam^{\sigma_U}}{Z_\weight^\tau(\boldlam)}, 
\quad \forall \sigma \in \Omega ~\text{s.t.}~ \sigma_\Lambda = \tau.
\]
Again this matches \cref{eq:gibbs} when there is no pinning.

\subsection{Spectral Independence}
\label{subsec:mixing_SI}

Let $P$ be the transition matrix of the Glauber dynamics with stationary distribution $\mu$, and let $P^t(X_0,\cdot)$ denote the distribution of the chain after $t$ steps starting from $X_0 \in \Omega_V$. The \emph{mixing time} of the Glauber dynamics is defined as
\[
T_{\mathrm{mix}}(P) = \max_{X_0 \in \Omega_V} \min\{t\ge 0: \norm{P^t(X_0,\cdot) - \mu}_{\mathrm{TV}} \le 1/4\},
\]
where $\norm{\cdot}_{\mathrm{TV}}$ denotes the total variation distance. 
Throughout this paper, we assume that the Glauber dynamics is ergodic for every conditional distribution $\mu^\tau$ where $\tau \in \Pinning$ is a pinning (such distribution $\mu$ is called totally-connected in \cite{CLV21,BCCPSV21}).

The notion of \emph{spectral independence} (see \cref{def:inf-matrix,def:SI}) was introduced in \cite{ALO20} and immediately becomes a powerful tool for establishing rapid mixing of the Glauber dynamics. 

\begin{theorem}[\cite{AL20,ALO20}]\label{thm:genericspecgaplocaltoglobal}
Let $\mu$ be an arbitrary distribution over $Q^V$ where $|V| = n$.
If $\mu$ is spectrally independent with constant $\eta$, then the Glauber dynamics mixes in $O(n^{\eta + 1} \log(1/\mu_{\min}))$ steps where $\mu_{\min} = \min_{\sigma \in \Omega_V} \mu(\sigma)$. 
\end{theorem}

The following mixing result for the Glauber dynamics is known for spin systems with nearest-neighbor interactions; 
it holds for Holant problems and tensor network contractions (see \cref{sec:binarysymmetric,sec:homtensornetwork} for definitions) as well since one can view these as spin systems defined on hypergraphs (also known as Markov random fields) and the proof approach of \cite{CLV21, BCCPSV21} still works when the underlying hypergraph has bounded maximum degree. 

\begin{theorem}[\cite{CLV21,BCCPSV21}]\label{thm:spinsystemlocaltoglobal}
Let $\mu$ be the Gibbs distribution of a spin system or a Holant problem or a tensor network contraction defined on an $n$-vertex graph of maximum degree $\Delta$. 
If $\mu$ is spectrally independent with constant $\eta$ and the marginal bound for $\mu$ is $\pp$, then the Glauber dynamics mixes in $C n\log n$ steps where $C = C(\Delta,\eta,\pp)$ is a constant independent of $n$.
\end{theorem}

\subsection{Stability}
\label{subsec:stable}

For $n$ sets $\Gamma_1,\dots,\Gamma_n$, let $\prod_{\ell=1}^n \Gamma_\ell = \Gamma_1 \times \dots \times \Gamma_n$ denote the Cartesian product of them.

\begin{definition}[Stability]
\label{def:stable}
For an integer $n \ge 1$ and $\mathcal{K} \subseteq \C^n$, 
we say a multivariate polynomial $P \in \C[z_1,\dots,z_n]$ is \emph{$\mathcal{K}$-stable} if $P(z_1,\dots,z_n) \neq 0$ whenever $(z_1,\dots,z_n) \in \mathcal{K}$. 
In particular, if $\mathcal{K} = \prod_{\ell = 1}^n \Gamma$ for some $\Gamma \subseteq \C$, then we simply say $P$ is \emph{$\Gamma$-stable}.
\end{definition}

\begin{theorem}[Hurwitz' Theorem]
\label{thm:Hurwitz'}
Let $n \ge 1$ be an integer and $\mathcal{K} \subseteq \C^n$ be an open connected set. 
Suppose that $\{f_m\}_{m=1}^\infty$ is a sequence of non-vanishing analytic functions on $\mathcal{K}$ that converges to $f$ uniformly on compact subsets of $\mathcal{K}$. Then $f$ is either non-vanishing on $\mathcal{K}$ or else identically zero.
\end{theorem}


\subsection{Complex Plane}
\label{subsec:plane_complex}
We refer to subsets of the complex plane as \emph{regions}. 
Note that this is slightly nonstandard in Complex Analysis, where a region (or domain) is more commonly defined as a non-empty open connected subset of $\C$. 

Let $\Gamma \subseteq \C$ be a region. 
Denote the \emph{complement} of $\Gamma$ by $\Gamma^\ccomp = \C \setminus \Gamma$, 
its \emph{interior} by $\interior{\Gamma}$, its \emph{closure} by $\closure{\Gamma}$, and its \emph{boundary} by $\boundary{\Gamma}$. 
We say $\Gamma$ is \emph{unbounded} if for any $M \in \R_+$ there exists $z \in \Gamma$ such that $|z| > M$; otherwise it is called \emph{bounded}.
For $z \in \C$ let $\dist(z,\Gamma) = \inf_{w \in \Gamma} |w-z|$ be the distance from $z$ to $\Gamma$ on the complex plane. 


For a region $\Gamma \subseteq \C$ and $z \in \C$, we define $\Gamma + z = \{w+z: w\in \Gamma\}$, $z\Gamma = \{zw: w \in \Gamma\}$, and $\Gamma^{-1} = (\Gamma \setminus \{0\})^{-1} = \{w^{-1}: w \in \Gamma \setminus \{0\}\}$. 
For $\Gamma_1,\Gamma_2 \subseteq \C$, let $\Gamma_1 \cdot \Gamma_2 = \{zw: z \in \Gamma_1, w \in \Gamma_2\}$ denote their Minkowski product; in particular, for $\Gamma \subseteq \C$ let $\Gamma^2 = \Gamma \cdot \Gamma = \{zw: z,w \in \Gamma\}$ (meanwhile we shall write $\prod_{\ell=1}^2 \Gamma = \Gamma \times \Gamma = \{(z,w): z,w \in \Gamma\}$ for the Cartesian product).

For $z \in \C$ and $r \in \R_+$, let $\D(z,r) = \{w \in \C: |w-z| < r\}$ denote the \emph{open disk} centered at $z$ with radius $r$, and let $\closure{\D}(z,r) = \{w \in \C: |w-z| \le r\}$ denote the \emph{closed disk}.
An (open or closed) \emph{polydisk} is a Cartesian product of (open or closed) disks. 
For $\eps \in \R_+$, let $\HP_\eps = \{x + iy: x < -\eps\}$ and $\closure{\HP}_\eps = \{x + iy: x \le -\eps\}$ be open and closed \emph{half-planes}. 

Let $\Gamma \subseteq \C$ be a non-empty open region. 
We say $w,z \in \Gamma$ are \emph{(path-)connected} in $\Gamma$ if there exists a continuous map $\gamma: [0,1] \to \Gamma$ such that $\gamma(0) = w$ and $\gamma(1) = z$. 
Observe that connectivity in $\Gamma$ is an equivalence relation, and we call each equivalence class a \emph{(path-)connected component} of $\Gamma$. 
The region $\Gamma$ is said to be \emph{(path-)connected} if every two points from $\Gamma$ are connected in $\Gamma$; namely, $\Gamma$ has a unique connected component which is itself. 
If $\Gamma$ is open then every connected component of $\Gamma$ is also open.

A non-empty open connected region $\Gamma \subseteq \C$ is called \emph{simply connected} if its complement in the Riemann sphere ($\C \cup \infty$) is also connected. 
A \emph{Jordan curve} (simple closed curve) is a continuous map $\gamma: [0,1] \to \C$ such that $\gamma(0) = \gamma(1)$ and the restriction of $\gamma$ to $[0,1)$ is injective. 
The Jordan curve theorem states that for a Jordan curve $\gamma$, the complement of its image on the complex plane consists of exactly two open connected components; one of these components is bounded called the \emph{interior} and the other is unbounded called the \emph{exterior}. 
A non-empty open connected region $\Gamma \subseteq \C$ is simply connected if and only if for every Jordan curve $\gamma$ whose image is contained in $\Gamma$, the interior of $\gamma$ is also contained in $\Gamma$.

\subsection{Complex Analysis}
\label{subsec:analysis_complex}

We present here a few useful theorems in Complex Analysis which can be found in many textbooks, see e.g.~\cite{Rudin}.
Throughout, we select the principal branch for the complex functions $z \mapsto \log z$ and $z \mapsto z^{1/d}$.

\begin{theorem}[Schwarz-Pick Theorem]
\label{thm:Schwarz-Pick}
Let $f: \D(0,1) \to \D(0,1)$ be a holomorphic function. Then
\[
|f'(0)| \le 1-|f(0)|^2 \le 1.
\]
\end{theorem}

For open regions $\Gamma_1,\Gamma_2 \subseteq \C$, a function $f: \Gamma_1 \to \Gamma_2$ is said to be \emph{biholomorphic} if $f$ is a bijective holomorphic function whose inverse is also holomorphic.

\begin{theorem}[Riemann Mapping Theorem]
\label{thm:Riemann_mapping}
Let $\Gamma \subseteq \C$ be a non-empty open simply connected region that is not $\C$. 
Then for any $z \in \Gamma$ there exists a unique biholomorphic mapping $f: \Gamma \to \D(0,1)$ such that
\[
f(z) = 0 \quad\text{and}\quad f'(z) \in \R_+.
\]
\end{theorem}

\begin{theorem}[Koebe's One-Quarter Theorem]
\label{thm:Koebe-1/4}
Let $\Gamma \subseteq \C$ and let $f: \D(0,1) \to \Gamma$ be an injective holomorphic function. Then
\[
\D\left( f(0), \frac{1}{4}|f'(0)| \right) \subseteq \Gamma.
\]
\end{theorem}

\begin{theorem}[Multivariate Open Mapping Theorem, {\cite[Theorem 1.8.1]{KW17}}]
\label{thm:open-map}
Let $n \ge 1$ be an integer and let $\mathcal{K} \subseteq \C^n$ be a non-empty open connected subset of $\C^n$. Let $f: \mathcal{K} \to \C$ be a non-constant holomorphic function. Then the image of $f$ is an open connected region. 
\end{theorem}

\section{Establishing Spectral Independence via Stability}
\label{sec:zerofree}

In this section, we deduce spectral independence of a distribution from the stability of the associated partition function, and thus prove \cref{thm:main_R+,thm:main_lambda_c,thm:main_arb}.

We first show that pinning preserves stability of the partition function if the zero-free region is unbounded and contains $0$ in its closure. 
Intuitively, the pinning $\sigma_v = \spb$ for $\spb \neq 0$ corresponds to taking $\lambda_{v,\spb} = \infty$ (which is achieved by taking derivative with respect to $\lambda_{v,\spb}$), and the pinning $\sigma_v = 0$ corresponds to taking $\lambda_{v,\spb} = 0$ for all $k \in Q_1$. 
Hence, under an arbitrary pinning the conditional partition function is just the original partition function specialized at $\infty$ and $0$ for specific external fields, and the fact that the closure of the zero-free region contains $\infty$ (i.e., unboundedness) and $0$ guarantees that after specialization the resulted partition function is still stable. 
This is formalized by the following lemma.

\begin{lemma}\label{lem:pinning-stable}
Let $\{\Gamma_{v,\spb} \subseteq \C: (v,\spb) \in \VSpairs_1 \}$ be a collection of non-empty open connected regions such that for every $(v,\spb) \in \VSpairs_1$ the region $\Gamma_{v,\spb}$ is unbounded and $0 \subseteq \closure{\Gamma_{v,\spb}}$.
If the multivariate partition function $Z_\weight$ is $\big( \prod_{(v,\spb) \in \VSpairs_1} \Gamma_{v,\spb} \big)$-stable, then for every pinning $\tau \in \Pinning$ the multivariate conditional partition function $Z_\weight^\tau$ is $\big( \prod_{(v,\spb) \in \VSpairs_1^\tau} \Gamma_{v,\spb} \big)$-stable. 
\end{lemma}

Next, we show the following theorem for deriving spectral independence under various assumptions in the multivariate setting, from which one can deduce the bounds on spectral independence given in \cref{thm:main_R+,thm:main_lambda_c,thm:main_arb}. 
\begin{theorem}\label{thm:multi-spin}
Let $\{\Gamma_{v,\spb} \subseteq \C: (v,\spb) \in \VSpairs_1 \}$ be a collection of non-empty open connected regions, 
and let $\boldlam: \VSpairs_1 \to \R_+$ such that $\lambda_{v,\spb} \in \R_+ \cap \Gamma_{v,\spb}$ for each $(v,\spb) \in \VSpairs_1$.
Suppose that for every pinning $\tau \in \Pinning$ the multivariate conditional partition function $Z_\weight^\tau$ is $\big( \prod_{(v,\spb) \in \VSpairs_1^\tau} \Gamma_{v,\spb} \big)$-stable. 
Then the Gibbs distribution $\mu = \mu_{\weight,\boldlam}$ with external fields $\boldlam$ is spectrally independent with constant
\begin{equation}\label{eq:feta}
\eta = \frac{2}{\pp\delta^2},
\end{equation}
where $\pp$ is the marginal bound for $\mu$ and
\[
\delta = \min_{(v,\spb) \in \VSpairs_1} \frac{1}{\lambda_{v,\spb}} \dist(\lambda_{v,\spb}, \boundary{\Gamma_{v,\spb}}).
\]
Furthermore:
\begin{enumerate}
\item \label{it:1} For each $v \in V$ let $\Gamma_v \subseteq \C$ be the connected component of the intersection $\bigcap_{\spb \in Q_1} \frac{1}{\lambda_{v,\spb}} \Gamma_{v,\spb}$ that contains $1$ (note that $1 \in \frac{1}{\lambda_{v,\spb}} \Gamma_{v,\spb}$ for all $(v,\spb)$). 
If for every $v\in V$ the region $\Gamma_v$ is unbounded and $0 \in \closure{\Gamma_v}$, then spectral independence holds with constant
\[
\eta = \frac{8}{\delta}.
\]
In particular, the statement is true if $\R_+ \subseteq \Gamma_{v,\spb}$ for each $(v,\spb) \in \VSpairs_1$.


\item \label{it:3} If there exists $\lambda^* \in \R_+$ such that $\lambda_{v,\spb} \in (0,\lambda^*) \subseteq \Gamma_{v,\spb}$ for every $(v,\spb) \in \VSpairs_1$, then spectral independence holds with constant
\[
\eta = \frac{8}{\delta}
    \min\left\{
        \frac{1-\pp}{\pp},\, \frac{\lambda_{\max}}{\pp (\lambda^*-\lambda_{\max})} + 1
    \right\},
\]
where $\lambda_{\max} = \max_{(v,\spb) \in \VSpairs_1} \lambda_{v,\spb}$.

\item \label{it:4} If there exists $\lambda^* \in \R_+$ such that $\lambda_{v,\spb} \in (\lambda^*,\infty) \subseteq \Gamma_{v,\spb}$ for every $(v,\spb) \in \VSpairs_1$, then spectral independence holds with constant
\[
\eta = \frac{8}{\delta}
    \min\left\{
        \frac{1-\pp}{\pp},\, \frac{\lambda^*}{\pp (\lambda_{\min}-\lambda^*)} + 1
    \right\},
\]
where $\lambda_{\min} = \min_{(v,\spb) \in \VSpairs_1} \lambda_{v,\spb}$.
\end{enumerate}
\end{theorem}

We prove \cref{thm:multi-spin} by upper bounding the absolute row sum of the influence matrix $\Psi_\mu^\tau$ for any pinning $\tau \in \Pinning$; namely, for each $(u,\spa) \in \VSpairs^\tau$ we bound the sum of absolute values of the influences from $(u,\spa)$ to all other pairs $(v,\spb) \in \VSpairs^\tau$, see \cref{lem:1-field}.
We accomplish this by strengthening and generalizing the proof strategy in \cite{AASV21}. 

At a high level, the work \cite{AASV21} views the sum of absolute influences as the derivative of some function $f$ produced by the conditional partition functions. 
The variables of $f$ are just the external fields of the partition function which lie in some zero-free region $\Gamma$ and the stability of the conditional partition functions guarantees that the image of $f$ is contained in some nice region $\Gamma'$. 
In \cite{AASV21}, the authors study sector-stability of the partition function for the binary state space $\{0,1\}^V$; in particular, both the zero-free region $\Gamma$ and the region $\Gamma'$ containing the image are sectors for their choice of $f$.
Then, by applying conformal mappings between the sector and the unit disk, the derivative of $f$ can be upper bounded using the Schwarz-Pick Theorem (\cref{thm:Schwarz-Pick}). 

However, here we are facing a more challenging situation since we are trying to establish spectral independence from an arbitrary zero-free region $\Gamma$ for any discrete product space $Q^V$. 
In fact, for us the regions $\Gamma$ and $\Gamma'$ are in abstract form and to apply the Schwarz-Pick Theorem we need to design good mappings from $\Gamma$ and $\Gamma'$ to the unit disk. 
This is achieved by utilizing tools from Complex Analysis, especially the Riemann Mapping Theorem (\cref{thm:Riemann_mapping}). See \cref{subsubsec:map1,subsubsec:map2} for details of this part.

We now provide the proofs of \cref{thm:main_R+,thm:main_lambda_c,thm:main_arb} from the introduction.

\begin{proof}[Proof of Theorem~\ref{thm:main_R+}]
Follows from \cref{lem:pinning-stable} and \cref{it:1} of \cref{thm:multi-spin}. 
\end{proof}

\begin{proof}[Proof of Theorem~\ref{thm:main_lambda_c}]
Follows from \cref{it:3,it:4} of \cref{thm:multi-spin}. 
\end{proof}

\begin{proof}[Proof of Theorem~\ref{thm:main_arb}]
Follows from \cref{eq:feta} of \cref{thm:multi-spin}. 
\end{proof}

After proving \cref{lem:pinning-stable} in \cref{subsec:pinning-stable}, we establish \cref{thm:multi-spin} in \cref{subsec:bound-SI} and prove \cref{lem:1-field}, a central lemma for bounding the absolute sum of influences, in \cref{subsec:proof-abs-sum}.

\subsection{Preservation of Stability under Pinnings}
\label{subsec:pinning-stable}

In this subsection we present the proof of \cref{lem:pinning-stable}.  

Let $\tau \in \Pinning$ be an arbitrary pinning on $\Lambda \subseteq V$ and let $U = V^\tau = V \setminus \Lambda$ be the set of unpinned vertices. 
We consider the conditional partition function $Z_\weight^\tau$ under the pinning $\tau$. 
As discussed earlier, one can view $Z_\weight^\tau$ as obtained from the original partition function $Z_\weight$ by specializing at $0$ and taking derivatives for certain variables provided by the pinning $\tau$. 
To be more precise, we define $\Lambda_0 = \{v \in \Lambda: \tau_v = 0\}$ to be the set of vertices pinned to spin $0$, and let $\Lambda_1 = \{v \in \Lambda: \tau_v \neq 0\}$ be those pinned to nonzero spins. We also define $\tau_1 = \tau_{\Lambda_1}$ to be the pinning restricted to vertices with nonzero spins. 
The key observation here is that
\begin{equation}\label{eq:pinning_cond}
Z_\weight^\tau(\boldlam) = \left( \frac{\partial}{\partial \boldlam_{\Lambda_1,\tau_1}} Z_\weight(\boldlam) \right) \Bigg\vert_{\boldlam_{\Lambda_0} = 0}
\end{equation}
where $\boldlam_{\Lambda_0} = 0$ represents plugging in $\lambda_{v,\spb} = 0$ for all $v \in \Lambda_0$ and $\spb \in Q_1$, and $\frac{\partial}{\partial \boldlam_{\Lambda_1,\tau_1}}$ represents taking derivatives $\frac{\partial}{\partial \lambda_{v,\tau_v}}$ for all $v \in \Lambda_1$. 
Hence, to establish \cref{lem:pinning-stable} it suffices to show that specialization at $0$ and differentiation preserves $\Gamma$-stability if the zero-free region $\Gamma$ is unbounded and $0 \in \closure{\Gamma}$.
This is actually true for any multi-affine polynomial, which is a polynomial whose monomials are all square-free. 

\begin{lemma}\label{lem:preserve-stable}
Let $n\ge 1$ be an integer and let $\Gamma_1, \dots, \Gamma_n \subseteq \C$ be non-empty open connected regions. 
Let $P \in \C[z_1,\dots,z_n]$ be a multi-affine polynomial and assume that $P$ is $(\prod_{\ell=1}^n \Gamma_\ell)$-stable. 
Then:
\begin{enumerate}
\item (Inversion) The polynomial $P_1(z_1,z_2,\dots,z_n) = z_1 P(\frac{1}{z_1},z_2,\dots,z_n)$ is $(\Gamma_1^{-1} \times \prod_{\ell=2}^n \Gamma_\ell)$-stable;
\item (Specialization) If $0 \in \closure{\Gamma_1}$, then the polynomial $P_2(z_2,\dots,z_n) = P(0,z_2,\dots,z_n)$ is either $(\prod_{\ell=2}^n \Gamma_\ell)$-stable or identically zero;
\item (Differentiation) If $\Gamma_1$ is unbounded, then the polynomial $P_3(z_2,\dots,z_n) = \frac{\partial}{\partial z_1} P(z_1,z_2,\dots,z_n)$ is either $(\prod_{\ell=2}^n \Gamma_\ell)$-stable or identically zero.
\end{enumerate}
\end{lemma}
\begin{proof}
Consider first the inversion property. 
Suppose for sake of contradiction that $P_1$ is not $(\Gamma_1^{-1} \times \prod_{\ell=2}^n \Gamma_\ell)$-stable. 
Then there exists $w_1 \in \Gamma_1^{-1}$ and $z_\ell \in \Gamma_\ell$ for $2 \le \ell \le n$ such that $P_1(w_1,z_2,\dots,z_n) = 0$. 
Note that $w_1 = 1/z_1$ for some $z_1 \in \Gamma_1 \setminus \{0\}$. 
It follows that
\[
0 = z_1 P_1(w_1,z_2,\dots,z_n) = z_1 w_1 P\left(\frac{1}{w_1},z_2,\dots,z_n\right) = P(z_1,z_2,\dots,z_n),
\]
contradicting to the stability of $P$. Hence, we have the desired stability for $P_1$. 

Next consider specialization. 
Since $\Gamma_1$ is open and $0 \in \closure{\Gamma_1}$, there exists a sequence of complex numbers $\{\zeta_m\}_{m=1}^\infty$ such that $\zeta_m \in \Gamma_1$ and $\lim_{m\to \infty} \zeta_m = 0$. 
Let $f_m(z_2,\dots,z_n) = P(\zeta_m,z_2,\dots,z_n)$ be a polynomial of degree $\le \deg(P)$ for each $m$. 
Then $f_m$ is $(\prod_{\ell=2}^n \Gamma_\ell)$-stable by the stability assumption of $P$. 
Furthermore, the sequence $\{f_m\}_{m=1}^\infty$ converges to $P_2$ coefficient-wise, and hence uniformly on compact subsets; see, e.g., Lemma 33 in \cite{AASV21}. 
Hurwitz' Theorem (\cref{thm:Hurwitz'}) then implies that $P_2$ is either $(\prod_{\ell=2}^n \Gamma_\ell)$-stable or identically zero, as claimed.

Last we consider differentiation. Since $\Gamma_1$ is open and unbounded, we deduce that the region $\Gamma_1^{-1} = \{1/z: z \in \Gamma_1 \setminus \{0\}\}$ is open and satisfies $0 \in \closure{\Gamma_1^{-1}}$. 
Recall that we have shown the inversion $P_1(z_1,z_2,\dots,z_n) = z_1 P(\frac{1}{z_1},z_2,\dots,z_n)$ is $(\Gamma_1^{-1} \times \prod_{\ell=2}^n \Gamma_\ell)$-stable. 
Now observe that, for a multi-affine polynomial $P$, the derivative $P_3$ of $P$ with respect to $z_1$ is the same as the specialization of $P_1$ at $z_1 = 0$:
\[
P_3(z_2,\dots,z_n) = \frac{\partial}{\partial z_1} P(z_1,z_2,\dots,z_n) = P_1(0,z_2,\dots,z_n).
\]
Hence, we immediately conclude from previous results that $P_3$ is either $(\prod_{\ell=2}^n \Gamma_\ell)$-stable or identically zero.
\end{proof}

\cref{lem:pinning-stable} is an immediate consequence of \cref{lem:preserve-stable}.

\begin{proof}[Proof of Lemma~\ref{lem:pinning-stable}]
Observe that the partition function \cref{eq:partition_func} is multi-affine. The lemma then follows from \cref{eq:pinning_cond,lem:preserve-stable}. 
Notice that the conditional partition functions are never identically zero since pinnings are extendable to valid full configurations. 
\end{proof}

\subsection{Bounds on Spectral Independence}
\label{subsec:bound-SI}

In this subsection we prove \cref{thm:multi-spin}. 

An important observation is that it is sufficient to assume $1 \in \Gamma_{v,\spb}$ for every $(v,\spb) \in \VSpairs_1$ and consider the Gibbs distribution with the all-one external fields $\boldone$. 
In general, given the external field $\lambda_{v,\spb} \in \Gamma_{v,\spb}$ for each $(v,\spb)$, we may reweight the configurations by 
\begin{equation}\label{eq:reweight}
\widetilde{\weight}(\sigma) = \weight(\sigma) \prod_{v \in V:\, \sigma_v \neq 0} \lambda_{v,\sigma_v}, \quad \forall \sigma \in \Omega
\end{equation}
and the new partition function is
\begin{equation}\label{eq:rewt_Z}
Z_{\widetilde{\weight}} (\boldlam) = \sum_{\sigma \in \Omega} \widetilde{\weight}(\sigma) \boldlam^\sigma.
\end{equation}
In particular, $Z_\weight(\boldlam) = Z_{\widetilde{\weight}}(\boldone)$ and $\mu_{\weight,\boldlam} = \mu_{\widetilde{\weight},\boldone}$ for the given $\boldlam = (\lambda_{v,\spb})$. 
In other words, we hide the external fields into the weight of configurations and under the new weights we are interested in the all-one external fields. 
This will simplify the notations. 
Note that, if for $\tau \in \Pinning$ the multivariate conditional partition function $Z_\weight^\tau$ is $\big( \prod_{(v,\spb) \in \VSpairs_1^\tau} \Gamma_{v,\spb} \big)$-stable, then the reweighted conditional partition function $Z_{\widetilde{\weight}}^\tau$ is $\big( \prod_{(v,\spb) \in \VSpairs_1^\tau} \widetilde{\Gamma}_{v,\spb} \big)$-stable where $\widetilde{\Gamma}_{v,\spb} = \frac{1}{\lambda_{v,\spb}} \Gamma_{v,\spb}$ for each $(v,\spb) \in \VSpairs_1$. 

In the rest of this section, we assume that $1 \in \Gamma_{v,\spb}$ and consider the case of all-one external fields.
The following lemma is an important step towards deducing \cref{thm:multi-spin}; it builds upon the proof strategy of \cite{AASV21} while generalizing their result.

\begin{lemma}\label{lem:1-field}
Consider the Gibbs distribution $\mu = \mu_{\weight,\boldone}$ with the all-one external fields $\boldone$. 
Let $\tau \in \Pinning$ be a fixed pinning and let $\{\Gamma_{v,\spb} \subseteq \C: (v,\spb) \in \VSpairs_1^\tau \}$ be a collection of non-empty open connected regions such that $1 \in \Gamma_{v,\spb}$ for each $(v,\spb) \in \VSpairs_1^\tau$. 
For every $v \in V^\tau$ let $\Gamma_v \subseteq \C$ be the connected component of the intersection $\bigcap_{0 \neq \spb \in \Omega_v^\tau} \Gamma_{v,\spb}$ that contains $1$. 
If the multivariate conditional partition function $Z_\weight^\tau$ is $\big( \prod_{(v,\spb) \in \VSpairs_1^\tau} \Gamma_{v,\spb} \big)$-stable, 
then the influence matrix $\Psi_\mu^\tau$ under the pinning $\tau$ satisfies 
\[
\eigmax(\Psi_\mu^\tau) \le \norm{\Psi_\mu^\tau}_\infty \le \min \left\{ \frac{2}{\pp\delta^2}, \frac{8}{\delta} \max_{\substack{\tau \in \Pinning\\(v,\spb) \in \VSpairs^\tau}} \dist\left( 1, \mathcal{C}_{v,\spb}^\tau \right) \right\}, 
\] 
where: $\pp$ is the marginal bound for $\mu$; 
\begin{align}
&\delta := \min_{(v,\spb) \in \VSpairs_1^\tau} \dist\left( 1, \boundary{\Gamma_{v,\spb}} \right); 
\quad
p_{v,\spb}^\tau := \mu^\tau(\sigma_v = \spb); \\
&\mathcal{C}_{v,\spb}^\tau 
:= -\frac{1}{p_{v,\spb}^\tau} \left( \Gamma_{v,\spb} - 1 \right)^{-1} 
= \left\{ -\frac{1}{p_{v,\spb}^\tau (z-1)} :\, z \in \Gamma_{v,\spb} \setminus \{1\} \right\},
\quad \text{for}~ \spb \neq 0; \\
&\mathcal{C}_{v,0}^\tau 
:= \frac{1}{p_{v,0}^\tau} \left(\left( \Gamma_v - 1 \right)^{-1} + 1 \right) 
= \left\{ \frac{z}{p_{v,0}^\tau (z-1)} :\, z \in \Gamma_v \setminus \{1\} \right\}.
\end{align}
\end{lemma}

The following two lemmas are helpful for bounding the distance $\dist( 1, \mathcal{C}_{v,\spb}^\tau )$ in \cref{lem:1-field}. 

\begin{lemma}\label{lem:C-nz-bound}
Let $\tau \in \Pinning$ and $(v,\spb) \in \VSpairs_1^\tau$.  
\begin{enumerate}
\item If $\Gamma_{v,\spb}$ is unbounded, then
\[
\dist\left( 1, \mathcal{C}_{v,\spb}^\tau \right) \le 1.
\]

\item Let $\alpha_{v,\spb} = \inf \left(\Gamma_{v,\spb} \cap \R_+\right)$ and $\beta_{v,\spb} = \sup \left(\Gamma_{v,\spb} \cap \R_+\right)$. Then
\[
\dist\left( 1, \mathcal{C}_{v,\spb}^\tau \right) \le 
	\min\left\{ \frac{\alpha_{v,\spb}}{p_{v,\spb}^\tau (1-\alpha_{v,\spb})} + \frac{1-p_{v,\spb}^\tau}{p_{v,\spb}^\tau},\, \frac{1}{p_{v,\spb}^\tau (\beta_{v,\spb}-1)} + 1 \right\},
\]
\end{enumerate}
with the convention that $\frac{1}{\infty} = 0$ if $\beta_{v,\spb} = \infty$. 
\end{lemma}

\begin{lemma}\label{lem:C-zero-bound}
Let $\tau \in \Pinning$ and $(v,0) \in \VSpairs^\tau$.  
\begin{enumerate}
\item If $0 \in \closure{\Gamma_v}$, then
\[
\dist\left( 1, \mathcal{C}_{v,0}^\tau \right) \le 1.
\]

\item Let $\alpha_v = \inf \left(\Gamma_v \cap \R_+\right)$ and $\beta_v = \sup \left(\Gamma_v \cap \R_+\right)$. Then
\[
\dist\left( 1, \mathcal{C}_{v,0}^\tau \right) \le 
	\min\left\{ \frac{\alpha_v}{p_{v,0}^\tau (1-\alpha_v)} + 1,\, \frac{1}{p_{v,0}^\tau (\beta_v-1)} + \frac{1-p_{v,0}^\tau}{p_{v,0}^\tau} \right\},	
\]
\end{enumerate}
with the convention that $\frac{1}{\infty} = 0$ if $\beta_v = \infty$. 
\end{lemma}

Combining \cref{lem:1-field,lem:C-nz-bound,lem:C-zero-bound}, we are able to establish \cref{thm:multi-spin}. 
The proofs of \cref{lem:C-nz-bound,lem:C-zero-bound} are technical and we postpone them to \cref{app:technicallemmas}.
The proof of \cref{lem:1-field} is presented in \cref{subsec:proof-abs-sum}.

\begin{proof}[Proof of Theorem~\ref{thm:multi-spin}]
As discussed at the beginning of this subsection, we can reweight the configurations by \cref{eq:reweight} and consider the all-one external fields under the new weights, so that \cref{lem:1-field} applies. 
In particular, for an arbitrary pinning $\tau \in \Pinning$ the reweighted conditional partition function $Z_{\widetilde{\weight}}^\tau$ given by \cref{eq:rewt_Z} is $\big( \prod_{(v,\spb) \in \VSpairs_1^\tau} \widetilde{\Gamma}_{v,\spb} \big)$-stable where $\widetilde{\Gamma}_{v,\spb} = \frac{1}{\lambda_{v,\spb}} \Gamma_{v,\spb}$ for each $(v,\spb) \in \VSpairs_1$. 
Thus, the first upper bound in \cref{lem:1-field} implies that the Gibbs distribution $\mu = \mu_{\weight,\boldlam} = \mu_{\widetilde{\weight},\boldone}$ is spectrally independent with constant $\eta = 2/(\pp \delta^2)$ 
where $\pp$ is the marginal bound for $\mu$ and 
\[
\delta = \min_{(v,\spb) \in \VSpairs_1} \dist\left( 1, \boundary{\widetilde{\Gamma}_{v,\spb}} \right)
= \min_{(v,\spb) \in \VSpairs_1} \frac{1}{\lambda_{v,\spb}} \dist\left( \lambda_{v,\spb}, \boundary{\Gamma_{v,\spb}} \right)
\]
as claimed. 

For each $v \in V$, let $\widetilde{\Gamma}_v = \Gamma_v \subseteq \C$ be the connected component of the intersection $\bigcap_{\spb \in Q_1} \widetilde{\Gamma}_{v,\spb} = \bigcap_{\spb \in Q_1} \frac{1}{\lambda_{v,\spb}} \Gamma_{v,\spb}$ that contains $1$. 
Then we further have the following. 
\begin{enumerate}
\item If for every $v\in V$ the region $\widetilde{\Gamma}_v$ is unbounded and $0$ is contained in the closure of $\widetilde{\Gamma}_v$, then the first part of \cref{lem:C-nz-bound,lem:C-zero-bound} implies that $\dist( 1, \mathcal{C}_{v,\spb}^\tau ) \le 1$ for all $\tau \in \Pinning$ and $(v,\spb) \in \VSpairs^\tau$. 
Hence, by the second upper bound in \cref{lem:1-field} spectral independence holds with constant $\eta = 8/\delta$. 

If $\R_+ \subseteq \Gamma_{v,\spb}$ for each $(v,\spb) \in \VSpairs_1$, then by definition $\widetilde{\Gamma}_v$ is unbounded and $0$ is contained in the closure of $\widetilde{\Gamma}_v$ for every $v \in V$; 
therefore, spectral independence holds with $\eta = 8/\delta$.

\item If there exists $\lambda^* \in \R_+$ such that $\lambda_{v,\spb} \in (0,\lambda^*) \subseteq \Gamma_{v,\spb}$ for every $(v,\spb) \in \VSpairs_1$, 
then one has 
\begin{align*}
\alpha_{v,\spb} &= \inf \left(\widetilde{\Gamma}_{v,\spb} \cap \R_+\right) = 0,
\quad
\beta_{v,\spb} = \sup \left(\widetilde{\Gamma}_{v,\spb} \cap \R_+\right) \ge \frac{\lambda^*}{\lambda_{v,\spb}}, \\
\alpha_v &= \inf \left(\widetilde{\Gamma}_v \cap \R_+\right) = 0,
\quad
\beta_v = \sup \left(\widetilde{\Gamma}_v \cap \R_+\right) \ge \frac{\lambda^*}{\lambda_{v,\max}},
\end{align*}
where $\lambda_{v,\max} = \max_{\spb \in Q_1} \lambda_{v,\spb} \le \lambda_{\max}$. 
Thus, we deduce from the second part of \cref{lem:C-nz-bound,lem:C-zero-bound} that for all $\tau \in \Pinning$ and $(v,\spb) \in \VSpairs^\tau$,
\begin{align*}
\dist\left( 1, \mathcal{C}_{v,\spb}^\tau \right) &\le 
	\min\left\{ \frac{1-p_{v,\spb}^\tau}{p_{v,\spb}^\tau},\, \frac{\lambda_{v,\spb}}{p_{v,\spb}^\tau (\lambda^*-\lambda_{v,\spb})} + 1 \right\} \\
	&\le
	\min\left\{ \frac{1-\pp}{\pp},\, \frac{\lambda_{\max}}{\pp (\lambda^*-\lambda_{\max})} + 1 \right\},
	\quad\text{for~} \spb \neq 0; \\
\dist\left( 1, \mathcal{C}_{v,0}^\tau \right) &\le 
	\min\left\{ 1,\, \frac{\lambda_{v,\max}}{p_{v,0}^\tau (\lambda^*-\lambda_{v,\max})} + \frac{1-p_{v,0}^\tau}{p_{v,0}^\tau} \right\} \le 1.	
\end{align*}
The second bound in \cref{lem:1-field} then yields the desired bound on spectral independence. 
Note that we may assume $\mu$ is supported on at least two configurations so that $(1-\pp)/\pp \ge 1$, namely $\pp \le 1/2$; otherwise $\mu$ is concentrated on a single configuration and spectral independence holds with constant $0$. 

\item If there exists $\lambda^* \in \R_+$ such that $\lambda_{v,\spb} \in (\lambda^*,\infty) \subseteq \Gamma_{v,\spb}$ for every $(v,\spb) \in \VSpairs_1$, 
then one has 
\begin{align*}
\alpha_{v,\spb} &= \inf \left(\widetilde{\Gamma}_{v,\spb} \cap \R_+\right) \le \frac{\lambda^*}{\lambda_{v,\spb}},
\quad
\beta_{v,\spb} = \sup \left(\widetilde{\Gamma}_{v,\spb} \cap \R_+\right) = \infty, \\
\alpha_v &= \inf \left(\widetilde{\Gamma}_v \cap \R_+\right) \le \frac{\lambda^*}{\lambda_{v,\min}},
\quad
\beta_v = \sup \left(\widetilde{\Gamma}_v \cap \R_+\right) = \infty,
\end{align*}
where $\lambda_{v,\min} = \min_{\spb \in Q_1} \lambda_{v,\spb}$. 
Thus, we deduce from the second part of \cref{lem:C-nz-bound,lem:C-zero-bound} that for all $\tau \in \Pinning$ and $(v,\spb) \in \VSpairs^\tau$,
\begin{align*}
\dist\left( 1, \mathcal{C}_{v,\spb}^\tau \right) &\le 
	\min\left\{ \frac{\lambda^*}{p_{v,\spb}^\tau (\lambda_{v,\spb}-\lambda^*)} + \frac{1-p_{v,\spb}^\tau}{p_{v,\spb}^\tau},\, 1 \right\} \le 1, 
	\quad\text{for~} \spb \neq 0; \\
\dist\left( 1, \mathcal{C}_{v,0}^\tau \right) &\le 
	\min\left\{ \frac{\lambda^*}{p_{v,0}^\tau (\lambda_{v,\min}-\lambda^*)} + 1,\, \frac{1-p_{v,0}^\tau}{p_{v,0}^\tau} \right\} \\
	&\le 
	\min\left\{ \frac{\lambda^*}{\pp (\lambda_{\min}-\lambda^*)} + 1,\, \frac{1-\pp}{\pp} \right\}. 
\end{align*}
The second bound in \cref{lem:1-field} then yields the bound on spectral independence as wanted. \qedhere
\end{enumerate}
\end{proof}

\begin{remark}
For \cref{it:3} of \cref{thm:multi-spin} our proof actually yields a more complicated but stronger constant for spectral independence:
\[
\eta = \frac{8}{\delta}  
	\max\left\{ 1, \max_{\substack{\tau \in \Pinning\\(v,\spb) \in \VSpairs_1^\tau}} 
	\left\{ 
		\min \left\{ \frac{1-p_{v,\spb}^\tau}{p_{v,\spb}^\tau},\, \frac{\lambda_{v,\spb}}{p_{v,\spb}^\tau (\lambda^*-\lambda_{v,\spb})} + 1 \right\} 
	\right\} 
	\right\}.
\]
For \cref{it:4}, the constant is:
\[
\eta = \frac{8}{\delta} 
	\max\left\{ 1, \max_{\substack{\tau \in \Pinning\\(v,0) \in \VSpairs^\tau}} 
	\left\{ 
		\min \left\{ \frac{1-p_{v,0}^\tau}{p_{v,0}^\tau},\, \frac{\lambda^*}{p_{v,0}^\tau (\lambda_{v,\min}-\lambda^*)} + 1 \right\} 
	\right\}
	\right\}.
\]
These two bounds are more robust in the sense of \cref{rmk:robust}, namely, when some external fields are close to $\lambda^*$ while others are close to $0$ (respectively, $\infty$).
\end{remark}

\subsection{Bounding the Absolute Sum of Influences: Proof of \texorpdfstring{\cref{lem:1-field}}{Lemma 3.4}}
\label{subsec:proof-abs-sum}

Let $\tau \in \Pinning$ be an arbitrary pinning and fix $\tau$. 
We will give an upper bound on the absolute row sum of the associated influence matrix $\Psi_\mu^\tau$ under $\tau$, which then provides an upper bound on the maximum eigenvalue of $\Psi_\mu^\tau$. 
In particular, for $(u,\spa) \in \VSpairs^\tau$ we define
\begin{align}
\widetilde{S}_\mu^\tau(u,\spa) := \sum_{(v,\spb) \in \VSpairs^\tau} \left| \Psi_\mu^\tau(u,\spa;v,\spb) \right| 
= \sum_{(v,\spb) \in \VSpairs^\tau: v \neq u} \left| \mu^\tau(\sigma_v = \spb \mid \sigma_u = \spa) - \mu^\tau(\sigma_v = \spb) \right| 
\end{align}
to be the absolute sum of influences in the row $(u,\spa)$, and define
\begin{align}
S_\mu^\tau(u,\spa) := \sum_{(v,\spb) \in \VSpairs_1^\tau} \left| \Psi_\mu^\tau(u,\spa;v,\spb) \right| 
= \sum_{(v,\spb) \in \VSpairs_1^\tau: v \neq u} \left| \mu^\tau(\sigma_v = \spb \mid \sigma_u = \spa) - \mu^\tau(\sigma_v = \spb) \right| 
\end{align}
to be the partial absolute sum of influences for pairs $(v,\spb)$ with $\spb \neq 0$ in the row $(u,\spa)$. 
Notice that one has $\widetilde{S}_\mu^\tau(u,\spa) \le 2S_\mu^\tau(u,\spa)$, because for each $(v,0) \in \VSpairs^\tau$ it follows from the triangle inequality that
\[
\left| \mu^\tau(\sigma_v = 0 \mid \sigma_u = \spa) - \mu^\tau(\sigma_v = 0) \right| 
\le
\sum_{\spb \in \Omega_v^\tau \setminus \{0\}} \left| \mu^\tau(\sigma_v = \spb \mid \sigma_u = \spa) - \mu^\tau(\sigma_v = \spb) \right|. 
\]
Hence,
\begin{equation}\label{eq:eigmax-bound}
\eigmax(\Psi_\mu^\tau) \le \norm{\Psi_\mu^\tau}_\infty = \max_{(u,\spa) \in \VSpairs^\tau} \widetilde{S}_\mu^\tau(u,\spa) \le 2\max_{(u,\spa) \in \VSpairs^\tau} S_\mu^\tau(u,\spa). 
\end{equation}

The rest of the proof aims to bound $S_\mu^\tau(u,\spa)$ for a fixed $(u,\spa) \in \VSpairs^\tau$. 
We consider two cases $\spa \neq 0$ and $\spa = 0$ separately, and prove the following. 
\begin{lemma}\label{lem:S-nz-bound}
For $\spa \neq 0$ we have
\[
S_\mu^\tau(u,\spa) \le \min\left\{ \frac{1}{\pp\delta^2}, \frac{4}{\delta} \dist\left( 1, \mathcal{C}_{u,\spa}^\tau \right) \right\}.
\]
\end{lemma}
\begin{lemma}\label{lem:S-zero-bound}
For $\spa = 0$ we have
\[
S_\mu^\tau(u,0) \le \min\left\{ \frac{1}{\pp\delta^2}, \frac{4}{\delta} \dist\left( 1, \mathcal{C}_{u,0}^\tau \right) \right\}.
\]
\end{lemma}

\cref{lem:1-field} follows immediately from these two lemmas.  

\begin{proof}[Proof of Lemma~\ref{lem:1-field}]
Combining \cref{eq:eigmax-bound,lem:S-nz-bound,lem:S-zero-bound}. 
\end{proof}

\subsubsection{Proof of \texorpdfstring{\cref{lem:S-nz-bound}}{Lemma 3.8}}
\label{subsubsec:map1}

Fix $(u,\spa) \in \VSpairs_1^\tau$. 
We follow the proof approach of \cite{AASV21} and view $S_\mu^\tau(u,\spa)$ as the derivative of a certain function related to the partition function; the lemma then follows from an application of the Schwarz-Pick Theorem (\cref{thm:Schwarz-Pick}) for bounding the derivative. 
For ease of notation we write
\[
\VSpairs' := \{ (v,\spb) \in \VSpairs_1^\tau: v \neq u \}
\quad\text{and}\quad
\mathcal{K} := \prod_{(v,\spb) \in \VSpairs'} \Gamma_{v,\spb}.
\]
Recall that $p_{u,\spa}^\tau = \mu^\tau(\sigma_u = \spa)$. 
Define the multivariate complex function $f: \mathcal{K} \to \C$ as
\begin{equation}\label{eq:f-func}
f(\boldlam) = \frac{1}{p_{u,\spa}^\tau} \frac{Z_\weight^{\tau \cup (u,\spa)} (\boldlam)}{Z_\weight^\tau (\boldlam \cup \boldone_u)}, 
\quad\text{for~} \boldlam \in \mathcal{K}
\end{equation}
where $\tau \cup (u,\spa) \in \Pinning$ is the pinning that combines $\tau$ and $\sigma_u = \spa$, and $\boldlam \cup \boldone_u$ is the vector of external fields that combines $\boldlam$ and the all-one fields $\boldone_u$ at $u$ (i.e., $\lambda_{u,\spa'} = 1$ for all $0 \neq \spa' \in \Omega_u^\tau$); notice that 
\[
Z_\weight^\tau (\boldlam \cup \boldone_u) = \sum_{\spa' \in \Omega_u^\tau} Z_\weight^{\tau \cup (u,\spa')} (\boldlam).
\]  
Note that $f$ is well-defined since by our assumption $Z_\weight^\tau (\boldlam \cup \boldone_u) \neq 0$ whenever each $\lambda_{v,\spb} \in \Gamma_{v,\spb}$.
The following claim summarizes several important properties of the function $f$.

\begin{claim}\label{clm:f}
Let $f: \mathcal{K} \to \C$ be the multivariate complex function defined by \cref{eq:f-func}. 
\begin{enumerate}
\item \label{it:f-i} The function $f$ is well-defined and holomorphic on $\mathcal{K}$, and $f(\boldone) = 1$. 
\item \label{it:f-ii} For every $(v,\spb) \in \VSpairs'$, 
\[
\left. \frac{\partial f}{\partial \lambda_{v,\spb}} \right|_{\boldlam = \boldone} = \Psi_\mu^\tau(u,\spa;v,\spb). 
\]
\item \label{it:f-iii} Suppose that $f \not\equiv 1$. 
Let $\mathcal{A} \subseteq \C$ be an open region defined as
\begin{align}
\mathcal{A} := -\frac{1}{p_{u,\spa}^\tau} \left( \Gamma_{u,\spa} - 1 \right)^{-1}.
\end{align}
Then $1 \notin \closure{\mathcal{A}}$. 
Let $\mathcal{A}_1$ be the connected component of $\closure{\mathcal{A}}^\ccomp$ which contains $1$. 
Then $\mathcal{A}_1$ is open and simply connected, and 
\[
\image(f) \subseteq \mathcal{A}_1. 
\]
\end{enumerate}
\end{claim}



If $f \equiv 1$, then \cref{it:f-ii} of \cref{clm:f} implies that $\Psi_\mu^\tau(u,\spa;v,\spb) = 0$ for all $(v,\spb) \in \VSpairs'$, and hence $S_\mu^\tau(u,\spa) = 0$. 
In the rest of the proof we assume that $f \not\equiv 1$.

Given \cref{clm:f}, in order to bound $S_\mu^\tau(u,\spa)$ it suffices to bound $\norm{\grad f(\boldone)}_1$. 
We do this by taking holomorphic functions $\varphi: \D(0,1) \to \mathcal{K}$, $\psi: \mathcal{A}_1 \to \D(0,1)$ and considering their composition with the function $f$. 
The bound on $\norm{\grad f(\boldone)}_1$ would then follow from the Schwarz-Pick Theorem which bounds the derivative of a holomorphic function from the open unit dist into itself. 

We now formalize this idea. 
Let $\varphi: \D(0,1) \to \mathcal{K}$ be a holomorphic vector-valued function such that for every $(v,\spb) \in \VSpairs'$, the $(v,\spb)$-coordinate function $\varphi_{v,\spb}: \D(0,1) \to \Gamma_{v,\spb}$ is holomorphic and satisfies $\varphi_{v,k}(0) = 1$ and $\varphi_{v,\spb}'(0) \in \R_+$ if $\Psi_\mu^\tau(u,\spa;v,\spb) \ge 0$ while $\varphi_{v,\spb}'(0) \in \R_-$ if $\Psi_\mu^\tau(u,\spa;v,\spb) \le 0$. 
Hence, $\varphi(0) = \boldone$ and $\varphi_{v,\spb}'(0) \Psi_\mu^\tau(u,\spa;v,\spb) \ge 0$ for all $(v,\spb)$. 
Meanwhile, for the region $\mathcal{A}_1$ given in \cref{it:f-iii} of \cref{clm:f}, let $\psi: \mathcal{A}_1 \to \D(0,1)$ be a holomorphic function such that $\psi'(1) \in \R_+$. 
We will specify our choice of $\varphi$ and $\psi$ soon. 
Also, we point out here that our assumptions $\varphi_{v,\spb}'(0) \in \R_+ /\R_-$ and $\psi'(1) \in \R_+$ would not cause strong restrictions; they can be easily satisfied by considering rotations $\varphi(e^{i \theta} z)$ and $e^{i \theta} \psi(z)$. 

Given such $\varphi$ and $\psi$, 
we define the holomorphic function $F: \D(0,1) \to \D(0,1)$ given by $F = \psi \circ f \circ \varphi$. 
Notice that $F(0) = \psi(1)$. 
The derivative $F'(0)$ at $0$ is real and can be bounded by
\begin{align}
F'(0) &= \psi'(1) \sum_{(v,\spb) \in \VSpairs'} \varphi_{v,\spb}'(0) \left. \frac{\partial f}{\partial \lambda_{v,\spb}} \right|_{\boldlam = \boldone} \nonumber\\
&= \psi'(1) \sum_{(v,\spb) \in \VSpairs'} \varphi_{v,\spb}'(0) \Psi_\mu^\tau(u,\spa;v,\spb) \nonumber\\
&\ge \psi'(1) \min_{(v,\spb) \in \VSpairs'} \left\{ \left| \varphi_{v,\spb}'(0) \right| \right\}  S_\mu^\tau(u,\spa), \label{eq:f_bound}
\end{align}
where the second equality follows from \cref{it:f-ii} of \cref{clm:f} and the inequality is due to our assumption that $\varphi_{v,\spb}'(0) \Psi_\mu^\tau(u,\spa;v,\spb) \ge 0$ for each $(v,\spb)$. 
The Schwarz-Pick Theorem (\cref{thm:Schwarz-Pick}) implies that $F'(0) \le 1$, and hence we obtain
\begin{equation}\label{eq:S-bound}
S_\mu^\tau(u,\spa) 
\le \frac{1}{\psi'(1)} \left( \min_{(v,\spb) \in \VSpairs'} \left|\varphi_{v,\spb}'(0)\right| \right)^{-1}. 
\end{equation}

It remains to choose $\varphi$ and $\psi$. Consider first the function $\varphi$. 
For each $(v,\spb) \in \VSpairs'$ we let
\begin{align}
\delta_{v,\spb} := \dist\left( 1, \boundary{\Gamma_{v,\spb}} \right)
\quad\text{and}\quad
\chi_{v,\spb} := \sgn(\Psi_\mu^\tau(u,\spa;v,\spb)) = 
\begin{cases}
+1, & \Psi_\mu^\tau(u,\spa;v,\spb) \ge 0;\\
-1, & \Psi_\mu^\tau(u,\spa;v,\spb) < 0.
\end{cases}
\end{align}
We then define $\varphi_{v,\spb}: \D(0,1) \to \Gamma_{v,\spb}$ by 
\begin{equation}\label{eq:phi_def}
\varphi_{v,\spb}(z) = 1 + \chi_{v,\spb} \delta_{v,\spb} z.
\end{equation}
Observe that $\varphi$ is holomorphic, $\varphi(0) = \boldone$, and $\varphi_{v,\spb}'(z) = \chi_{v,\spb} \delta_{v,\spb}$ has the same sign as $\Psi_\mu^\tau(u,\spa;v,\spb)$ for each $(v,\spb)$. 
Recall that $\delta = \min_{(v,\spb) \in \VSpairs_1} \dist\left( 1, \boundary{\Gamma_{v,\spb}} \right)$, and thus
\begin{equation}\label{eq:phi-bound}
\min_{(v,\spb) \in \VSpairs'} \left|\varphi_{v,\spb}'(0)\right| = \min_{(v,\spb) \in \VSpairs'} \delta_{v,\spb} \ge \delta.
\end{equation}

Next, we decide $\psi$. We will actually give two choices of $\psi$, denoted by $\psi_1$ and $\psi_2$ respectively, which correspond to the two bounds in \cref{lem:S-nz-bound}. 

We first consider the simpler choice $\psi_1$. 
Let $\delta_{u,\spa} = \dist\left( 1, \boundary{\Gamma_{u,\spa}} \right) \ge \delta$, and so $\D(1,\delta_{u,\spa}) \subseteq \Gamma_{u,\spa}$. 
Then, the region $\mathcal{A}$ from \cref{it:f-iii} of \cref{clm:f} satisfies
\[
\mathcal{A} = -\frac{1}{p_{u,\spa}^\tau} \left( \Gamma_{u,\spa} - 1 \right)^{-1}
\supseteq  -\frac{1}{p_{u,\spa}^\tau} \left( \D(1,\delta_{u,\spa}) - 1 \right)^{-1}
= \frac{1}{p_{u,\spa}^\tau \delta_{u,\spa}} \closure{\D}(0, 1)^\ccomp. 
\]
It follows that
\[
\mathcal{A}_1 \subseteq \closure{\mathcal{A}}^\ccomp \subseteq \frac{1}{p_{u,\spa}^\tau \delta_{u,\spa}} \D\left( 0, 1 \right).
\]
We can define $\psi_1: \mathcal{A}_1 \to \D(0,1)$ as
\[
\psi_1(z) = p_{u,\spa}^\tau \delta_{u,\spa} z. 
\]
Then, $\psi_1$ is holomorphic, $\psi_1'(z) = p_{u,\spa}^\tau \delta_{u,\spa} \in \R_+$, and 
\begin{equation}\label{eq:psi1-bound}
\frac{1}{\psi_1'(1)} = \frac{1}{p_{u,\spa}^\tau \delta_{u,\spa}} \le \frac{1}{\pp\delta}.
\end{equation}
Combining \cref{eq:S-bound,eq:phi-bound,eq:psi1-bound}, we obtain
\[
S_\mu^\tau(u,\spa) \le \frac{1}{\pp\delta^2}. 
\]
This shows the first bound in \cref{lem:S-nz-bound}. 

Next, we define $\psi_2$. 
Since $\emptyset \neq \mathcal{A}_1 \subsetneq \C$ is open and simply connected by \cref{it:f-iii} of \cref{clm:f}, 
the Riemann Mapping Theorem (\cref{thm:Riemann_mapping}) implies that there exists a (unique) biholomorphic mapping $\psi_2: \mathcal{A}_1 \to \D(0,1)$ such that $\psi_2(1) = 0$ and $\psi_2'(1) \in \R_+$. 
Write $h = \psi_2^{-1}$, which is a bijective holomorphic function from $\D(0,1)$ to $\mathcal{A}_1$ satisfying $h(0) = 1$. 
Then, Koebe's One-Quarter Theorem (\cref{thm:Koebe-1/4}) shows that
\[
\frac{1}{4}|h'(0)| \le \dist(1, \boundary{\mathcal{A}_1}) \le \dist(1, \mathcal{A}) = \dist\left( 1, \mathcal{C}_{u,\spa}^\tau \right). 
\]
It follows that
\begin{equation}\label{eq:psi2-bound}
\frac{1}{\psi_2'(1)} = h'(0) \le 4 \dist\left( 1, \mathcal{C}_{u,\spa}^\tau \right). 
\end{equation}
Combining \cref{eq:S-bound,eq:phi-bound,eq:psi2-bound}, we get
\[
S_\mu^\tau(u,\spa) \le \frac{4}{\delta} \dist\left( 1, \mathcal{C}_{u,\spa}^\tau \right),
\]
which is the second bound in \cref{lem:S-nz-bound}. 


\begin{remark}
The proof of \cref{lem:S-nz-bound} (and also \cref{lem:S-zero-bound} in \cref{subsubsec:map2}) leaves the possibility of further improvements on the spectral independence bounds for specific problems. 
Here in the proof we are given regions $\mathcal{K}$ and $\mathcal{A}_1$ in abstract forms and the choices of $\varphi$ and $\psi$ may not be optimal for specific instances; in particular, the Riemann Mapping Theorem only shows the existence of a biholomorphic mapping and there is no guarantee that such a choice is the best possible. 
Hence, for specific problems and specific zero-free regions, one may be able to pick $\varphi$ and $\psi$ in a smarter way to achieve a better bound on spectral independence. 
\end{remark}

It remains to prove \cref{clm:f}. 
The following lemma is helpful to us.

\begin{lemma}\label{lem:simply-connected}
Let $\mathcal{S} \subseteq \C$ be a non-empty open connected region such that $\mathcal{S}$ is unbounded and $\closure{\mathcal{S}} \neq \C$. 
If $\mathcal{S}_1$ is a connected component of $\closure{\mathcal{S}}^\ccomp$, then $\mathcal{S}_1$ is open and simply connected. 
\end{lemma}
\begin{proof}
Clearly $\mathcal{S}_1$ is open and connected. 
If $\mathcal{S}_1$ is not simply connected, then there exists a Jordan curve (simple closed curve) $\gamma$ in $\mathcal{S}_1$ whose interior region contains a point $z_0 \notin \mathcal{S}_1$. 
Note that we can actually find a point $z$ from the interior of $\gamma$ such that $z \in \closure{\mathcal{S}}$;
if not, then the whole interior of $\gamma$ is contained in $\closure{\mathcal{S}}^\ccomp$ and thus $z_0 \notin \mathcal{S}_1$ is connected to $\mathcal{S}_1$ in $\closure{\mathcal{S}}^\ccomp$, contradicting to the assumption that $\mathcal{S}_1$ is a connected component of $\closure{\mathcal{S}}^\ccomp$. 
Since the interior of $\gamma$ is open, this further implies that the interior of $\gamma$ contains a point $z \in \mathcal{S}$. 
Meanwhile, since $\mathcal{S}$ is unbounded the exterior of $\gamma$ contains a point $w \in \mathcal{S}$. 
Now, as $\mathcal{S}$ is connected there exists a path $p$ in $\mathcal{S}$ connecting $z$ and $w$. 
Note that $p$ must intersect with $\gamma$, because the interior and exterior of $\gamma$ are disconnected. 
This yields a contradiction since $\gamma \subseteq \mathcal{S}_1 \subseteq \mathcal{S}^\ccomp$ while $p \subseteq \mathcal{S}$. 
\end{proof}

Notice that the assumption of $\mathcal{S}$ being unbounded is necessary; for example, if $\mathcal{S} = \D(0,1)$ then $\closure{\mathcal{S}}^\ccomp = \C \setminus \closure{\D}(0,1) = \{z \in \C: |z| > 1\}$ is open and connected, but not simply connected.

We complete the proof of \cref{lem:S-nz-bound} with the proof of \cref{clm:f}. 

\begin{proof}[Proof of Claim~\ref{clm:f}]
1. Since $Z_\weight^\tau(\boldlam \cup \boldone_u) \neq 0$ whenever $\boldlam \in \mathcal{K}$ by our stability assumption, the function $f$ is well-defined and holomorphic on $\mathcal{K}$. 
Also, by definition we have $f(\boldone) = 1$.

\medskip
\noindent 2. 
Let $(v,\spb) \in \VSpairs'$. Then one has
\[
\frac{\partial f}{\partial \lambda_{v,\spb}} 
= \frac{1}{p_{u,\spa}^\tau} \left( 
	\frac{1}{Z_\weight^\tau (\boldlam \cup \boldone_u)} \left( \frac{\partial}{\partial \lambda_{v,\spb}} Z_\weight^{\tau \cup (u,\spa)} (\boldlam) \right)
	- 
	\frac{Z_\weight^{\tau \cup (u,\spa)} (\boldlam)}{Z_\weight^\tau (\boldlam \cup \boldone_u)^2} \left( \frac{\partial}{\partial \lambda_{v,\spb}} Z_\weight^\tau (\boldlam \cup \boldone_u) \right)
  \right). 
\]
Suppose $\tau$ is a pinning on $\Lambda \subseteq V$ and let $U = V^{\tau \cup (u,\spa)} = V \setminus \Lambda \setminus \{u\}$ be the set of unpinned vertices under the pinning $\tau \cup (u,\spa)$. 
We deduce that, 
\begin{align*}
\frac{\partial}{\partial \lambda_{v,\spb}} Z_\weight^{\tau \cup (u,\spa)} (\boldlam) 
&= \sum_{\sigma \in \Omega: \, \sigma_\Lambda = \tau, \sigma_u = \spa} \weight(\sigma) \cdot \frac{\partial}{\partial \lambda_{v,\spb}} \boldlam^{\sigma_U} \\
&= \sum_{\sigma \in \Omega: \, \sigma_\Lambda = \tau, \sigma_u = \spa, \sigma_v = \spb} \weight(\sigma) \boldlam^{\sigma_{U \setminus \{v\}}} \\
&= Z_\weight^{\tau \cup (u,\spa) \cup (v,\spb)} (\boldlam).
\end{align*}
Similarly,
\[
\frac{\partial}{\partial \lambda_{v,\spb}} Z_\weight^\tau (\boldlam \cup \boldone_u)
= Z_\weight^{\tau \cup (v,\spb)} (\boldlam \cup \boldone_u).
\]
We then get
\begin{align*}
\left. \frac{\partial f}{\partial \lambda_{v,\spb}} \right|_{\boldlam = \boldone}
&= \frac{1}{p_{u,\spa}^\tau} \left( 
	\frac{Z_\weight^{\tau \cup (u,\spa) \cup (v,\spb)} (\boldone)}{Z_\weight^\tau (\boldone)} 
	- 
	\frac{Z_\weight^{\tau \cup (u,\spa)} (\boldone) \cdot Z_\weight^{\tau \cup (v,\spb)} (\boldone)}{Z_\weight^\tau (\boldone)^2} 
  \right) \\
&= \frac{1}{\mu^\tau(\sigma_u = \spa)} \left( \mu^\tau(\sigma_u = \spa, \sigma_v = \spb) - \mu^\tau(\sigma_u = \spa) \mu^\tau(\sigma_v = \spb) \right) \\
&= \Psi_\mu^\tau(u,\spa;v,\spb),  
\end{align*}
as claimed.

\medskip
\noindent 3. 
We first show that $\image(f) \subseteq \mathcal{A}^\ccomp$. 
Suppose for sake of contradiction that $f(\boldlam) \in \mathcal{A}$ for some $\boldlam \in \mathcal{K}$. 
Then there exists $1 \neq y \in \Gamma_{u,\spa}$ such that 
\[
- \frac{1}{p_{u,\spa}^\tau (y-1)} = f(\boldlam) = \frac{1}{p_{u,\spa}^\tau} \frac{Z_\weight^{\tau \cup (u,\spa)} (\boldlam)}{Z_\weight^\tau (\boldlam \cup \boldone_u)}. 
\]
It follows that
\[
Z_\weight^\tau (\boldlam \cup \boldlam_u)
= y Z_\weight^{\tau \cup (u,\spa)} (\boldlam) + \sum_{\spa \neq \spa' \in \Omega_u^\tau} Z_\weight^{\tau \cup (u,\spa')} (\boldlam) 
= 0,
\]
where $\boldlam_u$ is the vector of external fields at $u$ defined by $\lambda_{u,\spa} = y$ and $\lambda_{u,\spa'} = 1$ for $\spa' \in \Omega_u^\tau \setminus \{0,\spa\}$. 
This contradicts our stability assumption that $Z_\weight^\tau (\boldlam \cup \boldlam_u) \neq 0$. 
Therefore, we have shown that $\image(f) \subseteq \mathcal{A}^\ccomp$. 

Now, since $\mathcal{K}$ is open and connected and $f$ is a non-constant holomorphic function, the Open Mapping Theorem (\cref{thm:open-map}) implies that $\image(f)$ is open and connected. 
Thus, we have $\image(f) \subseteq \interior{(\mathcal{A}^\ccomp)} = \closure{\mathcal{A}}^\ccomp$; note that in particular $1 \in \closure{\mathcal{A}}^\ccomp$. 
Furthermore, since $\image(f)$ is connected one has $\image(f) \subseteq \mathcal{A}_1$, the connected component of $\closure{\mathcal{A}}^\ccomp$ containing $1$. 
The region $\mathcal{A}_1$ is open and connected by definition. 
It remains to show that $\mathcal{A}_1$ is simply connected, which follows immediately from \cref{lem:simply-connected} and the fact that $\mathcal{A}$ is connected and unbounded. 
\end{proof}

\subsubsection{Proof of \texorpdfstring{\cref{lem:S-zero-bound}}{Lemma 3.9}}
\label{subsubsec:map2}

The proof of \cref{lem:S-zero-bound} is similar to that of \cref{lem:S-nz-bound}. 
We will use the same notations and only emphasize a few key steps that differ. 

Recall that
\[
\VSpairs' = \{ (v,\spb) \in \VSpairs_1^\tau: v \neq u \}
\quad\text{and}\quad
\mathcal{K} = \prod_{(v,\spb) \in \VSpairs'} \Gamma_{v,\spb}.
\]
Define the multivariate complex function $g: \mathcal{K} \to \C$ as
\begin{equation}\label{eq:g-func}
g(\boldlam) = \frac{1}{p_{u,0}^\tau} \frac{Z_\weight^{\tau \cup (u,0)} (\boldlam)}{Z_\weight^\tau (\boldlam \cup \boldone_u)}, 
\quad\text{for~} \boldlam \in \mathcal{K}
\end{equation}
where $\tau \cup (u,0) \in \Pinning$ is the pinning combining $\tau$ and $\sigma_u = 0$, and $\boldlam \cup \boldone_u$ is the vector of external fields that combines $\boldlam$ and $\boldone_u$. 
The following claim is analogous to \cref{clm:f} and summarizes key properties of the function $g$.

\begin{claim}\label{clm:g}
Let $g: \mathcal{K} \to \C$ be the multivariate complex function defined by \cref{eq:g-func}. 
\begin{enumerate}
\item \label{it:g-i} The function $g$ is well-defined and holomorphic on $\mathcal{K}$, and $g(\boldone) = 1$. 
\item \label{it:g-ii} For every $(v,\spb) \in \VSpairs'$, 
\[
\left. \frac{\partial f}{\partial \lambda_{v,\spb}} \right|_{\boldlam = \boldone} = \Psi_\mu^\tau(u,0;v,\spb). 
\]
\item \label{it:g-iii} Suppose that $g \not\equiv 1$. 
Let $\mathcal{B} \subseteq \C$ be an open region defined as
\begin{align}
\mathcal{B} := \frac{1}{p_{u,0}^\tau} \left(\left( \Gamma_u - 1 \right)^{-1} + 1 \right).
\end{align}
Then $1 \notin \closure{\mathcal{B}}$. 
Let $\mathcal{B}_1$ be the connected component of $\closure{\mathcal{B}}^\ccomp$ which contains $1$. 
Then $\mathcal{B}_1$ is open and simply connected, and 
\[
\image(g) \subseteq \mathcal{B}_1.
\]
\end{enumerate}
\end{claim}

We may assume that $g \not\equiv 1$ since otherwise $S_\mu^\tau(u,0) = 0$ and the lemma is trivial. 
Again we choose holomorphic functions $\varphi: \D(0,1) \to \mathcal{K}$, $\psi: \mathcal{B}_1 \to \D(0,1)$ and consider the holomorphic function $G: \D(0,1) \to \D(0,1)$ defined as $G = \psi \circ g \circ \varphi$. 
Just as in the proof of \cref{lem:S-nz-bound}, we let $\varphi: \D(0,1) \to \mathcal{K}$ be a holomorphic vector-valued function such that for every $(v,\spb) \in \VSpairs'$, the $(v,\spb)$-coordinate function $\varphi_{v,\spb}: \D(0,1) \to \Gamma_{v,\spb}$ is holomorphic and satisfies $\varphi_{v,k}(0) = 1$ and $\varphi_{v,\spb}'(0) \in \R_+$ if $\Psi_\mu^\tau(u,\spa;v,\spb) \ge 0$ while $\varphi_{v,\spb}'(0) \in \R_-$ if $\Psi_\mu^\tau(u,\spa;v,\spb) \le 0$. 
Meanwhile, let $\psi: \mathcal{B}_1 \to \D(0,1)$ be a holomorphic function such that $\psi'(1) \in \R_+$. 
Hence, we have $G(0) = \psi(1)$, and by \cref{clm:g} $G'(0) \in \R$ can be bounded by
\[
G'(0) 
\ge \psi'(1) \min_{(v,\spb) \in \VSpairs'} \left\{ \left| \varphi_{v,\spb}'(0) \right| \right\}  S_\mu^\tau(u,0), 
\]
which is analogous to \cref{eq:f_bound}.
We then deduce the analog of \cref{eq:S-bound} from the Schwarz-Pick Theorem (\cref{thm:Schwarz-Pick}):
\begin{equation}\label{eq:S-bound-zero}
S_\mu^\tau(u,0) 
\le \frac{1}{\psi'(1)} \left( \min_{(v,\spb) \in \VSpairs'} \left|\varphi_{v,\spb}'(0)\right| \right)^{-1}. 
\end{equation}

We specify next our choice of $\varphi$ and $\psi$. 
The function $\varphi$ is the same one as in the proof of \cref{lem:S-nz-bound}, and is given by \cref{eq:phi_def}. 
In particular, \cref{eq:phi-bound} still holds. 
We also give two choices of the function $\psi$, denoted by $\psi_3$ and $\psi_4$ respectively, corresponding to the two bounds in \cref{lem:S-zero-bound}. 

Consider first $\psi_3$. 
Recall that $\Gamma_u \subseteq \C$ is defined to be the connected component of the intersection $\bigcap_{0 \neq \spa \in \Omega_u^\tau} \Gamma_{u,\spa}$ that contains $1$. 
Let $\delta_u = \dist\left( 1, \boundary{\Gamma_u} \right) \ge \delta$ and thus $\D(1,\delta_u) \subseteq \Gamma_u$. 
Then we have
\[
\mathcal{B} = \frac{1}{p_{u,0}^\tau} \left(\left( \Gamma_u - 1 \right)^{-1} + 1 \right)
\supseteq  \frac{1}{p_{u,0}^\tau} \left( \D(0,\delta_u)^{-1} + 1 \right)
= \frac{1}{p_{u,0}^\tau \delta_u} \closure{\D}(\delta_u, 1)^\ccomp,
\]
and hence
\[
\mathcal{B}_1 \subseteq \closure{\mathcal{B}}^\ccomp \subseteq \frac{1}{p_{u,0}^\tau \delta_u} \D\left( \delta_u, 1 \right).
\]
We define $\psi_3: \mathcal{B}_1 \to \D(0,1)$ as
\[
\psi_3(z) = p_{u,0}^\tau \delta_u z - \delta_u. 
\]
Observe that $\psi_3$ is holomorphic, $\psi_3'(z) = p_{u,0}^\tau \delta_u \in \R_+$, and 
\begin{equation}\label{eq:psi3-bound}
\frac{1}{\psi_3'(1)} = \frac{1}{p_{u,0}^\tau \delta_u} \le \frac{1}{\pp\delta}.
\end{equation}
Combining \cref{eq:S-bound-zero,eq:phi-bound,eq:psi3-bound}, we obtain
\[
S_\mu^\tau(u,0) \le \frac{1}{\pp\delta^2}. 
\]
This shows the first bound in \cref{lem:S-zero-bound}. 

Finally, we define $\psi_4$. 
Since $\emptyset \neq \mathcal{B}_1 \subsetneq \C$ is open and simply connected by \cref{clm:g}, 
there exists a (unique) biholomorphic mapping $\psi_4: \mathcal{B}_1 \to \D(0,1)$ such that $\psi_4(1) = 0$ and $\psi_4'(1) \in \R_+$ by the Riemann Mapping Theorem (\cref{thm:Riemann_mapping}). 
Let $h = \psi_4^{-1}$ be the holomorphic mapping from $\D(0,1)$ to $\mathcal{B}_1$ with $h(0) = 1$. 
We deduce from the Koebe's One-Quarter Theorem (\cref{thm:Koebe-1/4}) that
\[
\frac{1}{4}|h'(0)| \le \dist(1, \boundary{\mathcal{B}_1}) \le \dist(1, \mathcal{B}) = \dist\left( 1, \mathcal{C}_{u,0}^\tau \right),  
\]
and hence
\begin{equation}\label{eq:psi4-bound}
\frac{1}{\psi_4'(1)} = h'(0) \le 4 \dist\left( 1, \mathcal{C}_{u,0}^\tau \right). 
\end{equation}
Combining \cref{eq:S-bound-zero,eq:phi-bound,eq:psi4-bound}, we get
\[
S_\mu^\tau(u,0) \le \frac{4}{\delta} \dist\left( 1, \mathcal{C}_{u,0}^\tau \right),
\]
which is the second bound in \cref{lem:S-zero-bound}.

\medskip 
We end this section with the proof of \cref{clm:g}.

\begin{proof}[Proof of Claim~\ref{clm:g}]
\cref{it:g-i} follows from the stability of the partition function 
and \cref{it:g-ii} can be deduced by direct calculations. 
We omit the details here and refer to the proof of \cref{clm:f}.

For \cref{it:g-iii}, again we first show that $\image(g) \subseteq \mathcal{B}^\ccomp$. 
Suppose for sake of contradiction that $g(\boldlam) \in \mathcal{B}$ for some $\boldlam \in \mathcal{K}$. 
Then there exists $1 \neq y \in \Gamma_u \subseteq \bigcap_{0 \neq \spa \in \Omega_u^\tau} \Gamma_{u,\spa}$ such that 
\[
\frac{y}{p_{u,0}^\tau (y-1)} = g(\boldlam) = \frac{1}{p_{u,0}^\tau} \frac{Z_\weight^{\tau \cup (u,0)} (\boldlam)}{Z_\weight^\tau (\boldlam \cup \boldone_u)}. 
\]
It follows that
\[
Z_\weight^\tau (\boldlam \cup y \boldone_u)
= Z_\weight^{\tau \cup (u,0)} (\boldlam) + \sum_{0 \neq \spa \in \Omega_u^\tau} y Z_\weight^{\tau \cup (u,\spa)} (\boldlam) 
= 0,
\]
where $y \boldone_u$ represents the vector of external fields at $u$ defined by $\lambda_{u,\spa} = y$ for all $0 \neq \spa \in \Omega_u^\tau$. 
This contradicts the stability assumption of the partition function. 
Therefore, $\image(g) \subseteq \mathcal{B}^\ccomp$. 
The Open Mapping Theorem (\cref{thm:open-map}) then implies that $\image(g) \subseteq \mathcal{B}_1$ which is the connected component of $\closure{\mathcal{B}}^\ccomp$ containing $1$. 
Meanwhile, notice that the region $\Gamma_u$ is open and connected since it is a connected component of the open set $\bigcap_{0 \neq \spa \in \Omega_u^\tau} \Gamma_{u,\spa}$, and so $\mathcal{B}$ is open, connected, and unbounded. 
Hence, \cref{lem:simply-connected} shows that $\mathcal{B}_1$ is open and simply connected. 
This completes the proof of the claim. 
\end{proof}

\section{Spectral Independence for Binary Symmetric Holant Problems}\label{sec:binarysymmetric}

Let $G = (V,E)$ be a graph of maximum degree $\Delta$. We consider the Holant problem in the binary symmetric case, which we now describe. Let $\{f_{v}\}_{v \in V}: \N \to \R_{\ge 0}$ be a family of functions, one for each vertex $v \in V$ in the input graph. One should think of each $f_{v}$ as representing a local constraint on the assignments to edges incident to $v$. Since we are restricting ourselves to the binary case, our configurations $\sigma$ will map edges to $\{0,1\}$. Furthermore, since we are restricting ourselves to the symmetric case, our local functions $f_{v}$ will only depend on the number of edges incident to $v$ which are mapped to $1$. With these $\{f_{v}\}_{v\in V}$ in hand, we may write the multivariate partition function as
\begin{equation}\label{eq:Holant_pf}
Z_{G}(\lambda) = \sum_{\sigma: E \to \{0,1\}} \prod_{v \in V} f_{v}(|\sigma_{E(v)}|) \prod_{e \in E} \lambda_e^{\one\{\sigma_e = 1\}},
\end{equation}
where $E(v)$ is the set of all edges adjacent to $v$, $\sigma_{E(v)}$ is the configuration restricted on $E(v)$, and $|\sigma_{E(v)}|$ is the number of edges in $E(v)$ with assignment $1$. 

This class of problems is already incredibly rich, and encompasses many classical objects studied in combinatorics and statistical physics including the following:
\begin{itemize}
\item \textit{Matchings/Monomer-Dimer Model:} Assume all $f_{v}$ are the same and given by the ``at-most-one'' function:
\[
f_{v}(k) = 
\begin{cases}
1, & \text{if~} k = 0, 1;\\
0, & \text{if~} k \ge 2.
\end{cases}
\]
Then $Z_{G}(\mathbf{1})$ yields the number of matchings (of any size) in $G$.
\item \textit{Weighted Edge Covers:} Assume all $f_{v}$ are the same and given by the weighted ``at-least-one'' function:
\[
f_{v}(k) = 
\begin{cases}
\rho, & \text{if~} k = 0;\\
1, & \text{if~} k \ge 1.
\end{cases}
\]
In the case $\rho = 0$, then $Z_{G}(\mathbf{1})$ yields the number of edge covers of $G$, that is, subsets of edges such that every vertex is incident to at least one selected edge.
\item \textit{Weighted Even Subgraphs:} In this case, all $f_{v}$ are the same and given by the weighted ``parity'' function. More specifically, for a fixed positive parameter $\rho > 0$, we have
\[
f_{v}(k) = 
\begin{cases}
1, & \text{if~$k$ is even};\\
\rho, & \text{if~$k$ is odd}.
\end{cases}
\]
In the case $\rho = 0$, then $Z_{G}(\mathbf{1})$ counts the number of even subgraphs, that is, subsets of edges such that all vertices have even degrees in the resulting subgraph. (Note that when $\rho = 0$, the Glauber dynamics is not ergodic.) 
\item \textit{Ising Model on Line Graphs:}
In this case, each $f_{v}$ depends on the degree of $v$. If $\beta > 0$ is some fixed parameter (independent of $v$), and $d = \deg(v)$, then we have
\begin{align*}
    f_{v}(k) &= \begin{cases}
        \beta^{\binom{k}{2}}\beta^{\binom{d-k}{2}},
        &\text{if~$0 \leq k \leq d$}; \\
        0, &\text{o/w.}
    \end{cases}
\end{align*}
\end{itemize}


In all of the above examples, prior works managed to show that the Glauber dynamics admits an inverse polynomial spectral gap (\cite{JSmatchings} for matchings, \cite{HLZ16} for edge covers, \cite{JSising} for weighted even subgraphs, and \cite{DHJM21} for the Ising model in the antiferromagnetic $\beta < 1$ regime). Furthermore, all of these results were obtained via the canonical paths method \cite{JSmatchings}, and its winding extension \cite{McQ}. However, one down-side behind these results is that the spectral gap bounds are suboptimal, and do not yield optimal mixing times nor sub-Gaussian concentration estimates. In contrast, by combining our framework with known zero-free regions for these models and the local-to-global mixing result of \cite{CLV21}, we obtain optimal mixing times and sub-Gaussian concentration results for these problems in the bounded-degree regime.

One of the convenient aspects of our approach is that establishing the required zero-free region for the complicated multivariate partition function can be boiled down to establishing stability for a bounded-degree univariate polynomial with coefficients coming from the local functions $f_{v}$. This was one of the main insights of \cite{Wag09, GLLZ21, BCR20}. More specifically, if $\Delta$ is the maximum degree of the input graph $G=(V,E)$, and $f_{v}: [d] \to \R_{\ge 0}$ is the local function for some vertex $v \in V$, where $d = \deg(v) \le \Delta$, then define the corresponding local polynomial at $v$ by
\begin{equation}\label{eq:Pfd}
P_{v}(z) = \sum_{k = 0}^d \binom{d}{k} f_{v}(k) z^k.
\end{equation}

A circular region on the complex plane is the interior or exterior of a disk, or an open half-plane. 
\cite{GLLZ21} showed using Asano--Ruelle contractions \cite{Asa70, Rue71} that in the case all $f_{v}$ are the same, and all $P_{v}$ are $\Phi$-stable for an open half-plane $\Phi \subseteq \C$, the multivariate partition function is $\Gamma$-stable where $\Gamma = \left[-(\Phi^{\ccomp})^{2}\right]^{\ccomp}$. 
This result actually holds for any circular region $\Phi \subseteq \C$ assuming that either $\Phi$ is convex or every local polynomial $P_{v}$ has degree $\deg(v)$; under these assumptions one can apply the famous Grace--Walsh--Szeg\"{o} Coincidence Theorem to the local polynomials, see \cite{GLLZ21,BB09}. 
A straightforward generalization of their techniques yields the following.
\begin{theorem}[\cite{GLLZ21}]\label{thm:holant_zf}
Let $G=(V,E)$ be a graph. Let $\{f_{v}\}_{v \in V} : \N \rightarrow \R_{\geq 0}$ be a family of local functions, and let $\{\Phi_{v}\}_{v \in V}$ be a family of circular regions containing $0$ such that for every $v \in V$, either $\Phi_{v}$ is convex or $f_{v}(\deg(v)) > 0$. If for every $v \in V$, the local polynomial $P_{v}$ is $\Phi_{v}$-stable, then the multivariate partition function $Z_{G}(\lambda)$ is $\prod_{e \in E} \Gamma_{e}$ stable, where for each edge $e = \{u,v\}$, $\Gamma_{e} = \left(-\Phi_{u}^{\ccomp} \cdot \Phi_{v}^{\ccomp}\right)^{\ccomp} \subseteq \C$.
\end{theorem}

Using \cref{thm:holant_zf}, \cite{GLLZ21} established zero-free regions for a large class of Holant problems satisfying generalized second-order recurrences, including matchings, weighted edge covers, and weighted even subgraphs. 
Our main theorems \cref{thm:edge_covers,thm:even_subgraphs,thm:anti_2-spin_edge} build upon these zero-free results as well as \cref{thm:main_R+,thm:spinsystemlocaltoglobal} (note that we can obtain spectral independence for matchings from \cref{thm:holant_zf,thm:main_R+}, which was already known in \cite{CLV21} with a better bound by correlation decay proofs). 
Zero-free regions were also established for weighted edge covers and the antiferromagnetic Ising model on line graphs in \cite{BCR20}, using techniques from \cite{Wag09}.

Before proving the main theorems, we will need the following simple lemma concerning the case where the regions $\Phi_{u}$ are half-planes.
Recall that $\HP_\eps = \{x + iy: x < -\eps\}$ and $\closure{\HP}_\eps = \{x + iy: x \le -\eps\}$ for $\eps \in \R_+$. 
\begin{lemma}[Lemma 5 in \cite{GLLZ21}]\label{lem:leftplaneprod}
For $\epsilon > 0$, let $\Gamma = ( -\closure{\HP}_\eps^2 )^{\ccomp}$ be a region. 
Then $\Gamma$ contains $\R_+$, and for every $\lambda \in \R_+$ we have $\dist(\lambda, \boundary{\Gamma}) = \lambda + \eps^2$ if $\lambda \in (0,\eps^2)$, and $\dist(\lambda, \boundary{\Gamma}) = 2\eps \sqrt{\lambda}$ if $\lambda \in [\eps^2,\infty)$. 
\end{lemma}

For completeness, we provide a proof in \cref{app:technicallemmas}. With these tools in hand, we deduce strong zero-free regions for the above examples. We use these to prove our main mixing results \cref{thm:edge_covers,thm:even_subgraphs,thm:anti_2-spin_edge}. 
Note that by \cref{lem:pinning-stable,thm:multi-spin}, one can in fact establish rapid mixing results for these models with non-uniform external fields, though we only state the uniform case for simplicity.

\begin{proof}[Proof of Theorem~\ref{thm:edge_covers}]
By \cref{thm:spinsystemlocaltoglobal}, it suffices to prove $\eta$-spectral independence for $\eta = O_{\Delta,\lambda,\rho}(1)$.
By \cref{thm:main_R+}, it suffices to prove that the multivariate partition function \cref{eq:Holant_pf} is $\Gamma$-stable, where $\Gamma \subseteq \C$ is an open connected region containing $\R_{+}$ and $\delta = \frac{1}{\lambda} \dist(\lambda, \boundary{\Gamma}) = \Omega_{\Delta,\lambda,\rho}(1)$.

It is more convenient for us to work with the model on complements of weighted edge covers, whose partition function is the inversion of that for weighted edge covers. 
For this, the local polynomial is given by
\[
P_{v}(z) = (1 + z)^{\deg(v)} - (1 - \rho) z^{\deg(v)},
\] 
which is $\closure{\HP}_{1/2}^{\ccomp}$-stable. 
Then by \cref{thm:holant_zf}, the inversion of the weighted edge cover partition function $Z_{G}(\lambda)$ is $\left( -\closure{\HP}_{1/2}^2 \right)^\ccomp$-stable, 
and therefore $Z_{G}(\lambda)$ is $\Gamma$-stable for
\[
\Gamma 
= \left[ \left( -\closure{\HP}_{1/2}^2 \right)^\ccomp \right]^{-1}
= [-\closure{\D}(-1, 1)^{2}]^{\ccomp}.
\]
This region $\Gamma$ is also derived in \cite{BCR20}. 
We remark that the region $-\closure{\D}(-1, 1)^{2}$ is cardioid-shaped, and its complement $\Gamma$ is an open connected region containing $\R_{+}$; see Lemma 3.9 and Figure 1 in \cite{BCR20}. 
Hence, we have $\R_+ \subseteq \Gamma$ and $\delta = \Omega_{\Delta,\lambda,\rho}(1)$ as wanted. 
\end{proof}

\begin{proof}[Proof of Theorem~\ref{thm:even_subgraphs}]
We may assume $\rho \in (0,1)$ since if $\rho=1$ then we get a trivial product distribution. 
Once again, by \cref{thm:spinsystemlocaltoglobal}, it suffices to prove $\eta$-spectral independence for $\eta = O_{\Delta,\lambda,\rho}(1)$,
and by \cref{thm:main_R+}, it suffices to prove that the multivariate partition function \cref{eq:Holant_pf} is $\Gamma$-stable, where $\Gamma \subseteq \C$ is an open connected region containing $\R_{+}$ and $\delta = \frac{1}{\lambda} \dist(\lambda, \boundary{\Gamma}) = \Omega_{\Delta,\lambda,\rho}(1)$.

For this, observe that the local polynomial is given by
\begin{align*}
    P_{v}(z) &= \sum_{k=0}^{\deg(v)} \binom{\deg(v)}{k} \left(\frac{1 + \rho}{2} + \frac{1 - \rho}{2} (-1)^{k}\right)z^{k} \\
    &= \frac{1 + \rho}{2} (1 + z)^{\deg(v)} + \frac{1 - \rho}{2}(1 - z)^{\deg(v)}.
\end{align*}
Since $0 < \rho < 1$, the roots of $P_{v}$ are given by $\frac{\omega - t_{v}}{\omega + t_{v}}$ where $\omega \in \C$ satisfies $\omega^{\deg(v)} = -1$, and $t_v \in \R_+$ is given by  
\[
t_{v} = \left(\frac{1 + \rho}{1 - \rho}\right)^{1/\deg(v)} > 1.
\]
It follows that $P_{v}$ is $\left[\closure{\D}\left(-\frac{t_{v}^{2} + 1}{t_{v}^{2} - 1}, \frac{2t_{v}}{t_{v}^{2} - 1}\right)\right]^{\ccomp}$-stable. Then by \cref{thm:holant_zf}, $Z_{G}(\lambda)$ is $\prod_{e \in E} \Gamma_{e}$-stable, where for each edge $e = uv \in E$,
\begin{equation*}
    \Gamma_e = \left[- \closure{\D}\left(-\frac{t_{u}^{2} + 1}{t_{u}^{2} - 1}, \frac{2t_{u}}{t_{u}^{2} - 1}\right) \cdot \closure{\D}\left(-\frac{t_{v}^{2} + 1}{t_{v}^{2} - 1}, \frac{2t_{v}}{t_{v}^{2} - 1}\right)\right]^{\ccomp}. 
\end{equation*}
In particular, $Z_{G}(\lambda)$ is $\Gamma$-stable for
\[
\Gamma = \left[- \closure{\D}\left( - \frac{t^2 + 1}{t^2 - 1}, \frac{2t}{t^2 - 1} \right)^2 \right]^{\ccomp} 
\subseteq \Gamma_e, \quad \forall e \in E, 
\quad \text{where}~ t = \left(\frac{1 + \rho}{1 - \rho}\right)^{1/\Delta} > 1.
\]

The region $\Gamma$ is open and connected. Observe that we have $\Gamma \supseteq \left( - \closure{\HP}_{\frac{t - 1}{t + 1}}^2 \right)^\ccomp$. 
Hence, by \cref{lem:leftplaneprod} we have $\R_+ \subseteq \Gamma$ and $\delta = \Omega_{\Delta,\lambda,\rho}(1)$ as wanted.
\end{proof}

\begin{proof}[Proof of Theorem~\ref{thm:anti_2-spin_edge}]
By \cref{thm:spinsystemlocaltoglobal} it suffices to prove $\eta$-spectral independence for $\eta = O_{\Delta,\beta,\gamma,\lambda}(1)$. By \cref{thm:main_R+} it suffices to prove that the multivariate partition function \cref{eq:Holant_pf} is $\Gamma$-stable, where $\Gamma \subseteq \C$ is an open connected region containing $\R_{+}$ and $\delta = \frac{1}{\lambda} \dist(\lambda, \boundary{\Gamma}) = \Omega_{\Delta,\beta,\gamma,\lambda}(1)$.

For this, observe that the local polynomial is given by 
\[
P_{v}(z) = \sum_{k=0}^{\deg(v)}\binom{\deg(v)}{k} \beta^{\binom{k}{2}}\gamma^{\binom{\deg(v)-k}{2}}z^{k}.
\]
By \cref{prop:antiferrotwospinedge} below (see \cref{subsec:antiferrotwospinedge} for the proof), all roots of these polynomials are strictly negative reals, i.e., they are contained in $(-\infty,-\eps_{\deg(v)}]$ for some constant $\eps_{\deg(v)} = \eps_{\deg(v)}(\beta,\gamma) > 0$ depending only on $\deg(v),\beta,\gamma$. 
Then by \cref{thm:holant_zf}, $Z_{G}(\lambda)$ is $\prod_{e \in E} \Gamma_{e}$-stable, where for each edge $e = uv \in E$, 
\[
\Gamma_{e} = \left( -\closure{\HP}_{\eps_{\deg(u)}} \cdot \closure{\HP}_{\eps_{\deg(v)}} \right)^{\ccomp}.
\]
In particular, $Z_{G}(\lambda)$ is $\Gamma$-stable for $\Gamma = ( -\closure{\HP}_\eps^2 )^{\ccomp}$ where $\eps = \min_{1 \le d \le \Delta} \eps_d$ depends only on $\Delta, \beta, \gamma$. 
The region $\Gamma$ is open and connected, and by \cref{lem:leftplaneprod} it contains $\R_+$ and we have $\delta = \Omega_{\Delta,\beta,\gamma,\lambda}(1)$ as wanted.
\end{proof}

\subsection{Stability for Antiferromagnetic Two-Spin Edge Models}\label{subsec:antiferrotwospinedge}
In this subsection, we analyze the roots of the local polynomial for antiferromagnetic two-spin edge models, which is needed in the proof of \cref{thm:anti_2-spin_edge} above. We generalize a result due to \cite{BCR20} which proves that the local polynomial for the antiferromagnetic edge Ising model has strictly negative real roots. We achieve this by generalizing their arguments to all antiferromagnetic two-spin edge models.
\begin{proposition}[Generalization of Lemma 4.3 in \cite{BCR20}]\label{prop:antiferrotwospinedge}
For every $\beta \ge 0,\, \gamma > 0$ with $\beta\gamma < 1$ and every positive integer $d \ge 1$, the univariate polynomial 
\[
P_{d}(z) = \sum_{k=0}^{d} \binom{d}{k} \beta^{\binom{k}{2}}\gamma^{\binom{d - k}{2}} z^{k}
\]
has strictly negative real roots. 
\end{proposition}
We prove this via an inductive approach, relying on the following decomposition of $P_{d}$.
\begin{lemma}\label{lem:antiferrotwospindecomp}
For every $\beta \ge 0,\, \gamma > 0$ and every positive integer $d \ge 1$, we have that
\begin{align*}
    P_{d+1}(z) = \gamma^{d} P_{d}(z / \gamma) + z P_{d}(\beta z).
\end{align*}
\end{lemma}
\begin{proof}
We have
\begin{align*}
    P_{d+1}(z) &= \sum_{k=0}^{d+1} \underset{= \binom{d}{k} + \binom{d}{k-1}}{\underbrace{\binom{d+1}{k}}}\beta^{\binom{k}{2}}\gamma^{\binom{d+1-k}{2}} z^{k} \\
    &= \sum_{k=0}^{d} \binom{d}{k} \beta^{\binom{k}{2}} \gamma^{\binom{d+1-k}{2}} z^{k} + \sum_{k=0}^{d} \binom{d}{k}\beta^{\binom{k+1}{2}} \gamma^{\binom{d-k}{2}} z^{k+1} \\
    &= \sum_{k=0}^{d} \binom{d}{k} \beta^{\binom{k}{2}} \gamma^{\binom{d-k}{2}} \gamma^{d-k}z^{k} + z \sum_{k=0}^{d} \binom{d}{k} \beta^{\binom{k}{2}} \gamma^{\binom{d-k}{2}} \beta^{k}z^{k} \\
    &= \gamma^{d} P_{d}(z/\gamma) + z P_{d}(\beta z). \qedhere
\end{align*}
\end{proof}

\begin{proof}[Proof of Proposition~\ref{prop:antiferrotwospinedge}]
If $\beta = 0$ then $P_{d}$ is linear and the proposition is immediate. We may assume $\beta > 0$. 
We prove via induction the following stronger claim: The roots $r_{1} > \dots > r_{d}$ of $P_{d}$ are distinct, real, and strictly negative, and further satisfy $r_{i} / r_{i+1} < \beta\gamma$. The cases $d = 0, 1$ are vacuous. When $d = 2$, the polynomial $P_{2}(z) = \beta z^{2} + 2z + \gamma$ has roots $(-1 \pm \sqrt{1 - \beta\gamma})/\beta$, which are distinct, real, and strictly negative since $\beta\gamma < 1$. One can also check that $r_{1} / r_{2} < \beta\gamma$ via a straightforward calculation. This establishes the base case.

Assume the stronger conclusion holds for some $d \geq 2$. By \cref{lem:antiferrotwospindecomp}, we may write $P_{d+1}(z) = \gamma^{d}P_{d}(z / \gamma) + zP_{d}(\beta z)$. If $r_{1} > \dots > r_{d}$ are the roots of $P_{d}$, then $\gamma r_{1} > \dots > \gamma r_{d}$ are the roots of $\gamma^{d}P_{d}(z / \gamma)$, and $0 = r_{0} / \beta > r_{1} / \beta > \dots > r_{d} / \beta$ are the roots of $z P_{d}(\beta z)$, where for convenience we define $r_{0} = 0$. First, we claim that the roots of $\gamma^{d}P_{d}(z / \gamma)$ interlace the roots of $zP_{d}(\beta z)$, i.e., 
\[
0 = r_0 / \beta > \gamma r_1 > r_1 / \beta > \gamma r_2 > \dots > r_{d-1} / \beta > \gamma r_d > r_d / \beta.  
\]
To see this, observe that $\gamma r_{i} > r_{i} / \beta$ since $\beta\gamma < 1$, and $r_{i-1} / \beta > \gamma r_i$ since $r_{i-1} / r_i < \beta \gamma$ by the induction hypothesis for $P_{d}$.

Now, we claim that for each $i=2,\dots,d$, the evaluations 
\[
P_{d+1}(\gamma r_{i}) = \gamma r_{i} P_{d}(\beta\gamma r_{i}) 
\quad\text{and}\quad
P_{d+1}(r_{i-1} / \beta) = \gamma^{d} P_{d}(r_{i-1} / \beta\gamma) 
\]
are nonzero and have different signs. 
Observe that $\beta\gamma r_{i}, r_{i-1} / \beta\gamma \in (r_i, r_{i-1})$; 
hence, the evaluations $P_{d}(\beta\gamma r_{i})$ and $P_{d}(r_{i-1} / \beta\gamma)$ are nonzero and have the same sign, and we deduce the claim by $r_i < 0$.
It then follows from the Intermediate Value Theorem that $P_{d+1}$ has a root $s_i \in (\gamma r_{i}, r_{i-1} / \beta)$ for each $i=2,\dots,d$. 

Moreover, $P_{d+1}$ also has a root $s_1 \in (\gamma r_1, 0)$ and a root $s_{d+1} \in (-\infty, r_d / \beta)$. 
Observe that the evaluations $P_{d+1}(\gamma r_1) = \gamma r_1 P_d(\beta \gamma r_1)$ and $P_{d+1}(0) = \gamma^d P_d(0)$ are nonzero and have different signs since $0 > \beta\gamma r_1 > r_1$, and the Intermediate Value Theorem implies a root $s_1 \in (\gamma r_1, 0)$.
Meanwhile, $P_{d+1}(r_d / \beta) = \gamma^d P_{d}(r_d / \beta\gamma)$ and $P_d(-\infty)$ are nonzero and have the same sign since $-\infty < r_d / \beta\gamma < r_d$. 
Also, $P_d(-\infty)$ and $P_{d+1}(-\infty)$ have different signs since the two polynomials differ in degree by $1$. 
This shows that $P_{d+1}(r_d / \beta)$ and $P_{d+1}(-\infty)$ are nonzero and have different signs, and the Intermediate Value Theorem shows the existence of a root $s_{d+1} \in (-\infty, r_d / \beta)$. 

To summarize, we prove that $P_{d+1}$ has roots $s_{1} > \dots > s_{d+1}$ which are distinct strictly negative real numbers and (taking $r_0 = 0$ and $r_{d+1} = -\infty$ for convenience) satisfy $s_{i} \in (\gamma r_{i}, r_{i-1} / \beta)$ for any $i=1,\dots,d+1$. 
To finish the induction, we need to show that $s_i / s_{i+1} < \beta \gamma$ for all $i=1,\dots,d$, which follows by $s_i / s_{i+1} < (\gamma r_i) / (r_i / \beta) = \beta\gamma$.
\end{proof}

\section{Spectral Independence for Weighted Graph Homomorphisms and Tensor Network Contractions}\label{sec:homtensornetwork}
In this section, we study spectral independence for general tensor network contractions and weighted graph homomorphisms. Unlike binary symmetric Holant problems, where rapid mixing of the Glauber dynamics was already known for our main examples such as matchings \cite{JSmatchings}, Ising model on line graphs \cite{DHJM21}, edge covers \cite{HLZ16}, and weighted even subgraphs \cite{JSising}, in the setting we consider here, rapid mixing for any local Markov chain was not known beyond simple and standard path coupling arguments. Prior works \cite{BS16, BS17, Reg18, PR17} had studied these problems but only from the perspective of deterministic approximation algorithms using Barvinok's polynomial interpolation method \cite{Bar17book}. While these algorithms run in polynomial time for bounded-degree graphs, the exponent typically depends on the maximum degree, and are more difficult to implement. 

Here, we show that the Glauber dynamics mixes in $O(n\log n)$ steps for these problems on bounded-degree graphs, yielding significantly faster and simpler algorithms for computing the partition function. We again reduce rapid mixing to spectral independence via \cref{thm:spinsystemlocaltoglobal}, and then reduce spectral independence to the existence of a sufficiently large zero-free region for the multivariate partition function via \cref{thm:multi-spin}. Fortunately, such zero-free regions were already obtained in prior works, as they are the entire basis for Barvinok's polynomial interpolation method. We leverage them here in a completely black-box manner.

\subsection{Weighted Graph Homomorphisms}
Here, we study weighted graph homomorphisms, which may also be viewed as spin systems on vertices. In the bounded-degree setting, we show that the Glauber dynamics on vertex configurations for these models mixes in $O(n\log n)$ steps, provided the weights are sufficiently close to~$1$. This is analogous to classical mixing results stating the Glauber dynamics mixes rapidly in the ``high-temperature'' regime.
\begin{theorem}[Spectral Independence for Weighted Graph Homomorphisms]\label{thm:specindgraphhom}
Fix a positive integer $q \geq 2$, let $G=(V,E)$ be a graph with maximum degree $\leq \Delta$, and for each edge $uv \in E$, let $A^{uv} \in \R_{\geq0}^{q \times q}$ be a (not necessarily symmetric) nonnegative matrix. There exists a universal constant $\gamma \approx 0.56$ independent of $q,G,\{A^{uv}\}_{uv \in E}$ such that if $\left|A^{uv}(j,k) - 1 \right| \leq \frac{\gamma}{\Delta + \gamma} - \epsilon$ for some $\epsilon > 0$, all $uv \in E$ and all $j,k \in [q]$, then the associated graph homomorphism distribution $\mu$ on vertex configurations $\sigma:V \rightarrow [q]$ given by
\begin{align*}
    \mu(\sigma) \propto \prod_{uv \in E} A^{uv}(\sigma(u), \sigma(v))
\end{align*}
is $\eta$-spectrally independent for some constant $\eta = \eta(\Delta,\epsilon)$. In particular, if $\Delta,\epsilon = \Theta(1)$, then the Glauber dynamics for sampling from $\mu$ mixes in $O(n\log n)$ steps. 
\end{theorem}
\begin{remark}\label{rmk:dobrushincompare}
A straightforward application of the classical Dobrushin uniqueness condition yields rapid mixing when $|A^{uv}(j,k) - 1| \lesssim \frac{1}{2\Delta}$ for all $uv \in E$ and all $j,k \in [q]$.
\end{remark}

The zero-free region for the graph homomorphism partition function was studied in \cite{BS17}. 
We state here a slightly more general theorem, the proof of which is included in \cref{subapp:zerofreehom} for completeness. 

\begin{theorem}[Zeros for Weighted Graph Homomorphisms; \cite{BS17}]\label{thm:zerosgraphhom}
Fix a positive integer $q \geq 2$, let $G=(V,E)$ be a graph with maximum degree $\leq \Delta$, and for each edge $e=uv \in E$, let $A^{uv} \in \C^{q \times q}$ be a (not necessarily symmetric or Hermitian) complex matrix. There exists a universal constant $\gamma \approx 0.56$ independent of $q,G,\{A^{uv}\}_{uv \in E}$ such that if $\left|A^{uv}(j,k) - 1\right| < \frac{\gamma}{\Delta + \gamma}$ for all $uv \in E$ and all $j,k \in [q]$, then for every $S \subseteq V$ and every $\phi : S \rightarrow [q]$, the graph homomorphism partition function 
\begin{align*}
    \sum_{\substack{\sigma:V \rightarrow [q] \\ \sigma \mid_{S} = \phi}} \prod_{uv \in E} A^{uv}(\sigma(u), \sigma(v))
\end{align*}
with pinning $\phi$ is nonzero. 
\end{theorem}

We give below the proof of \cref{thm:specindgraphhom}. 

\begin{proof}[Proof of Theorem~\ref{thm:specindgraphhom}]
By \cref{thm:multi-spin}, it suffices to prove that the multivariate partition function
\begin{align}\label{eq:graphhommultivariate}
    \sum_{\substack{\sigma:V \rightarrow [q] \\ \sigma \mid_{S} = \phi}} \prod_{uv \in E} A^{uv}(\sigma(u), \sigma(v)) \prod_{v \in V} \lambda_{v,\sigma(v)}
\end{align}
is nonzero in the polydisk $\mathcal{D} = \left\{\lambda \in \C^{V\times [q]} : \left|\lambda_{v,k} - 1\right| < c, \forall v \in V, \forall k \in [q]\right\}$ for all pinnings $\phi$, where $c = c(\Delta,\epsilon) > 0$ is some constant depending only on $\Delta,\epsilon$ but not $G$. 
Define a new set of matrices $\{\tilde{A}^{uv}\}_{uv \in E}$ by 
\begin{align*}
    \tilde{A}^{uv}(j,k) &= A^{uv}(j,k) \cdot \lambda_{u,j}^{1/\deg(u)} \cdot \lambda_{v,k}^{1/\deg(v)}, \quad\forall uv \in E, \forall j,k \in [q].
\end{align*}
Note that the partition function for $G,\{\tilde{A}^{uv}\}_{uv \in E}$ is precisely given in \cref{eq:graphhommultivariate}.

Since $|A^{uv}(j,k) - 1| \leq \frac{\gamma}{\Delta + \gamma} - \epsilon$, there exists our desired $c(\Delta,\epsilon) > 0$ such that $|\lambda_{u,j} - 1|, |\lambda_{v,k} - 1| < c(\Delta,\epsilon)$ implies $|\tilde{A}^{uv}(j,k) - 1| < \frac{\gamma}{\Delta + \gamma}$, for all $uv \in E$ and all $j,k \in [q]$. It follows from \cref{thm:zerosgraphhom} that the multivariate partition function \cref{eq:graphhommultivariate} is nonzero. As this holds for all $\lambda \in \mathcal{D}$, we are done.
\end{proof}

\subsection{Tensor Network Contractions}
Here, we study general tensor network contractions, which is a partition function of a distribution over configurations on edges of a graph. Tensor networks are heavily studied in quantum computing \cite{MS08, AL10, Oru14} and are also used to model Holant problems \cite{CHL10, CLX09, CLX11}. In the bounded-degree setting, we also show that the Glauber dynamics on edge configurations for these models mixes in $O(n\log n)$ steps, provided the weights are sufficiently close to $1$. Again, this is analogous to classical mixing results stating the Glauber dynamics mixes rapidly in the ``high-temperature'' regime.

To state our main result, let us first define tensor network contraction. Given a graph $G=(V,E)$ and a collection of local functions $\{f_{v}:[q]^{E(v)} \rightarrow \R_{\geq0}\}_{v \in V}$ on configurations on edges, we define the associated tensor network distribution $\mu$ over edge configurations $\sigma:E \rightarrow [q]$ to be given by
\begin{align}\label{eq:tensornetworkdistr}
    \mu(\sigma) \propto \prod_{v \in V} f_{v}(\sigma \mid_{E(v)}).
\end{align}
The associated partition function, known as a tensor network contraction, is given by
\begin{align*}
    \sum_{\sigma:E \rightarrow [q]} \prod_{v \in V} f_{v}(\sigma \mid_{E(v)}).
\end{align*}
The name ``tensor network'' comes from the fact that each $f_{v}$ may be viewed as a tensor with axes corresponding to edges in $E(v)$ and indexed by $[q]$. This is a vast generalization of the Holant problems considered in \cref{sec:binarysymmetric} (see, for instance, \cref{eq:Holant_pf}), where $q = 2$ and each local function $f_{v}$ is symmetric. Zeros for tensor network contractions were analyzed in \cite{Reg18} in the symmetric case.
\begin{theorem}[Spectral Independence for Tensor Network Distribution]\label{thm:specindtensornetwork}
Fix a positive integer $q \geq 2$, let $G=(V,E)$ be a graph with maximum degree $\leq \Delta$, and for each vertex $v \in V$, let $f_{v}:[q]^{E(v)} \rightarrow \R_{\geq0}$ 
be a nonnegative function on configurations of edges incident to $v$. There exists a universal constant $\gamma \approx 0.56$ independent of $q,G,\{f_{v}\}_{v \in V}$ such that if $|f_{v}(\alpha) - 1| \leq \frac{\gamma}{\Delta+1+\gamma} - \epsilon$ for some $\epsilon > 0$, all $v \in V$ and all $\alpha:E(v) \rightarrow [q]$, then the tensor network distribution $\mu$ on edge configurations $\sigma:E \rightarrow [q]$ given by \cref{eq:tensornetworkdistr} is $\eta$-spectrally independent for some constant $\eta = \eta(\Delta,\epsilon)$. In particular, if $\Delta,\epsilon = \Theta(1)$, then the Glauber dynamics for sampling from $\mu$ mixes in $O(n\log n)$ steps.
\end{theorem}

\begin{remark}
A straightforward application of the classical Dobrushin uniqueness condition yields rapid mixing when $|f_{v}(\alpha) - 1| \lesssim \frac{1}{4(\Delta-1)}$ for all $v \in V$ and all $\alpha:E(v) \rightarrow [q]$.
\end{remark}

To establish this spectral independence, we need a sufficiently large zero-free region. This was proved by \cite{Reg18} in the symmetric case, where each local function $f_{v}$ depends only on the number of incident edges that are mapped to each color in $[q]$. It turns out using nearly identical arguments, one can obtain the following more general theorem. We provide a proof in \cref{subapp:zerofreetensor} for completeness.
\begin{theorem}[Zeros of Tensor Network Contractions; \cite{Reg18}]\label{thm:zerostensornetwork}
Fix a positive integer $q \geq 2$, let $G=(V,E)$ be a graph with maximum degree $\leq \Delta$, and for each vertex $v \in V$, let $f_{v}: [q]^{E(v)} \rightarrow \C$ be a complex function on configurations of edges incident to $v$. There exists a universal constant $\gamma \approx 0.56$ independent of $q,G,\{f_{v}\}_{v \in V}$ such that if $\left|f_{v}(\alpha) - 1\right| < \frac{\gamma}{\Delta+1+\gamma}$ for all $v \in V$ and all $\alpha :E(v) \rightarrow [q]$, then for every $F \subseteq E$ and every $\phi : F \rightarrow [q]$, the tensor network contraction 
\begin{align*}
    \sum_{\substack{\sigma:E \rightarrow [q] \\ \sigma \mid_{F} = \phi}} \prod_{v \in V} f_{v}(\sigma \mid_{E(v)})
\end{align*}
with pinning $\phi$ is nonzero.
\end{theorem}

We give below the proof of \cref{thm:specindtensornetwork}.

\begin{proof}[Proof of Theorem~\ref{thm:specindtensornetwork}]
By \cref{thm:multi-spin}, it suffices to prove that the multivariate partition function
\begin{align}\label{eq:multivartensornetwork}
    \sum_{\substack{\sigma:E \rightarrow [q] \\ \sigma \mid_{F} = \phi}} \prod_{v \in V} f_{v}(\sigma \mid_{E(v)}) \prod_{e \in E} \lambda_{e,\sigma(e)}
\end{align}
is nonzero whenever $\lambda$ lies in the polydisk $\mathcal{D} = \{\lambda \in \C^{E \times [q]} : |\lambda_{e,k} - 1| < c, \forall e \in E, \forall k \in [q]\}$ for all pinnings $\phi$, where $c=c(\Delta,\epsilon) > 0$ is some constant depending only on $\Delta,\epsilon$ but not $G$. 
Define a new set of local constraint functions $\{\tilde{f}_{v}\}_{v \in V}$ by
\begin{align*}
    \tilde{f}_{v}(\alpha) = f_{v}(\alpha) \cdot \prod_{e \in E(v)} \lambda_{e,\alpha(e)}^{1/2}, \quad \forall v \in V, \forall \alpha : E(v) \rightarrow [q].
\end{align*}
Note that the partition function for $G,\{\tilde{f}_{v}\}_{v \in V}$ is precisely given in \cref{eq:multivartensornetwork}.

Since $|f_{v}(\alpha) - 1| \leq \frac{\gamma}{\Delta+1+\gamma} - \epsilon$, there exists our desired $c(\Delta,\epsilon) > 0$ such that $|\lambda_{e,k} - 1| < c(\Delta,\epsilon)$ for all $e \in E(v)$ implies  $|\tilde{f}_{v}(\alpha) - 1| < \frac{\gamma}{\Delta+1+\gamma}$, for all $v \in V$ and all $\alpha:E(v) \rightarrow [q]$. It follows from \cref{thm:zerostensornetwork} that the multivariate partition function \cref{eq:multivartensornetwork} is nonzero. As this holds for all $\lambda \in \mathcal{D}$, we are done.
\end{proof}

\section{Arbitrary Measures on the Discrete Cube}\label{sec:fourier}
In this section, we state a general result for mixing of an arbitrary measure on the discrete cube $\{-1,1\}^{n}$. For this, we fix an arbitrary potential $f:\{-1,1\}^{n} \rightarrow \R$. A standard result from analysis of Boolean functions says that $f$ admits a unique representation as a multilinear polynomial $f(x) = \sum_{S \subseteq [n]} \hat{f}(S) \prod_{i \in S} x_{i}$. This representation is known as the Fourier-Walsh transform of $f$ (see \cite{Odo14} and references therein), and the coefficients $\hat{f}(S)$ are known as the Fourier coefficients. \cite{Bar17boolean} showed that when the Fourier coefficients, as well as the degree $\deg f$ of $f$ as a multilinear polynomial, are sufficiently small, then one has a zero-free disk for the corresponding partition function $\sum_{x \in \{-1,1\}^{n}} e^{f(x)}$. We convert this via \cref{thm:multi-spin} into a corresponding statement for the spectral independence of the distribution. Since we do not assume that $f$ arises from a spin system (or, more generally, tensor network) on a bounded-degree graph, we only obtain a spectral gap bound with a relatively large exponent using \cref{thm:genericspecgaplocaltoglobal} proved in \cite{AL20,ALO20}. 
\begin{theorem}\label{thm:fourierspecind}
Let $f:\{-1,1\}^{n} \rightarrow \R$ and $\epsilon > 0$ be given, and assume that
\begin{align*}
    \sqrt{\deg f} \cdot L(f) \le C - \epsilon,
\end{align*}
where $C \approx 0.55$ is an absolute constant, and $L(f) := \max_{i \in [n]} \sum_{S \subseteq [n] : S \ni i} \left|\hat{f}(S)\right|$. Further assume that the associated Gibbs distribution $\mu$ on $\{-1,1\}^{n}$ given by
\begin{align*}
    \mu(x) \propto \exp(f(x))
\end{align*}
is $b$-marginally bounded for some $b > 0$. Let $\xi := \frac{2C}{\sqrt{\deg f}} - 2L(f)$. Then $\mu$ is $\eta$-spectrally independent where $\eta$ is a constant depending only on $b$ and $\xi$. In particular, $\mu$ is $O(1)$-spectrally independent if~$\epsilon, \deg(f),b = \Theta(1)$.
\end{theorem}
\begin{remark}
One may also view $L(f)$ as bounding the Lipschitz constant of $f$.
\end{remark}
\begin{remark}
A standard calculation using Dobrushin uniqueness condition yields that the Glauber dynamics is rapidly mixing when
\begin{align*}
    \max_{i \in [n]} \sum_{S \subseteq [n] : S \ni i} (|S| - 1) \cdot \left|\hat{f}(S)\right| < 1,
\end{align*}
which can be weakened to $(\deg(f) - 1) \cdot L(f) < 1$. These bounds are in general not comparable with the above due to the square root. While this bound is stronger when $\deg(f)$ is small, the above is stronger when most of the Fourier mass of $f$ is on high-degree monomials.
\end{remark}
\begin{remark}
A standard notion in analysis of Boolean functions is also that of ``influence'', which to avoid confusion with the notion of pairwise influence used to define spectral independence, we refer to as ``voter influence''. This terminology is consistent with the traditional applications of analysis of Boolean functions to social choice theory and voting systems; see \cite{Odo14} and references therein. A standard result in analysis of Boolean functions says that the ``voter influence'' of coordinate $i$ is precisely
\begin{align*}
    \sum_{S \subseteq [n] : S \ni i} \left|\hat{f}(S)\right|^{2}.
\end{align*}
Hence, while we do not establish a formal connection between small ``voter influence'' and strong spectral independence guarantees, our result \cref{thm:fourierspecind} says this is true at least morally.
\end{remark}

We need the following zero-free result from \cite{Bar17boolean}.

\begin{theorem}[\cite{Bar17boolean}]\label{thm:zerosfourier}
Let $f:\{-1,1\}^{n} \rightarrow \C$ be given, and assume that
\begin{align*}
    \sqrt{\deg(f)} \cdot L(f) < C,
\end{align*}
where $C \approx 0.55$ is an absolute constant, and $L(f) := \max_{i \in [n]} \sum_{S \subseteq [n] : S \ni i} \left|\hat{f}(S)\right|$. Then for every $S \subseteq [n]$ and every pinning $\phi : S \rightarrow \{-1,1\}$, we have that the partition function of the associated Gibbs measure on $\{-1,1\}^{n}$ with pinning $\phi$ is nonzero:
\begin{align*}
    \sum_{x \in \{-1,1\}^{n} : x\mid_{S} = \phi} \exp(f(x)) \neq 0.
\end{align*}
\end{theorem}

We now prove \cref{thm:fourierspecind}. 

\begin{proof}[Proof of Theorem~\ref{thm:fourierspecind}]
By \cref{thm:multi-spin}, it suffices to prove that the multivariate partition function
\begin{align}\label{eq:multivarfourier}
    \sum_{x \in \{-1,1\}^{n} : x\mid_{S} = \phi} \exp(f(x)) \prod_{i \in [n] : x_{i} = 1} \lambda_{i}
\end{align}
is nonzero whenever $\lambda$ lies in the set $\mathcal{D} = \{\lambda \in \C^{n} : |\lambda_{i} - 1| < c, \forall i \in [n]\}$ for all pinnings $\phi$, where $c = c(\xi) > 0$ is a constant depending only on $\xi$ but not $n$. Define a new function $g:\{-1,1\}^{n} \rightarrow \C$ by
\begin{align*}
    g(x) = f(x) + \sum_{i=1}^{n} \frac{1 + x_{i}}{2} \log \lambda_{i} = f(x) + \sum_{i \in [n]: x_{i} = 1} \log \lambda_{i}.
\end{align*}
Then $\exp(g(x)) = \exp(f(x)) \prod_{i \in [n]: x_{i} = 1} \lambda_{i}$ and the partition function $\sum_{x \in \{-1,1\}^{n} : x\mid_{S} = \phi} \exp(g(x))$ associated with $g$ is precisely our desired multivariate partition function \cref{eq:multivarfourier}. Our goal is to apply \cref{thm:zerosfourier} to $g$ and deduce our desired stability statement.

First, it is clear from the definition of $g$ that the Fourier coefficients of $g$ are given by
\begin{align*}
    \hat{g}(S) &= \begin{cases}
        \hat{f}(S), &\quad\text{if } |S| > 1; \\
        \hat{f}(i) + \frac{1}{2} \log \lambda_{i}, &\quad\text{if } S = \{i\} \text{ for some } i \in [n]; \\
        \hat{f}(\emptyset) + \frac{1}{2}\sum_{i=1}^{n} \log\lambda_{i}, &\quad\text{if } S = \emptyset.
    \end{cases}
\end{align*}
It follows that
\begin{align*}
    L(g) \leq L(f) + \frac{1}{2} \max_{i \in [n]} \left|\log \lambda_{i}\right|.
\end{align*}
Note that $\deg(g) = \deg(f)$ (unless $\deg(f) \le 1$, in which case spectral independence and rapid mixing is trivial). Hence, if $\lambda \in \C^{n}$ satisfies $|\log \lambda_{i}| < \xi$ for all $i \in [n]$, then rearranging yields precisely that $\sqrt{\deg(g)} \cdot L(g) < C$ and the zero-freeness follows from \cref{thm:zerosfourier}. 
Furthermore, it is clear that the set $\{\lambda \in \C^{n} : |\log \lambda_{i}| < \xi, \forall i \in [n]\}$ contains $\mathcal{D}$ for a value of $c(\xi) > 0$ which depends only on $\xi$, just by continuity of the logarithm.
\end{proof}

\section{Future Directions}
One open problem of our work is to obtain spectral independence for sampling $b$-matchings and $b$-edge-covers for general $b$ on all graphs. It was shown in \cite{HLZ16} that the Glauber dynamics for sampling $b$-matchings (respectively, $b$-edge-covers) mixes rapidly when $b \leq 7$ (respectively, $b \leq 2$); their proof approach relies on the canonical paths technique and in particular, the winding method \cite{McQ}. In a recent paper \cite{CG24}, spectral independence was established for $b$-matchings and $b$-edge-covers for all $b$ on bounded-degree graphs.


\printbibliography

@book{KW17,
  title={Several complex variables},
  author={Korevaar, Jacob and Wiegerinck, Jan},
  year={2017},
  publisher={Korteweg-de Vries Institute for Mathematics},
  url={https://staff.science.uva.nl/j.j.o.o.wiegerinck/edu/scv/scvboek.pdf}
}

@book{Rudin,
  title={Real and Complex Analysis},
  author={Rudin, Walter},
  isbn={9780071002769},
  lccn={86000007},
  series={Mathematics series},
  year={1987},
  publisher={McGraw-Hill},
}

@inproceedings{RW99,
  title={Sampling spin configurations of an {I}sing system},
  author={Randall, Dana and Wilson, David},
  booktitle={Proceedings of the 10th Annual ACM-SIAM Symposium on Discrete Algorithms (SODA)},
  pages={959--960},
  year={1999},
  url={https://dl.acm.org/doi/pdf/10.5555/314500.314945}
}

@inproceedings{ChenLV21,
  author       = {Zongchen Chen and
                  Kuikui Liu and
                  Eric Vigoda},
  title        = {Spectral Independence via Stability and Applications to Holant-Type
                  Problems},
  booktitle    = {62nd {IEEE} Annual Symposium on Foundations of Computer Science, (FOCS)},
  pages        = {149--160},
  publisher    = {{IEEE}},
  year         = {2021},
  doi          = {10.1109/FOCS52979.2021.00023},
}

@article{McQ,
author={Colin McQuillan},
title={Approximating {H}olant problems by winding},
journal={arXiv preprint arXiv:1301.2880},
year={2013},
doi={10.48550/arXiv.1301.2880}
}

@article{BB21,
author={Barvinok, Alexander and Barvinok, Nicholas},
title={More on zeros and approximation of the {I}sing partition function},
volume={9}, 
journal={Forum of Mathematics, Sigma}, 
year={2021}, 
pages={e46},
doi={10.1017/fms.2021.40}
}

@article{BD20,
title={Testing for Dense Subsets in a Graph via the Partition Function},
author={Barvinok, Alexander and Della Pella, Anthony},
year={2020},
journal={SIAM Journal on Discrete Mathematics},
volume={34},
issue={1},
pages={308--327},
doi={10.1137/19m1247413}
}

@article{Bar17perm,
title={Approximating permanents and hafnians},
author={Barvinok, Alexander},
year={2017},
journal={Discrete Analysis},
numpages={34},
volume={2},
doi={10.19086/da.1244}
}

@article{Bar15,
 author = {Barvinok, Alexander},
 title = {Computing the Partition Function for Cliques in a Graph},
 year = {2015},
 pages = {339--355},
 doi = {10.4086/toc.2015.v011a013},
 publisher = {Theory of Computing},
 journal = {Theory of Computing},
 volume = {11},
 number = {13},
 URL = {https://theoryofcomputing.org/articles/v011a013},
}

@article{LY,
  title={Statistical theory of equations of state and phase transitions. {II}. {L}attice gas and {I}sing model},
  author={Lee, Tsung-Dao and Yang, Chen-Ning},
  journal={Physical Review},
  volume={87},
  number={3},
  pages={410--419},
  year={1952},
  publisher={APS},
  doi={10.1103/physrev.87.410}
}

@inproceedings{LLL,
title={A Simple {FPTAS} for Counting Edge Covers},
author={Chengyu Lin and Jingcheng Liu and Pinyan Lu},
year={2014},
booktitle = {Proceedings of the 25th Annual ACM-SIAM Symposium on Discrete Algorithms (SODA)},
pages={341--348},
doi={10.1137/1.9781611973402.25}
}

@inproceedings{BR09,
  author    = {Ivona Bez{\'{a}}kov{\'{a}} and
               William A. Rummler},
  title     = {Sampling Edge Covers in 3-Regular Graphs},
  booktitle = {Proceedings of the 34th International
               Symposium on Mathematical Foundations of Computer Science (MFCS)},
  pages     = {137--148},
  year      = {2009},
doi = {10.1007/978-3-642-03816-7_13}
  }

@inproceedings{AASV21,
  title={Fractionally Log-Concave and Sector-Stable Polynomials: Counting Planar Matchings and More},
  author={Alimohammadi, Yeganeh and Anari, Nima and Shiragur, Kirankumar and Vuong, Thuy-Duong},
  booktitle={Proceedings of the 53rd Annual ACM Symposium on Theory of Computing (STOC)},
  year={2021},
  doi={10.1145/3406325.3451123}
}

@inproceedings{AL20,
  title={Improved analysis of higher order random walks and applications},
  author={Alev, Vedat Levi and Lau, Lap Chi},
  booktitle={Proceedings of the 52nd Annual ACM Symposium on Theory of Computing (STOC)},
  pages={1198--1211},
  year={2020},
  doi={10.1145/3357713.3384317}
}

@article{BS16,
title = {Computing the partition function for graph homomorphisms with multiplicities},
journal = {Journal of Combinatorial Theory, Series A},
volume = {137},
pages = {1--26},
year = {2016},
author = {Alexander Barvinok and Pablo Sober\'{o}n},
doi = {10.1016/j.jcta.2015.08.001}
}

@article{BS17,
title = {Computing the partition function for graph homomorphisms},
journal = {Combinatorica},
volume = {37},
pages = {633--650},
year = {2017},
author = {Alexander Barvinok and Pablo Sober\'{o}n},
doi = {10.1007/s00493-016-3357-2}
}

@article{Bar16,
title = {Computing the Permanent of (Some) Complex Matrices},
journal = {Foundations of Computational Mathematics},
volume = {16},
pages = {329--342},
year = {2016},
author = {Alexander Barvinok},
doi = {10.1007/s10208-014-9243-7}
}

@article{Reg18,
title = {Zero-Free Regions of Partition Functions with Applications to Algorithms and Graph Limits},
journal = {Combinatorica},
volume = {38},
pages = {987--1015},
year = {2018},
author = {Guus Regts},
doi = {10.1007/s00493-016-3506-7}
}

@article{Wag09,
title = {Weighted enumeration of spanning subgraphs with degree constraints},
journal = {Journal of Combinatorial Theory, Series B},
volume = {99},
number = {2},
pages = {347--357},
year = {2009},
author = {David G. Wagner},
doi = {10.1016/j.jctb.2008.07.007}
}

@INPROCEEDINGS{GM07,
author = {Antoine Gerschenfeld and Andrea Montanari},
booktitle = {Proceedings of the 48th Annual IEEE Symposium on Foundations of Computer Science (FOCS)},
title = {Reconstruction for Models on Random Graphs},
year = {2007},
pages = {194--204},
doi={10.1109/focs.2007.58}
}

@article{BCR20,
title={Some applications of {W}agner's weighted subgraph counting polynomial},
author={Bencs, Ferenc and Csikv\'{a}ri, P\'{e}ter and Regts, Guus},
journal={The Electronic Journal of Combinatorics},
volume={28},
number={4},
year={2021},
pages={P4.14},
doi={10.37236/10185}
}

@article{GLLZ21,
author = {Guo, Heng and Liao, Chao and Lu, Pinyan and Zhang, Chihao},
title = {Zeros of {H}olant Problems: Locations and Algorithms},
year = {2021},
volume = {17},
number = {1},
journal = {ACM Transactions on Algorithms},
pages={1--25},
doi = {10.1137/1.9781611975482.137}
}

@article{DHJM21,
title={Polynomial-time approximation algorithms for the antiferromagnetic {I}sing model on line graphs},
author={Dyer, Martin and Heinrich, Marc and Jerrum, Mark and M\"{u}ller, Haiko},
year={2021},
journal={Combinatorics, Probability and Computing},
publisher={Cambridge University Press},
pages={1--17},
doi={10.1017/s0963548321000080}
}

@inproceedings{HLZ16,
author = {Huang, Lingxiao and Lu, Pinyan and Zhang, Chihao},
title = {Canonical Paths for {MCMC}: From Art to Science},
year = {2016},
booktitle = {Proceedings of the 27th Annual ACM-SIAM Symposium on Discrete Algorithms (SODA)},
pages = {514--527},
numpages = {14},
doi = {10.1137/1.9781611974331.ch38}
}

@article{JSising,
author = {Jerrum, Mark and Sinclair, Alistair},
title = {Polynomial-Time Approximation Algorithms for the {I}sing Model},
year = {1993},
volume = {22},
number = {5},
journal = {SIAM Journal on Computing},
pages = {1087--1116},
doi = {10.1137/0222066}
}

@article{JSmatchings,
 author = {Jerrum, Mark and Sinclair, Alistair},
 title = {Approximating the Permanent},
 journal = {SIAM Journal on Computing},
 volume = {18},
 number = {6},
 year = {1989},
 pages = {1149--1178},
 doi = {10.1137/0218077}
}

@article{CLV20,
  title={Rapid mixing of {G}lauber dynamics up to uniqueness via contraction},
  author={Chen, Zongchen and Liu, Kuikui and Vigoda, Eric},
  journal={SIAM Journal on Computing},
  volume={52},
  number={1},
  pages={196--237},
  year={2023},
  publisher={SIAM},
  doi={10.1137/20m136685x}
}

@inproceedings{ALO20,
    author={Anari, Nima and Liu, Kuikui and {Oveis Gharan}, Shayan},
    booktitle = {Proceedings of the 61st Annual IEEE Symposium on Foundations of Computer Science (FOCS)},
    pages = {1319--1330},
    title = {Spectral Independence in High-Dimensional Expanders and Applications to the Hardcore Model},
    year = {2020},
    doi = {10.1109/focs46700.2020.00125}
}

@article{CLV21,
author = {Chen, Zongchen and Liu, Kuikui and Vigoda, Eric},
title = {Optimal Mixing of {G}lauber Dynamics: {E}ntropy Factorization via High-Dimensional Expansion},
journal = {SIAM Journal on Computing},
pages = {STOC21-104--153},
doi = {10.1137/21M1443340}
}

@inproceedings{Wei06,
 author = {Weitz, Dror},
 title = {Counting Independent Sets Up to the Tree Threshold},
 booktitle = {Proceedings of the 38th Annual ACM Symposium on Theory of Computing (STOC)},
 year = {2006},
 pages = {140--149},
 numpages = {10},
 doi = {10.1145/1132516.1132538}
}

@inproceedings{Liu21,
  author={Liu, Kuikui},
  title={From Coupling to Spectral Independence and Blackbox Comparison with the Down-Up Walk},
  booktitle =	{Approximation, Randomization, and Combinatorial Optimization. Algorithms and Techniques (APPROX/RANDOM 2021)},
  pages =	{32:1--32:21},
  year =	{2021},
  volume =	{207},
  doi =		{10.4230/LIPIcs.APPROX/RANDOM.2021.32}
}

@inproceedings{BCCPSV21,
author={Blanca, Antonio and Caputo, Pietro and Chen, Zongchen and Parisi, Daniel and \v{S}tefankovi\v{c}, Daniel and Vigoda, Eric},
title={On Mixing of {M}arkov Chains: Coupling, Spectral Independence, and Entropy Factorization},
  booktitle = {Proceedings of the 33rd Annual {ACM-SIAM} Symposium on Discrete Algorithms
               (SODA)},
  pages     = {3670--3692},
    year      = {2022},
doi = {10.1137/1.9781611977073.145}
}

@article{PR17,
title = {Deterministic polynomial-time approximation algorithms for partition functions and graph polynomials},
journal = {SIAM Journal on Computing},
volume={46},
number={6},
year={2017},
pages={1893--1919},
author = {Patel, Viresh and Regts, Guus},
doi={10.1137/16m1101003}
}

@BOOK{Bar17book,
  TITLE = {Combinatorics and Complexity of Partition Functions},
  AUTHOR = {Barvinok, Alexander},
  YEAR = {2016},
  PUBLISHER = {Springer Algorithms and Combinatorics},
  volume={30},
  doi={10.1007/978-3-319-51829-9}
}

@article{Bar17boolean,
title={Computing the Partition Function of a Polynomial on the {B}oolean Cube},
author={Barvinok, Alexander},
journal={A Journey Through Discrete Mathematics},
year={2017},
pages={135--164},
doi={10.1007/978-3-319-44479-6_7}
}

@book{Odo14,
author = {O'Donnell, Ryan},
title = {Analysis of Boolean Functions},
year = {2014},
isbn = {1107038324},
publisher = {Cambridge University Press},
address = {USA},
doi = {10.1017/CBO9781139814782}
}

@inproceedings{LLZ14,
title={{FPTAS} for Counting Weighted Edge Covers},
author={Liu, Jingcheng and Lu, Pinyan and Zhang, Chihao},
booktitle={Proceedings of the 22nd Annual European Symposium on Algorithms (ESA)},
year={2014},
pages={654--665},
doi={10.1007/978-3-662-44777-2_54}
}

@article{CHL10,
  title={From {H}olant to \#{CSP} and back: Dichotomy for {H}olant$^c$ problems},
  author={Cai, Jin-Yi and Huang, Sangxia and Lu, Pinyan},
  journal={Algorithmica},
  volume={64},
  number={3},
  pages={511--533},
  year={2012},
  publisher={Springer},
  doi={10.1007/s00453-012-9626-6}
}

@article{Oru14,
title = {A practical introduction to tensor networks: Matrix product states and projected entangled pair states},
journal = {Annals of Physics},
volume = {349},
pages = {117--158},
year = {2014},
author = {Rom\'{a}n Or{\'{u}}s},
doi = {10.1016/j.aop.2014.06.013}
}

@inproceedings{CLX09,
author = {Cai, Jin-Yi and Lu, Pinyan and Xia, Mingji},
title = {{H}olant Problems and Counting {CSP}},
year = {2009},
booktitle = {Proceedings of the 41st Annual ACM Symposium on Theory of Computing (STOC)},
pages = {715--724},
doi={10.1145/1536414.1536511}
}

@article{MS08,
title={Simulating Quantum Computation by Contracting Tensor Networks},
author={Markov, Igor L. and Shi, Yaoyun},
journal={SIAM Journal on Computing},
volume={38},
issue={3},
pages={963--981},
year={2008},
doi={10.1137/050644756}
}

@article{CLX11,
title={Computational Complexity of {H}olant Problems},
author={Cai, Jin-Yi and Lu, Pinyan and Xia, Mingji},
year={2011},
journal={SIAM Journal on Computing},
volume={40},
issue={4},
pages={1101--1132},
doi={10.1137/100814585}
}

@article{AL10,
author = {Arad, Itai and Landau, Zeph},
title = {Quantum Computation and the Evaluation of Tensor Networks},
year = {2010},
volume = {39},
number = {7},
journal = {SIAM Journal on Computing},
pages = {3089--3121},
doi = {10.1137/080739379}
}

@article{LSS19,
  title={The {I}sing partition function: Zeros and deterministic approximation},
  author={Liu, Jingcheng and Sinclair, Alistair and Srivastava, Piyush},
  journal={Journal of Statistical Physics},
  volume={174},
  number={2},
  pages={287--315},
  year={2019},
  publisher={Springer},
  doi={10.1007/s10955-018-2199-2}
}

@article{Rue71,
  title={Extension of the {L}ee--{Y}ang circle theorem},
  author={Ruelle, David},
  journal={Physical Review Letters},
  volume={26},
  number={6},
  pages={303--304},
  year={1971},
  publisher={APS},
  doi={10.1103/physrevlett.26.303}
}

@article{Asa70,
  title = {Theorems on the Partition Functions of the {H}eisenberg Ferromagnets},
  author = {Asano, Taro},
  journal = {Journal of the Physical Society of Japan},
  volume = {29},
  number = {2},
  pages = {350--359},
  numpages = {10},
  year = {1970},
  doi = {10.1143/jpsj.29.350}
}

@inproceedings{CGSV21,
  title={Rapid Mixing for Colorings via Spectral Independence},
  author={Chen, Zongchen and Galanis, Andreas and {\v{S}}tefankovi{\v{c}}, Daniel and Vigoda, Eric},
  booktitle = {Proceedings of the 32nd Annual ACM-SIAM Symposium on Discrete Algorithms (SODA)},
  pages = {1548--1557},
  year={2021},
  doi={10.1137/1.9781611976465.94}
}

@article{FGYZ21,
  title={Rapid mixing from spectral independence beyond the {B}oolean domain},
  author={Feng, Weiming and Guo, Heng and Yin, Yitong and Zhang, Chihao},
  journal={ACM Transactions on Algorithms (TALG)},
  volume={18},
  number={3},
  pages={1--32},
  year={2022},
  publisher={ACM New York, NY},
  doi={10.1145/3531008}
}

@article{GJ09,
  title={Random even graphs},
  author={Grimmett, Geoffrey and Janson, Svante},
  journal={The Electronic Journal of Combinatorics},
  volume={16},
  number={1},
  pages={R46},
  year={2009},
  doi={10.37236/135}
}

@article{BB09,
  title={The {L}ee-{Y}ang and {P}{\'o}lya-{S}chur programs. {I}. linear operators preserving stability},
  author={Borcea, Julius and Br{\"a}nd{\'e}n, Petter},
  journal={Inventiones mathematicae},
  volume={177},
  number={3},
  pages={541},
  year={2009},
  publisher={Springer},
  doi={10.1007/s00222-009-0189-3}
}

@article{ES88,
  title={Generalization of the {F}ortuin-{K}asteleyn-{S}wendsen-{W}ang representation and {M}onte {C}arlo algorithm},
  author={Edwards, Robert G and Sokal, Alan D},
  journal={Physical review D},
  volume={38},
  number={6},
  pages={2009},
  year={1988},
  publisher={APS},
  doi={10.1103/physrevd.38.2009}
}

@article{FK72,
  title={On the random-cluster model: {I}. {I}ntroduction and relation to other models},
  author={Fortuin, Cornelius Marius and Kasteleyn, Piet W},
  journal={Physica},
  volume={57},
  number={4},
  pages={536--564},
  year={1972},
  publisher={Elsevier},
  doi={10.1016/0031-8914(72)90045-6}
}

@article{FGW23,
  title={{S}wendsen-{W}ang dynamics for the ferromagnetic {I}sing model with external fields},
  author={Feng, Weiming and Guo, Heng and Wang, Jiaheng},
  journal={Information and Computation},
  volume={294},
  pages={105066},
  year={2023},
  publisher={Elsevier},
  doi={10.1016/j.ic.2023.105066}
}

@inproceedings{CG24,
  title={Fast sampling of $b$-matchings and $b$-edge covers},
  author={Chen, Zongchen and Gu, Yuzhou},
  booktitle={Proceedings of the 2024 Annual ACM-SIAM Symposium on Discrete Algorithms (SODA)},
  pages={4972--4987},
  year={2024},
  doi={10.1137/1.9781611977912.178}
}

\appendix
 
\crefalias{section}{appsec}
\crefalias{subsection}{appsec}

\section{Proofs of Zero-Free Results}\label{app:zerofree}
In this section, we supply proofs of the main zero-free statements used in \cref{sec:homtensornetwork}. As noted earlier, for technical reasons, we need straightforward generalizations of prior results which do not make symmetry assumptions. We manage to adapt previous arguments without much additional effort, which we provide here for completeness.

The main idea in these zero-free proofs is to do induction by conditioning on the assignment of fewer and fewer vertices (respectively, edges) for weighted homomorphisms (respectively, tensor networks). However, one needs to strengthen the inductive hypothesis beyond simple zero-freeness. To the best of our knowledge, this type of argument was first pioneered by Barvinok, and has had a wide range of applications; see \cite{Bar15, Bar17perm, BD20, BB21} for applications besides those discussed in this paper.

The crucial tool is the following geometric lemma, which provides a kind of ``reverse triangle inequality''. The version below is due to Boris Bukh; a weaker version, with $\cos(\theta/2)$ replaced by $\sqrt{\cos \theta}$, was known due to \cite{Bar16}. See \cite{Bar17book} for a proof.
\begin{lemma}[Angle Lemma]\label{lem:angle}
Let $x_{1},\dots,x_{n} \in \C$ be nonzero complex numbers viewed as vectors in $\R^{2}$. Suppose there is an angle $0 \leq \theta < 2\pi/3$ such that for all $i,j$, the angle between $x_{i},x_{j}$ is at most $\theta$. Then we have the lower bound $\left|\sum_{i=1}^{n} x_{i}\right| \geq \cos(\theta/2) \sum_{i=1}^{n} \left|x_{i}\right|$.
\end{lemma}

\subsection{Proofs for Weighted Graph Homomorphisms}\label{subapp:zerofreehom}
Our goal in this subsection is to prove \cref{thm:zerosgraphhom}, i.e., that the weighted graph homomorphism partition function
\begin{align*}
    Z_{\phi}^{S}(A) = \sum_{\substack{\sigma:V \rightarrow [q] \\ \sigma\mid_{S} = \phi}} \prod_{uv \in E} A^{uv}(\sigma(u), \sigma(v))
\end{align*}
is nonzero in a large polydisk around $1$, where $S \subseteq V, \phi:S \rightarrow [q]$, and we view $Z_{\phi}^{S}(A)$ as a polynomial with variables $\{A^{uv}(j,k)\}_{uv \in E, j,k \in [q]}$. For convenience, for a $\delta > 0$, define
\begin{align*}
    \mathcal{U}(\delta) = \{A = \{A^{uv}\}_{uv \in E} : |A^{uv}(j,k) - 1| < \delta, \forall uv \in E, \forall j,k \in [q]\}.
\end{align*}
Additionally, for a partial configuration $\phi:S \rightarrow [q]$, a vertex $u \in V \setminus S$ and a spin $j \in [q]$, we write $\phi_{u,j}:S \cup \{u\} \rightarrow [q]$ for the unique extension of $\phi$ with $\phi_{u,j}(u) = j$.

We will need the following lemmas to implement an inductive approach.
\begin{lemma}[Lemma 3.3 from \cite{BS17}]\label{lem:homderivtoangle}
Let $\tau, \delta > 0$, and suppose $A \in \mathcal{U}(\delta)$. Let $S \subseteq V$, $\phi : S \rightarrow [q]$, $u \in V \setminus S$ be arbitrary. Assume the following hold:
\begin{enumerate}
    \item[(1)] $Z_{\phi_{u,j}}^{S \cup \{u\}}(A) \neq 0$ for every $u \in V \setminus S$ and every $j \in [q]$;
    \item[(2)] For every $u \in V \setminus S$ and every $j \in [q]$, we have
    \[
    |Z_{\phi_{u,j}}^{S \cup \{u\}}(A)| \geq \frac{\tau}{\Delta} \sum_{v \sim u} \sum_{k \in [q]} |A^{uv}(j,k)| \cdot \left|\frac{\partial}{\partial A^{uv}(j,k)} Z_{\phi_{u,j}}^{S \cup \{u\}}(A)\right|.
    \]
\end{enumerate}
Then for every $u \in V \setminus S$ and every $j,k \in [q]$, the angle between $Z_{\phi_{u,j}}^{S \cup \{u\}}(A)$ and $Z_{\phi_{u,k}}^{S \cup \{u\}}(A)$ in $\C$ is at most $\frac{2\delta \Delta}{\tau(1-\delta)}$.
\end{lemma}
\begin{proof}
By assumption (1), the relevant partition functions are nonzero, and so the logarithm is well-defined when applied to these partition functions and we may bound the angle between $Z_{\phi_{u,j}}^{S \cup \{u\}}(A)$ and $Z_{\phi_{u,k}}^{S \cup \{u\}}(A)$ by
\begin{align}\label{eq:homcondlogdiff}
    \left|\log Z_{\phi_{u,j}}^{S \cup \{u\}}(A) - \log Z_{\phi_{u,k}}^{S \cup \{u\}}(A)\right|.
\end{align}
The strategy is to write $Z_{\phi_{u,k}}^{S \cup \{u\}}(A)$ as $Z_{\phi_{u,j}}^{S \cup \{u\}}(\tilde{A})$ for some $\tilde{A} \in \mathcal{U}(\delta)$ which differs from $A$ by a small number of coordinates, and then apply the Fundamental Theorem of Calculus and assumption (2). For every $v \sim u$, we set $\tilde{A}^{uv}(j,c) = A^{uv}(k,c)$ for every $c \in [q]$, and $\tilde{A}^{uv}(\ell,c) = A^{uv}(\ell,c)$ for all $\ell \neq j$. For all other edges $vw \in E$, we set $\tilde{A}^{vw} = A^{vw}$.

It is clear that $Z_{\phi_{u,k}}^{S \cup \{u\}}(A) = Z_{\phi_{u,j}}^{S \cup \{u\}}(\tilde{A})$. By the Fundamental Theorem of Calculus, we may upper bound \cref{eq:homcondlogdiff} by
\begin{align*}
    &\max_{B \in \mathcal{U}(\delta)} \sum_{v \sim u} \sum_{c \in [q]} \left|\frac{\partial}{\partial A^{uv}(j,c)} \log Z_{\phi_{u,j}}^{S \cup \{u\}}(B)\right| \cdot \underset{\leq 2\delta \text{ since } A,\tilde{A} \in \mathcal{U}(\delta)}{\underbrace{|A^{uv}(j,c) - \tilde{A}^{uv}(j,c)|}} \\
    \leq{}& \frac{2\delta}{1 - \delta} \max_{B \in \mathcal{U}(\delta)} \underset{\leq \Delta/\tau \text{ by assumption (2)}}{\underbrace{\sum_{v \sim u} \sum_{c \in [q]} |A^{uv}(j,c)| \cdot \frac{1}{|Z_{\phi_{u,j}}^{S \cup \{u\}}(B)|} \cdot \left|\frac{\partial}{\partial A^{uv}(j,c)} Z_{\phi_{u,j}}^{S \cup \{u\}}(B)\right|}} \\
    \leq{}& \frac{2\delta\Delta}{\tau(1-\delta)}. \qedhere
\end{align*}
\end{proof}

\begin{lemma}[Lemma 3.4 from \cite{BS17}]\label{lem:homangletoderiv}
Let $0 \leq \theta < 2\pi/3$, $\delta > 0$, and suppose $A \in \mathcal{U}(\delta)$. Let $S \subseteq V$, $\phi:S \rightarrow [q]$ be arbitrary. Assume the following hold:
\begin{enumerate}
    \item[(1)] $Z_{\phi_{u,j}}^{S \cup \{u\}}(A) \neq 0$ for every $u \in V \setminus S$ and every $j \in [q]$;
    \item[(2)] The angle between $Z_{\phi_{u,j}}^{S \cup \{u\}}(A)$ and $Z_{\phi_{u,k}}^{S \cup \{u\}}(A)$ in $\C$ is at most $\theta$, for every $u \in V \setminus S$ and every $j,k \in [q]$.
\end{enumerate}
Then for every $u \in S$, we have the lower bound
\begin{align*}
    |Z_{\phi}^{S}(A)| \geq \frac{\cos(\theta/2)}{\Delta} \sum_{v \sim u} \sum_{k \in [q]} |A^{uv}(\phi(u), k)| \cdot \left|\frac{\partial}{\partial A^{uv}(\phi(u),k)} Z_{\phi}^{S}(A)\right|.
\end{align*}
\end{lemma}
\begin{proof}
If $v \in S$ as well, then there is a unique $k \in [q]$ for which $\frac{\partial}{\partial A^{uv}(\phi(u),k)} Z_{\phi}^{S}(A) \neq 0$, namely $k = \phi(v)$. In this case, $A^{uv}(\phi(u),k) \cdot \frac{\partial}{\partial A^{uv}(\phi(u),k)} Z_{\phi}^{S}(A) = Z_{\phi}^{S}(A)$. Otherwise, $v \notin S$ and $\frac{\partial}{\partial A^{uv}(\phi(u),k)} Z_{\phi}^{S}(A) = \frac{1}{A^{uv}(\phi(u),k)} \cdot Z_{\phi_{v,k}}^{S \cup \{v\}}(A)$, where $\phi_{v,k}$ is the unique extension of $\phi$ mapping $v$ to $k$.

Combining these two observations, we obtain
\begin{align*}
    &\sum_{v \sim u} \sum_{k \in [q]} |A^{uv}(\phi(u), k)| \cdot \left|\frac{\partial}{\partial A^{uv}(\phi(u),k)} Z_{\phi}^{S}(A)\right| \\
    ={}& |N(u) \cap S| \cdot |Z_{\phi}^{S}(A)| + \sum_{v \sim u : v \notin S} \sum_{k \in [q]} |Z_{\phi_{v,k}}^{S \cup \{v\}}(A)| \\
    \leq{}& |N(u) \cap S| \cdot |Z_{\phi}^{S}(A)| + \frac{1}{\cos(\theta/2)}\underset{= |N(u) \setminus S| \cdot |Z_{\phi}^{S}(A)|}{\underbrace{\left|\sum_{v \sim u : v \notin S} \sum_{k \in [q]} Z_{\phi_{v,k}}^{S \cup \{v\}}(A)\right|}} \tag{\cref{lem:angle}} \\
    \leq{}& \frac{\Delta}{\cos(\theta/2)} \cdot |Z_{\phi}^{S}(A)|.
\end{align*}
Rearranging yields the desired result.
\end{proof}

With these lemmas in hand, we can now prove the main zero-free result.
\begin{proof}[Proof of Theorem~\ref{thm:zerosgraphhom}]
Let $0 < \theta < 2\pi/3$ be a parameter to be determined later, set $\tau = \cos(\theta/2)$, and let $\delta > 0$ satisfy $\theta = \frac{2\delta\Delta}{\tau(1-\delta)}$; in particular, $\delta = \frac{\frac{1}{2\Delta}\theta\cos(\theta/2)}{1 + \frac{1}{2\Delta}\theta\cos(\theta/2)}$. We show by descending induction on $|S|$ that the following three statements are all true:
\begin{enumerate}
    \item[(i)] For every $S \subseteq V$, $\phi:S \rightarrow [q]$ and $A \in \mathcal{U}(\delta)$, we have $Z_{\phi}^{S}(A) \neq 0$.
    \item[(ii)] For every $S \subseteq V$, $u \in V \setminus S$, $\phi:S \rightarrow [q]$, $A \in \mathcal{U}(\delta)$ and $j,k \in [q]$, the angle between $Z_{\phi_{u,j}}^{S \cup \{u\}}(A)$ and $Z_{\phi_{u,k}}^{S \cup \{u\}}(A)$ in $\C$ is at most $\theta$.
    \item[(iii)] For every $S \subseteq V$, $u \in S$, $A \in \mathcal{U}(\delta)$, we have the inequality
    \begin{align*}
        |Z_{\phi}^{S}(A)| \geq \frac{\cos(\theta/2)}{\Delta} \sum_{v \sim u} \sum_{k \in [q]} |A^{uv}(\phi(u), k)| \cdot \left|\frac{\partial}{\partial A^{uv}(\phi(u),k)} Z_{\phi}^{S}(A)\right|.
    \end{align*}
\end{enumerate}
The base case $S = V$ is easily verified since $Z_{\phi}^{S}(A) = \prod_{uv \in E} A^{uv}(\phi(u),\phi(v))$, a product of nonzero complex numbers.

Now, let $S \subseteq V$ with $|S| < |V|$.
\begin{description}
    \item[Proof of (i)] Let $u \in V \setminus S$, which exists since $|S| < |V|$. It follows that (i) holds for $S \cup \{u\}$ by the inductive hypothesis. Since $Z_{\phi}^{S}(A) = \sum_{k \in [q]} Z_{\phi_{u,k}}^{S \cup \{u\}}(A)$, \cref{lem:angle} applied to $Z_{\phi}^{S}(A)$ yields (i) assuming that (ii) holds. We prove (ii) below.
    \item[Proof of (ii)] Let $u \in V \setminus S$, which exists since $|S| < |V|$. Then (i) and (iii) hold for $S \cup \{u\}$ by the inductive hypothesis. (ii) then follows by \cref{lem:homderivtoangle}.
    \item[Proof of (iii)] Let $u \in S$. Then (i) holds for $S \cup \{u\}$ by the inductive hypothesis. Since (ii) holds for $S$ (as proved earlier), we may then apply \cref{lem:homangletoderiv}, yielding (iii) for $S$. 
\end{description}
Now, we choose $0 < \theta < 2\pi/3$. As we wish to maximize the size of our zero-free region, namely $\delta$, we need to maximize $\theta\cos(\theta/2)$. As shown in \cite{Reg18}, the maximum is attained when $2/\theta = \tan(\theta/2)$, which has solution $\theta^{*} \approx 1.72067$ and has objective value $x^{*} = \theta^{*}\cos(\theta^{*}/2) \approx 1.12219$. This yields $\delta = \frac{\frac{x^{*}}{2}}{\Delta + \frac{x^{*}}{2}}$ as claimed.
\end{proof}

\subsection{Proofs for Tensor Network Contractions}\label{subapp:zerofreetensor}
Our goal in this subsection is to prove \cref{thm:zerostensornetwork}, i.e., that the tensor network partition function
\begin{align*}
    Z_{\phi}^{F}(h) = \sum_{\substack{\sigma:E \rightarrow [q] \\ \sigma \mid_{F} = \phi}} \prod_{v \in V} h_{v}(\sigma \mid_{E(v)})
\end{align*}
is nonzero in a large polydisk around $1$, where $F \subseteq E, \phi:F \rightarrow [q]$, and we view $Z_{\phi}^{F}(\cdot)$ as a polynomial with variables $\{h_{v}(\alpha)\}_{v,\alpha}$. We prove the following stronger result.

\begin{theorem}[Generalization of Theorem 6 from \cite{Reg18}]\label{thm:zerofreetensorprecise}
Let $G=(V,E)$ be a graph of maximum degree $\leq\Delta$. Then for every $F \subseteq E$, $\phi:F \rightarrow [q]$, $\eta > 0$, and $0 \leq \theta < 2\pi/3$, the function $Z_{\phi}^{F}(h)$ is nonzero whenever $h \in \prod_{v \in V} S_{v}(\delta,\eta)$, where
\begin{align*}
    S_{v}(\delta,\eta) = \left\{h_{v} : [q]^{E(v)} \rightarrow \C : \substack{|h_{v}(\alpha) - h_{v}(\beta)| < \delta, \;\forall \alpha,\beta :E(v) \rightarrow [q], \\[3pt] |h_{v}(\alpha)| \geq \eta, \;\forall \alpha:E(v) \rightarrow [q]}\right\}
\end{align*}
and $\delta = \eta \cdot \min\left\{1, \frac{\theta \cos(\theta/2)}{\Delta + 1}\right\}$.
\end{theorem}
Before we prove this result, let us see how this gives \cref{thm:zerostensornetwork}.
\begin{proof}[Proof of Theorem~\ref{thm:zerostensornetwork}]
Observe that $S_{v}(\delta,\eta)$ contains a disk around $1$ of radius $\min\{\delta/2,$ $1-\eta\}$. Using \cref{thm:zerofreetensorprecise} and given that $\delta = \eta \cdot \min\left\{1, \frac{\theta\cos(\theta/2)}{\Delta+1}\right\}$, where $0 < \theta < 2\pi/3$, our goal is to maximize $\theta\cos(\theta/2)$ over $0 < \theta < 2\pi/3$ to obtain the largest zero-free disk. As shown in \cite{Reg18}, this maximum is attained when $2/\theta = \tan(\theta/2)$, which has solution $\theta^{*} \approx 1.72067$ and has objective value $x^{*} = \theta^{*}\cos(\theta^{*}/2) \approx 1.12219$. Given this, to obtain the largest possible radius disk, we equalize $1 - \eta$ and $\delta/2 = \eta \cdot \frac{x^{*}}{2(\Delta + 1)}$. Solving, we obtain $\eta = \frac{1}{1 + \frac{x^{*}}{2(\Delta+1)}}$, yielding radius $\frac{\frac{x^{*}}{2(\Delta + 1)}}{1 + \frac{x^{*}}{2(\Delta + 1)}}$ as desired.
\end{proof}
It remains to prove \cref{thm:zerofreetensorprecise}. We will need the following lemmas to implement an inductive approach.
\begin{lemma}[Lemma 8 from \cite{Reg18}]\label{lem:derivativetoangle}
Let $\tau > 0$, $F \subseteq E$, $\phi:F \rightarrow [q]$ and $u\in V$ be arbitrary. Suppose for all $h \in \prod_{v \in V} S_{v}(\delta,\eta)$ and all $\psi:F \cup E(u) \rightarrow [q]$ extending $\phi$, the following hold:
\begin{enumerate}
    \item[(1)] $Z_{\psi}^{F \cup E(u)}(h) \neq 0$;
    \item[(2)] For all $v \in N(u) \cup \{u\}$, we have
    \begin{align*}
        |Z_{\psi}^{F \cup E(u)}(h)| \geq \tau \sum_{\substack{\alpha:E(v) \rightarrow [q] \\ \text{compatible with }\psi}} |h_{v}(\alpha)| \cdot \left|\frac{\partial}{\partial h_{v}(\alpha)} Z_{\psi}^{F \cup E(u)}(h)\right|.
    \end{align*}
\end{enumerate}
Then for all extensions $\psi,\tilde{\psi} : F \cup E(u) \rightarrow \C$ of $\phi$, the angle between $Z_{\psi}^{F \cup E(u)}$ and $Z_{\tilde{\psi}}^{F \cup E(u)}(h)$ in $\C$ is at most $\frac{\delta(\Delta+1)}{\tau\eta}$. 
\end{lemma}
\begin{proof}
By assumption (1), the relevant partition functions are nonzero, and so the logarithm is well-defined when applied to these partition functions and we may bound the angle between $Z_{\psi}^{F \cup E(u)}(h)$ and $Z_{\tilde{\psi}}^{F \cup E(u)}(h)$ by 
\begin{align}\label{eq:tensorcondlogdiff}
    \left|\log Z_{\psi}^{F \cup E(u)}(h) - \log Z_{\tilde{\psi}}^{F \cup E(u)}(h)\right|.
\end{align}
The strategy is to write $Z_{\tilde{\psi}}^{F \cup E(u)}(h)$ as $Z_{\psi}^{F \cup E(u)}(\tilde{h})$ for some $\tilde{h} \in \prod_{v \in V} S_{v}(\delta,\eta)$ which differs from $h$ by a small number of coordinates, and then apply the Fundamental Theorem of Calculus and assumption (2). Let $v \in V$. We consider three cases.
\begin{itemize}
    \item $\mathbf{v \notin N(u) \cup \{u\}}$: In this case, $\psi,\tilde{\psi}$ agree on $E(v)$ and so we may simply take $h_{v} = \tilde{h}_{v}$.
    \item $\mathbf{v \in N(u)}$: In this case, $\psi,\tilde{\psi}$ differ only on the single edge $uv$. If $\alpha:E(v) \rightarrow [q]$ agrees with $\psi$ on $uv$, then let $\alpha':E(v) \rightarrow [q]$ be given by replacing $\alpha(uv) = \psi(uv)$ with $\tilde{\psi}(uv)$, and take $\tilde{h}_{v}(\alpha) = h_{v}(\alpha')$. Otherwise, just set $\tilde{h}_{v}(\alpha) = h_{v}(\alpha)$. (Note that it does not really matter what we set $\tilde{h}_{v}(\alpha)$ to since $Z_{\psi}^{F \cup E(u)}(h)$ only has the term $h_{v}(\alpha)$ when $\alpha$ agrees with $\psi$ on $uv$. However, we wish to minimize the number of coordinates in which $h,\tilde{h}$ differ.)
    \item $\mathbf{v = u}$: In this case, just set $\tilde{h}_{v}(\psi\mid_{E(v)}) = h_{v}(\tilde{\psi} \mid_{E(v)})$ and $\tilde{h}_{v}(\alpha) = h_{v}(\alpha)$ for all $\alpha \neq \psi\mid_{E(v)}$. 
\end{itemize}
It is clear that $Z_{\tilde{\psi}}^{F \cup E(u)}(h) = Z_{\psi}^{F \cup E(u)}(\tilde{h})$. By the Fundamental Theorem of Calculus, we may upper bound \cref{eq:tensorcondlogdiff} by 
\begin{align*}
     &\max_{x \in \prod_{v \in V} S_{v}(\delta,\eta)} \sum_{v \in N(u) \cup \{u\}} \sum_{\substack{\alpha : E(v) \rightarrow [q] \\ \text{compatible with } \psi}} \left|\frac{\partial}{\partial h_{v}(\alpha)} \log Z_{\psi}^{F \cup E(u)}(x)\right| \cdot \left|h_{v}(\alpha) - \tilde{h}_{v}(\alpha)\right| \\
     \leq{}& \frac{\delta}{\eta} \max_{x \in \prod_{v \in V} S_{v}(\delta,\eta)} \underset{\leq \Delta + 1}{\underbrace{\sum_{v \in N(u) \cup \{u\}}}} \underset{\leq 1/\tau \text{ by assumption } (2)}{\underbrace{\sum_{\substack{\alpha : E(v) \rightarrow [q] \\ \text{compatible with } \psi}} |h_{v}(\alpha)| \cdot \frac{1}{|Z_{\psi}^{F \cup E(u)}(x)|} \cdot\left|\frac{\partial}{\partial h_{v}(\alpha)} Z_{\psi}^{F \cup E(u)}(x)\right|}} \tag{Definition of $S_{v}(\delta,\eta)$} \\
     \leq{}& \frac{\delta(\Delta + 1)}{\tau\eta}. \qedhere
\end{align*}
\end{proof}

\begin{lemma}[Lemma 9 from \cite{Reg18}]\label{lem:angletoderivative}
Let $0 \leq \theta < 2\pi/3$, $u \in V$, $F \subseteq E$ satisfying $F \supseteq E(u)$, and $\phi:F \rightarrow [q]$. Suppose for all $v \in N(u) \cup \{u\}$, all $h \in \prod_{v \in V} S_{v}(\delta,\eta)$, and all extensions $\psi,\tilde{\psi}:F \cup E(v) \rightarrow [q]$ of $\phi$, the following hold:
\begin{enumerate}
    \item[(1)] $Z_{\psi}^{F \cup E(v)}(h) \neq 0$;
    \item[(2)] The angle between $Z_{\psi}^{F \cup E(v)}(h)$ and $Z_{\tilde{\psi}}^{F \cup E(v)}(h)$ in $\C$ is at most $\theta$.
\end{enumerate}
Then for all $v \in N(u) \cup \{u\}$ and all $h \in \prod_{v \in V} S_{v}(\delta,\eta)$, we have
\begin{align*}
    |Z_{\phi}^{F}(h)| \geq \cos(\theta/2) \sum_{\substack{\alpha:E(v) \rightarrow [q] \\ \text{compatible with }\phi}} |h_{v}(\alpha)| \cdot \left|\frac{\partial}{\partial h_{v}(\alpha)} Z_{\phi}^{F}(h)\right|.
\end{align*}
\end{lemma}
\begin{proof}
The conclusion is trivially true if $v = u$, since by the assumption $E(u) \subseteq F$, there is only one $\alpha :E(v) \rightarrow [q]$ compatible with $\phi$, namely $\phi\mid_{E(u)}$ itself. In this case, $h_{v}(\alpha)$ divides $Z_{\phi}^{F}(h)$ and we can replace $\cos(\theta/2)$ by $1$.

Suppose $v \in N(u)$. Since $Z_{\phi}^{F}(h) = \sum_{\substack{\psi:F \cup E(v) \rightarrow [q] \\ \psi\mid_{F} = \phi}} Z_{\psi}^{F \cup E(v)}(h)$, assumptions (1) and (2) make \cref{lem:angle} applicable, yielding
\begin{align*}
    |Z_{\phi}^{F}(h)| &\geq \cos(\theta/2) \sum_{\substack{\psi:F \cup E(v) \rightarrow [q] \\ \psi\mid_{F} = \phi}} |Z_{\psi}^{F \cup E(v)}(h)| \\
    &= \cos(\theta/2) \sum_{\substack{\alpha:E(v) \rightarrow [q] \\ \text{compatible with }\psi}} |h_{v}(\alpha)| \cdot \left|\frac{\partial}{\partial h_{v}(\alpha)} Z_{\phi}^{F}(h)\right|
\end{align*}
as desired.
\end{proof}
With these lemmas in hand, we may now proceed with the proof of \cref{thm:zerofreetensorprecise}.
\begin{proof}[Proof of Theorem~\ref{thm:zerofreetensorprecise}]
Let $\eta > 0$ and $0 \leq \theta < 2\pi/3$ be arbitrary, and take $\tau = \cos(\theta/2)$, $\delta = \eta \cdot \min \left\{1, \frac{\theta\tau}{\Delta + 1}\right\}$. We show by descending induction on $|F|$ that the following three statements are all true:
\begin{enumerate}
    \item[(i)] For every $F \subseteq E$, $\phi:F \rightarrow [q]$ and $h \in \prod_{v \in V} S_{v}(\delta,\eta)$, we have $Z_{\phi}^{F}(h) \neq 0$.
    \item[(ii)] For every $F \subseteq E$, $u \in V$, $\phi:F \rightarrow [q]$, $h \in \prod_{v \in V} S_{v}(\delta,\eta)$ and $\psi,\tilde{\psi}:F \cup E(u) \rightarrow [q]$ extending $\phi$, the angle between $Z_{\psi}^{F \cup E(u)}(h)$ and $Z_{\tilde{\psi}}^{F \cup E(u)}(h)$ in $\C$ is at most $\theta$.
    \item[(iii)] For every $F \subseteq E$, $u \in V$ satisfying $E(u) \subseteq F$, $\phi:F \rightarrow [q]$, $h \in \prod_{v \in V} S_{v}(\delta,\eta)$ and $v \in N(u) \cup \{u\}$, we have the inequality
    \begin{align*}
        |Z_{\phi}^{F}(h)| \geq \cos(\theta/2) \sum_{\substack{\alpha:E(v) \rightarrow [q] \\ \text{compatible with } \phi}} |h_{v}(\alpha)| \cdot \left|\frac{\partial}{\partial h_{v}(\alpha)} Z_{\phi}^{F}(h)\right|.
    \end{align*}
\end{enumerate}
The base case $F = E$ is easily verified since $Z_{\phi}^{F}(h) = \prod_{v \in V} h_{v}(\phi \mid_{E(v)})$, a product of nonzero complex numbers.

Now, let $F \subseteq E$ with $|F| < |E|$.
\begin{description}
    \item[Proof of (i)] Let $v \in V$ with $E(v) \not\subseteq F$. Since $|F \cup E(v)| > |F|$, (i) holds for $F \cup E(v)$ by the inductive hypothesis. Since $Z_{\phi}^{F}(h) = \sum_{\substack{\psi:F \cup E(v) \rightarrow [q] \\ \psi\mid_{F} = \phi}} Z_{\psi}^{F \cup E(v)}(h)$, \cref{lem:angle} applied to $Z_{\psi}^{F \cup E(v)}(h)$ yields (i) assuming that (ii) holds. We prove (ii) below.
    \item[Proof of (ii)] Let $u \in V$ and $\phi:F \rightarrow [q]$. If $E(u) \subseteq F$, then the claim is trivially true since $\psi = \tilde{\psi} = \phi$. Otherwise, assume $E(u) \not\subseteq F$ and let $\psi,\tilde{\psi}:F \cup E(u) \rightarrow [q]$ extend $\phi$. Since $|F \cup E(u)| > |F|$, (i) and (iii) hold for $F \cup E(u)$ by the inductive hypothesis. Applying \cref{lem:derivativetoangle} to $F \cup E(u)$ then yields (ii).
    \item[Proof of (iii)] Let $u \in V$ with $E(u) \subseteq F$. Without loss of generality, we may assume such an $u$ exists since otherwise, there is nothing to prove. Let $v \in N(u) \cup \{u\}$. If $E(v) \subseteq F$, then (iii) trivially holds with $\cos(\theta/2)$ replaced by $1$, since there is only one term in the summation, namely $\alpha = \phi\mid_{E(v)}$. Hence, assume $E(v) \not\subseteq F$. In this case, $|F \cup E(v)| > |F|$ and so (i) holds for $F \cup E(v)$ by the inductive hypothesis. Since (ii) for $F$ holds (as proved earlier), we may then apply \cref{lem:angletoderivative}, yielding (iii) for $F$. \qedhere
\end{description}
\end{proof}

\section{Proofs of Technical Lemmas}\label{app:technicallemmas}

\begin{proof}[Proof of  Lemma~\ref{lem:C-nz-bound}]
By definition we have
\[
\dist\left( 1, \mathcal{C}_{v,\spb}^\tau \right) = \inf_{1 \neq z \in \Gamma_{v,\spb}} \left| -\frac{1}{p_{v,\spb}^\tau (z-1)}  - 1 \right|.
\]
If $\Gamma_{v,\spb}$ is unbounded, then there exists a sequence $\{z_n\}$ such that $1\neq z_n \in \Gamma_{v,\spb}$ and $\lim_{n \to \infty} |z_n| = \infty$. 
Therefore,
\[
\dist\left( 1, \mathcal{C}_{v,\spb}^\tau \right) \le \liminf_{n \to \infty} \left| -\frac{1}{p_{v,\spb}^\tau (z_n-1)}  - 1 \right| 
\le 1 + \liminf_{n \to \infty} \frac{1}{p_{v,\spb}^\tau |z_n-1|} = 1.
\]
This shows the first part.

For the second part, observe that $\alpha_{v,\spb} < 1 < \beta_{v,\spb}$ since $\Gamma_{v,\spb}$ is open and $1 \in \Gamma_{v,\spb}$. 
Hence, we obtain
\[
\dist\left( 1, \mathcal{C}_{v,\spb}^\tau \right) 
\le \inf_{x \in \Gamma_{v,\spb} \cap (0,1)} \left| -\frac{1}{p_{v,\spb}^\tau (x-1)}  - 1 \right| 
= \frac{1}{p_{v,\spb}^\tau (1-\alpha_{v,\spb})} - 1
= \frac{\alpha_{v,\spb}}{p_{v,\spb}^\tau (1-\alpha_{v,\spb})} + \frac{1-p_{v,\spb}^\tau}{p_{v,\spb}^\tau},
\]
and also
\[
\dist\left( 1, \mathcal{C}_{v,\spb}^\tau \right) 
\le \inf_{x \in \Gamma_{v,\spb} \cap (1,\infty)} \left| -\frac{1}{p_{v,\spb}^\tau (x-1)}  - 1 \right|
= \frac{1}{p_{v,\spb}^\tau (\beta_{v,\spb}-1)} + 1.
\]
The second part follows.
\end{proof}

\begin{proof}[Proof of Lemma~\ref{lem:C-zero-bound}]
By definition we have
\[
\dist\left( 1, \mathcal{C}_{v,0}^\tau \right) = \inf_{1 \neq z \in \Gamma_v} \left| \frac{z}{p_{v,0}^\tau (z-1)} - 1 \right|.
\]
If $0 \in \closure{\Gamma_v}$, then there exists a sequence $\{z_n\}$ such that $1\neq z_n \in \Gamma_{v,\spb}$ and $\lim_{n \to \infty} z_n = 0$. 
Therefore,
\[
\dist\left( 1, \mathcal{C}_{v,0}^\tau \right) \le \liminf_{n \to \infty} \left| \frac{z_n}{p_{v,0}^\tau (z_n-1)} - 1 \right| 
\le 1 + \liminf_{n \to \infty} \frac{|z_n|}{p_{v,0}^\tau |z_n-1|} = 1.
\]
This shows the first part.

For the second part, observe that $\alpha_v < 1 < \beta_v$ since $\Gamma_v$ is open and $1 \in \Gamma_{v,\spb}$. 
Hence, we obtain
\[
\dist\left( 1, \mathcal{C}_{v,0}^\tau \right) 
\le \inf_{x \in \Gamma_v \cap (0,1)} \left| \frac{x}{p_{v,0}^\tau (x-1)} - 1 \right| 
= \frac{\alpha_v}{p_{v,0}^\tau (1-\alpha_v)} + 1,
\]
and also
\[
\dist\left( 1, \mathcal{C}_{v,0}^\tau \right) 
\le \inf_{x \in \Gamma_v \cap (1,\infty)} \left| \frac{x}{p_{v,0}^\tau (x-1)} - 1 \right| 
= \frac{\beta_v}{p_{v,0}^\tau (\beta_v-1)} - 1
= \frac{1}{p_{v,0}^\tau (\beta_v-1)} + \frac{1-p_{v,0}^\tau}{p_{v,0}^\tau}.
\]
The second part follows.
\end{proof}

\begin{proof}[Proof of Lemma~\ref{lem:leftplaneprod}]
It was shown in \cite{GLLZ21} that 
\[
\Gamma = \left( - \closure{\HP}_\eps^2 \right)^\ccomp 
= \left\{\rho e^{i\theta} : \rho(1-\cos\theta) < 2\eps^2,\, 0 \le \theta < 2\pi\right\}.
\]
To make this more interpretable, we rewrite the set in Cartesian coordinates. If $z = \rho e^{i\theta}$, then by Euler's formula we may write $z = x + iy$ where $x = \rho \cos\theta$ and $y = \rho \sin\theta$. 
We then obtain
\begin{align*}
    & \rho(1-\cos\theta) < 2\eps^2 \\
    \Longleftrightarrow\qquad & \rho < x + 2\eps^2 \\
    \Longleftrightarrow\qquad & x^2 + y^2 < (x + 2\eps^2)^2 \\
    \Longleftrightarrow\qquad & y^2 < 4\eps^2(x+\eps^2).
\end{align*}
Therefore, we see that
\[
\Gamma = \left\{ x + iy: y^2 < 4\eps^2(x+\eps^2) \right\},
\]
which clearly contains $\R_{+}$.

Furthermore, for $\lambda \in \R_+$ we have
\begin{align*}
    \dist(\lambda, \boundary{\Gamma}) 
    &= \inf_{z \in \boundary{\Gamma}} |z - \lambda| \\
    &= \inf_{(x,y) \in \R^2:\, y^2 = 4\eps^2(x+\eps^2)} \sqrt{ (x-\lambda)^2 + y^2 } \\
    &= \inf_{x \in [-\eps^2, \infty)} \sqrt{ (x-\lambda)^2 + 4\eps^2(x+\eps^2) } \\
    &= 
    \begin{cases}
    \lambda + \eps^2, & \lambda \in (0, \eps^2); \\
    2\eps \sqrt{\lambda}, & \lambda \in [\eps^2, \infty).
    \end{cases}
\end{align*}
This establishes the lemma.
\end{proof}

\end{document}